\documentclass[aps,prd,longbibliography, twocolumn,lengthcheck,superscriptaddress,nopreprintnumbers,reprint]{revtex4-1}
\usepackage{natbib,units}
\usepackage{graphicx}
\usepackage{xcolor}
\usepackage{dcolumn}
\usepackage{bm}
\usepackage{amssymb}
\usepackage{amsmath}
\usepackage{verbatim}
\usepackage[utf8]{inputenc}
\usepackage{tikz,hyperref}
\usepackage{dcolumn}
\usepackage{amsmath,amssymb}
\usepackage{dsfont}
\usepackage{animate,media9}
\usepackage{slashed}
\usepackage{braket}
\usepackage{subfigure}
\usepackage{slashed}
\usepackage{hyperref}
\usepackage[section]{placeins}

\newcommand{\mnras}{MNRAS}
\newcommand{\aj}{Astrophys. J.}
\newcommand{\apjl}{ApJ Letters}

\newcommand{\jcap}{JCAP}

\newcommand{\physrep}{Physics Reports}

\begin{document}

\title{Exploring Self-Interacting Dark Matter Halos with Diverse Baryonic Distributions: \\ A Parametric Approach}

\author{Daneng Yang}
\email{yangdn@pmo.ac.cn}
\affiliation{Purple Mountain Observatory, Chinese Academy of Sciences, Nanjing 210023, China}
\affiliation{Department of Physics and Astronomy, University of California, Riverside, California 92521, USA}


\begin{abstract}
Galaxies residing in dark matter halos exert significant gravitational effects that alter halo structure and dynamics.
The complexity of these interactions escalates with the diversity of galactic structures and the variability in dark matter halo profiles under self-interacting dark matter (SIDM) models. 
This work extends the parametric model for dark matter-only halos presented in Ref.~\cite{yang:2023jwn} to incorporate baryons. 
We adapt this model to consistently represent the SIDM halo density profile over time, highlighting the role of a gravothermal phase in characterizing the state of an SIDM halo. 
Given this phase, the density profile in SIDM is determined by a fictitious progenitor --- consisting of an NFW halo influenced by a baryonic potential --- that has evolved to its present state. 
In the temporal dimension, the model incorporates a form factor that rescales the evolution time in the dark matter-only case, thereby enabling the introduction of a universal phase.
In the radial dimension, the halo density profile is parametrized to reflect the influences of baryons. 
We calibrate the model through N-body simulations with baryon potentials to fit various stellar-to-halo mass ratios and size-mass relationships. 
Our parametric approach is numerically efficient, enabling the exploration of SIDM effects across a diverse set of halos, as exemplified by a case study using an illustrative sample that spans five orders of magnitude in the mass range.
We also demonstrate that the effects of evolution history and the specific SIDM model can be separated from the current states of galaxies and halos, leaving the task of identifying consistent SIDM models to dedicated post-processing analyses. 
\end{abstract}

\maketitle
\tableofcontents

\section{Introduction}

Cold dark matter (CDM) aligns well with observations at scales larger than megaparsecs. However, discrepancies arise at smaller scales in the inner structures of dark matter (DM) halos, where the density profiles can be either cored or cuspy~\cite{1994ApJ...427L...1F,spergel9909386,deBlok:2001hbg,Gentile:2004tb,2006ApJ...652..306S,2009MNRAS.397.1169D,denaray09123518,2011AJ....141..193O,oman150401437,Salucci:2018hqu,2019MNRAS.490.5451D,DeLaurentis:2022nrv}. 
The cored cases may exhibit fainter or more diffuse profiles than most models predicted, while the cuspy cases can display steeper inner-density slopes.
To explain these variations, researchers have elaborated on both baryonic feedback models and self-interacting dark matter (SIDM) models. Baryonic effects diversify the inner halo structures because it alters the structure by leading to compact stellar distributions through dissipation while simultaneously dispersing these distributions through feedback. The effectiveness of these processes varies with scale and depends on complex mechanisms including star formation and supernova feedback~\cite{Navarro:1996bv,Gnedin:2001ec,Read:2004xc,2010Natur.463..203G,DiCintio:2014xia,Chan:2015tna,2016MNRAS.456.3542T,2020mnras.497.2393l,Maccio:2020svl,2020mnras.495...58s}. Similarly, SIDM contributes to diversity in inner halo structures through elastic scatterings that facilitate energy transfer within halos. This results in core formation and core collapse stages in the thermodynamic evolution of self-gravitating halos, known as gravothermal evolution~\cite{Tulin170502358,Adhikari220710638,Spergel:1999mh,Kochanek:2000pi,kamada:2016euw,ren180805695,santos-santos191109116,spergel9909386,correa:2022dey,yang:2022mxl,balberg0110561,balberg:2002ue,koda11013097,nishikawa:2019lsc,sameie:2019zfo,zavala:2019sjk,kaplinghat:2019svz,kahlhoefer:2019oyt,correa:2020qam,turner:2020vlf,slone:2021nqd,silverman:2022bhs,correa:2022dey,yang:2022mxl,minor:2020hic,yang:2021kdf,gilman:2021sdr,gilman:2022ida,nadler:2023nrd,pollack:2014rja,choquette:2018lvq,feng:2020kxv,feng210811967,meshveliani:2022rih,fischer:2023lvl,Dhanasingham:2023thg,Zhang:2024ggu,Tran:2024vxy}.
In probing SIDM, the diversity from baryon physics naturally integrate through gravitational interactions, resulting in a nonlinear and complex combined evolution. Nonetheless, understanding the interplay between baryons and dark matter is vital for making realistic SIDM predictions, as we depend on visible baryonic matter to probe the elusive dark matter structures~\cite{Vogelsberger:2014pda,Robertson:2018anx,Robertson:2020pxj,yang:2021kdf,Rose:2022mqj,Mastromarino:2022hwx,Rahimi:2022ymu,zhong:2023yzk,2017MNRAS.468.2283C,Fry:2015rta,kong:2024zyw,Fischer:2023eet}. 

To date, there have been only a handful of hydrodynamical cosmological simulations that incorporate SIDM physics~\cite{Robertson:2017mgj,2017MNRAS.472.2945R,2018ApJ...853..109E,despali:2018zpw,Robertson:2018anx}. These simulations have revealed a complex interplay dependent on mass, galaxy type, and accretion history. 
Meanwhile, theoretical studies have used conducting fluid models, semi-analytic models, and idealized N-body simulations to model the combined effects of a baryonic potential and gravothermal evolution~\cite{balberg:2002ue,koda11013097,essig:2018pzq,nishikawa:2019lsc,feng:2020kxv,yang:2022zkd,jiang:2022aqw,yang:2023stn,zhong:2023yzk,Gad-Nasr:2023gvf}. 
It has been illuminated that galaxy formation can adiabatically contract dark matter, increase the DM scattering rate, and thus accelerate the gravothermal evolution. Some empirical relations have also been proposed to enhance our understanding of these complex processes~\cite{elbert:2016dbb,sameie:2018chj,feng:2020kxv,yang:2023stn,zhong:2023yzk}.

In this study, we explore the gravothermal evolution of DM halos induced by SIDM in the presence of baryons, utilizing a parametric approach. Within this framework, each halo is posited to reside in a gravothermal state characterized by a dimensionless {\it gravothermal phase}, $\tau$, defined as the ratio of the elapsed time since formation to the core collapse time ($t/t_c$). 
The halo's density profile is then determined from parameters that describe its characteristics under the CDM scenario, accounting for both halo and baryonic properties. 
The model for the density profile is parametric, with time-dependent parameters represented as functions of $\tau$. These parameters are calibrated against high-resolution N-body simulations, as detailed in Ref.~\cite{yang:2023jwn}.
This method provides an efficient way to achieve accurate results. It is founded on the universal characteristics of gravothermal evolution and integrates methods to manage differential SIDM cross sections and halos with realistic accretion histories~\cite{outmezguine:2022bhq,yang:2023jwn,yang220503392,yang:2022zkd}. 

Building upon the model of Ref.~\cite{yang:2023jwn} for dark matter-only halos, we derive and calibrate a new formula for the core collapse time, making it applicable for a broad range of stellar-to-halo mass and size ratios. 
We also introduce procedures to modify the SIDM density profile in DM-only scenarios to account for the presence of baryons.
Unlike the DM-only case, which has an initial Navarro-Frenk-White (NFW) profile~\cite{1997apj...490..493n}, our new profile is based on the Dekel-Zhao (DZ) profile~\cite{Zhao:1995cp,2017MNRAS.468.1005D,2020MNRAS.499.2912F}, which has been shown applicable for halos in hydrodynamical simulations, with the effects of stellar feedback and adiabatic contraction accounted for. 
We validate the model through eight controlled N-body simulations, each employing distinct yet representative sets of baryon and halo parameters. 

\begin{figure}[htbp]
  \centering
  \includegraphics[width=8cm]{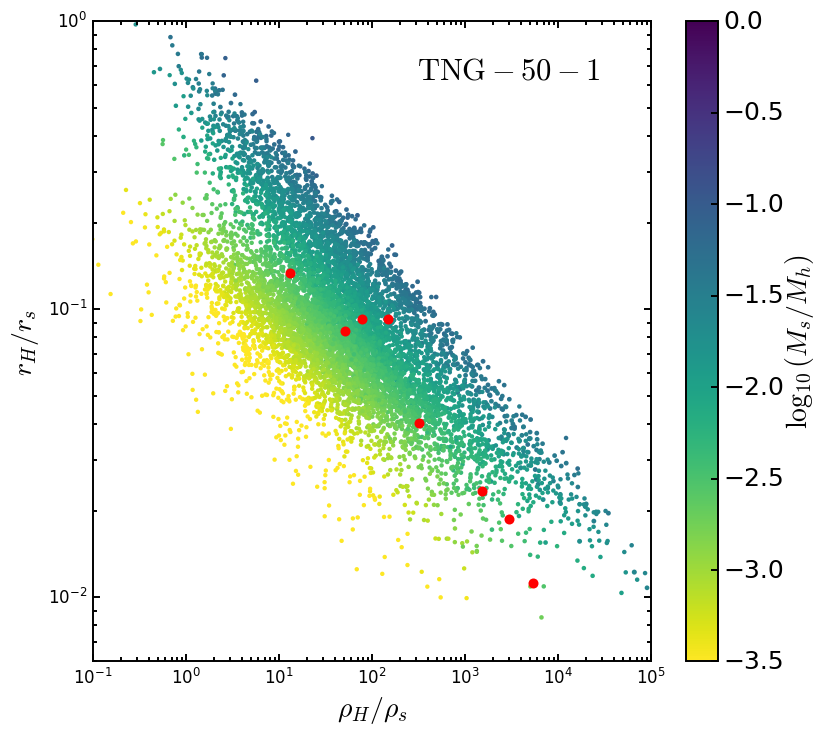}
  \caption{\label{fig:TNG} The diverse baryonic structures in isolated halos from the TNG-50-1 simulation, illustrated by the distribution of $\rho_H/\rho_s$ versus $r_H/r_s$~\cite{nelson:2018uso,nelson:2019jkf,pillepich:2019bmb}. Here, $\rho_s$ and $r_s$ represent the NFW scale density and radius of the halo component, respectively, while $\rho_H$ and $r_H$ denote the corresponding Hernquist parameters for the stellar component. The color coding represents the stellar-to-halo mass ratio on a logarithmic scale, indicating that higher mass halos tend to have higher $\rho_H/\rho_s$ values on average. The red dots correspond to the simulated benchmark halos listed in Table~\ref{tab:simbm}, which are used to calibrate the analytic model. }
\end{figure}

In Fig.~\ref{fig:TNG}, we highlight the diverse baryonic structures across a wide range of samples depicted on the plane of scale density ratio $\rho_H/\rho_s$ versus scale radius ratio $r_H/r_s$. Here, the halos and their stellar components are modeled using NFW and Hernquist~\cite{1990ApJ...356..359H} profiles, characterized by scale parameters $\rho_s$, $r_s$ for halos and $\rho_H$, $r_H$ for stellar components. The color codes on the plot correspond to the stellar-to-halo mass ratio, represented on a logarithmic scale, revealing a trend where higher $M_s/M_h$ halos move up along the $\ln(y)\propto \ln(x)$ direction, exhibiting larger $\rho_H/\rho_s$ and $r_H/r_s$ ratios. Red dots indicate benchmark halos from Table~\ref{tab:simbm}, utilized for validating the model. These halos cover a broad spectrum in the $\rho_H/\rho_s$ and $r_H/r_s$ plane, demonstrating the model's broad applicability.

\begin{table*}[htbp]
\begin{center}
\begin{tabular}{c|c|c|c|c|c|c|c|c|c}
\hline
\hline
Benchmark  & $M_h ({\rm M_{\odot}})$ & $\rho_s ({\rm M_{\odot}/kpc^3})$ & $r_s ({\rm kpc})$ & $M_s/M_h$  & $\rho_H/\rho_s$  &  $r_H/r_s$  & $\sigma/m$ ($\rm cm^2/g$)  & $t_{c,b}~\rm (Gyr)$  & $t_{c,b}/t_{c,0}$ \\ 
\hline
{\it DM11}           & $10^{11}$ & $6.89\times 10^6$ &  $9.1$       & 0                &  0         &  0         & 10 &    66  &  1.0       \\
{\it DM11+baryon}    & $10^{11}$ & $6.89\times 10^6$ &  $9.1$       & 0.010      & 51.7             &  0.0844     & 10 &    25 & 0.38      \\
{\it DM11+baryon R2} & $10^{11}$ & $6.89\times 10^6$ &  $9.1$       & 0.010      & 13.0             &  0.134      & 10 &    32 & 0.49       \\
{\it DM11+baryon M2} & $10^{11}$ & $6.89\times 10^6$ &  $9.1$       & 0.020      & 77.5             &  0.0929     & 10 &    12 & 0.18       \\
\hline
{\it DM12+baryon}   & $10^{12}$ & $1.23\times 10^7$ & $15.5$  & 0.034 &  149  & 0.0926  & 1 &  9.0 & 0.055     \\
{\it DM13 compact} &  $10^{13}$ & $1.43\times 10^7$ & 42.1  & 0.013   & 2972 & 0.0187  & 1 &  9.0  &  0.19   \\
{\it DM13+baryon}  & $2.2\times 10^{13}$ & $1.17\times 10^6$ & 115  & 0.010  & 1544  & 0.0234  & 1 &  173 & 0.23   \\
{\it DM13+baryon2}  & $3\times 10^{13}$ & $5.64\times 10^6$ & 67.1  & 0.0074  & 321  & 0.0402  & 1 &  42 & 0.35   \\
{\it DM13 extreme}     & $3\times 10^{13}$ & $7.80\times 10^6$ & 70.0  & 0.0043  & 5442 & 0.0112  & 1 &  27   &  0.38   \\
\hline
\end{tabular}
\caption{ Properties of N-body simulated benchmarks:  
The first four simulations, introduced in Ref.\cite{zhong:2023yzk}, correspond to the initial block of cases in Table 1 therein. The final two benchmarks, {\it DM13+baryon2} and {\it DM13 extreme} are analogous to J1636+4707 and JWST-ER1, as described in Ref.~\cite{kong:2024zyw}. 
Additionally, the {\it DM12+baryon}, {\it DM13 compact}, and {\it DM13+baryon} benchmarks represent further simulations conducted to broaden the range of benchmark scenarios. 
The core collapse times are computed considering $\sigma/m=10~\rm cm^2/g$ for the {\it DM11} cases in the first block and $\sigma/m=1~\rm cm^2/g$ for the {\it DM12} and {\it DM13} cases, to be consistent with the curves in Fig.~\ref{fig:valid2}. 
We use $t_{c,b}$ and $t_{c,0}$ to represent the core collapse times in the {\it DM+baryon} and {\it DM-only} scenarios. The halo masses are calculated as the mass enclosed within radii where the density is 200 times the critical density ($R_{\rm 200}$), using the scale density ($\rho_s$) and scale radius ($r_s$) of the NFW profile. The stellar masses are total masses computed using the scale density ($\rho_H$) and scale radius ($r_H$) of the Hernquist profile.
\label{tab:simbm} }
\end{center}
\end{table*}

Recognizing the mutual gravitational influence between dark matter and baryons, we employ the parametric model to investigate the conditions under which the effects of SIDM on baryons are pronounced.
Our findings indicate that SIDM can notably affect baryons in cases where the maximum core sizes exceed the baryons' scale radii. This core formation tends to expand the baryon distribution, whereas core collapse generally compresses it. Conversely, when the maximum core size is smaller than the baryon scale radius, the baryonic potential strongly suppresses distinct core formation within the halo. As a result, the baryon profile becomes more compact as the halo contracts. 
Our study on this topic is simplified; therefore, we leave the modeling of consistent coevolution of dark matter and baryons to future work.

To demonstrate the effectiveness of our model, we present two illustrative applications. We first apply the model to a population of sampled isolated halos, utilizing generic relationships between halos and galaxies, and focusing on the distribution of maximum circular velocity ($V_{\rm max}$) and its corresponding radius ($R_{\rm max}$). 
Next, we illustrate how the gravothermal evolution induced by SIDM is primarily captured in the gravothermal phases, which can, in principle, be reconstructed from data. 
The CDM and SIDM density profiles are linked via the gravothermal phases, and by reversing these phases to zero, the outcomes are expected to align with our existing knowledge of $\Lambda$CDM.
Notably, these gravothermal phases are independent of the halos' accretion history as well as the underlying SIDM model. By post-processing these results, we can discern SIDM interpretations based on stellar accretion histories, aiding in the identification of the preferred SIDM model that coherently explains all observations. 
These applications highlight the model's adaptability and effectiveness in addressing the complex interplay between SIDM halos and their baryonic counterparts.

We organize the paper as follows. 
Section \ref{sec:samples} provides a summary of the simulation samples utilized in this study. 
In Section \ref{sec:gravo}, we first discuss the theoretical basis for the parametric model, following that we introduce a core collapse time formula that incorporates baryons. 
Section \ref{sec:modeling} presents parametric SIDM models based on a contracted $\beta4$ and Cored-DZ profiles. The latter is more intricate but consistently incorporates the baryon-induced contraction for halos hosting massive galaxies. 
Section \ref{sec:valid} is dedicated to the validation of the proposed model.
In Section \ref{sec:sidmfeedback}, we investigate the effect of SIDM on the baryon distributions using the proposed model. 
We provide two example applications in Section \ref{sec:applications}: applying the parametric model to sampled halos, and a post-processing of the gravothermal state for the SIDM interpretations. 
Finally, Section \ref{sec:conclusion} summarizes the paper and discusses directions for future research.

\section{Simulation data}
\label{sec:samples}

As illustrated in Fig.~\ref{fig:TNG}, there is a significant diversity in the baryonic structures within halos. The range of the $\rho_H/\rho_s$ ratio, in particular, spans approximately five orders of magnitude. For a better presentation of physically relevant cases, we selected isolated halos with masses greater than $10^{10} \rm M_{\odot}$. This criterion is based on the consideration that the stellar-to-halo mass ratio reaches its peak at halos of around $10^{12} \rm M_{\odot}$, and the overall halo population increases towards the lower mass, inversely proportional to the square of the mass. 

Given this diversity, it is crucial to test our model against a broad spectrum of scenarios. To this end, we make use of eight N-body simulations where baryons are represented using a Hernquist potential. See Table~\ref{tab:simbm} for details of their properties. 
Although our sample size is limited, these examples span a substantial range in both $\rho_H/\rho_s$ and $r_H/r_s$ ratios, ensuring a comprehensive validation of our model across different baryonic structures and halo configurations. We depict these samples as red dots in Fig.~\ref{fig:TNG}.

Our simulations are idealized in the treatment of baryon potentials, which are fixed at all times and take the form of the Hernquist potential. 
While this setup overlooked baryonic dynamics, it fits into the context of the integral approach in Ref.~\cite{yang:2023jwn}, where each SIDM halo is assumed to arise from a fictitious CDM halo. 
In application, the effect of baryon dynamics, including its interplay with the gravothermal evolution, is encoded into the evolution of the potential and can be {\it consistently} incorporated in the integral approach, given the parametric model with baryons. 

Among the eight samples analyzed, four were previously utilized in Ref.~\cite{zhong:2023yzk}. In these simulations, each consists of $4\times 10^6$ particles, with a force softening length set to $h=0.13$ kpc. The Hernquist baryonic potentials in these simulations are introduced instantaneously at $t=0$ Gyr, and analyses have shown that this approach yields results comparable to those obtained by gradually developing the potential over 4 Gyr—a period significantly shorter than the core collapse time for the simulated halos.

The remaining samples are simulated first in CDM, with the baryon potential increased linearly in $\rho_H$ over 4 Gyr. The snapshots at $5$ Gyr are subsequently used as initial conditions for the SIDM simulations.  
The simulation parameters vary in the samples. 
The {\it DM13+baryon} and {\it DM13+baryon2} samples are simulated using $2\times 10^6$ particles with a force softening length $h=2.8~$kpc, while the other samples are simulated using $10^6$ particles with $h=0.42$~kpc.
The {\it DM13+baryon2} and {\it DM13 extreme} samples correspond to the J1636+4707 and JWST-ER1 systems, respectively, as detailed in Ref.~\cite{kong:2024zyw}. These samples have been re-simulated in this study to reach the deep core collapse phases.

\section{Normalized gravothermal evolution with baryons}
\label{sec:gravo}

In this section, we first review the theoretical rationale that supports the construction of parametric models. Following that, we present a new core collapse formula that incorporates the effects of baryons. 

\subsection{The universal gravothermal evolution}

We start with CDM halos that are spherically symmetric and isolated. Their density profiles are known to be well-described by the NFW profile, which can be written in terms of two scale parameters $\rho_s$ and $r_s$ as $\rho_{\rm NFW} = \rho_s/\left( (r/r_s) (1+r/r_s)^2 \right)$. This profile is self-similar in nature: all halos exhibit the same density profile once normalized by the $\rho_s$ and $r_s$. 
In SIDM, scatterings introduce an ``arrow of time'' which breaks the self-similarity of the profile. 
Intriguingly, such self-similarity is replaced by so-called ``universality'', after incorporating the time dimension into the modeling and with explicit SIDM dependence absorbed into the evolution time~\cite{outmezguine:2022bhq,yang:2023jwn,zhong:2023yzk}.

To see how this is achieved, we examine the three fluid equations describing the halo evolution: the continuity equation, the momentum conservation equation, and the transport equation.
In the context of SIDM, the continuity equation remains unaffected since the interactions are elastic. 
The momentum and transport equations will both be affected by viscosity through the viscous tensor, denoted as $\Pi_{ij}^{\rm vis}$, which is proportional to $\eta (\nabla_i u_j - \nabla_j u_i)$, with $\eta$ representing the viscosity coefficient and $\mathbf{u}$ denoting the average particle velocity in a small given volume.
However, in spherical halos that maintain isotropic velocity distributions, the average velocities in both polar and azimuthal directions become negligible, eliminating viscosity’s impact. 
Under these conditions, the transport equation emerges as particularly crucial and drives the gravothermal evolution.
This equation is typically represented as:
\begin{equation}
\frac{3\rho}{2 m} \frac{D T}{D t} = -\mathbf{\nabla}\cdot \mathbf{q} - P \mathbf{\nabla}\cdot \mathbf{u},
\end{equation}
where $D/Dt = (\partial/\partial t + \mathbf{u\cdot \nabla} )$ refers to the material derivative, $T \equiv m\nu^2$ denotes the ``temperature,'' $P \equiv \rho \nu^2$ indicates the ``pressure,'' and $\mathbf{q} = -\kappa \mathbf{\nabla} T$ describes the heat flux, linked to thermal conductivity $\kappa$.
In the spherical coordinates, this equation simplifies to
\begin{eqnarray}
\label{eq:trans}
\frac{\partial}{\partial r} \left( r^2 \kappa m  \frac{\partial \nu^2}{\partial r} \right) = r^2 \rho \nu^2 \frac{D}{D t} \ln \frac{\nu^3}{\rho}, 
\end{eqnarray}
where the left-hand-side depends on SIDM through the heat conductivity $\kappa$, and the right-hand-side introduces evolution through the material derivative $D/Dt$.
Suppose that $\kappa$ depends on the scattering cross section per mass through a function $f(\sigma/m)$, which has no $r$ or $t$ dependencies,
one can absorb the SIDM dependence into the evolution time by redefining $t$ as $t f(\sigma/m)$.
Notably, the continuity and the momentum equations primarily represent mass conservation and hydrostatic equilibrium, both of which are inherently independent of time.
Therefore, the evolution of a dark matter halo's density profile, starting from an initial NFW profile, follows a unique trajectory. 
This uniqueness implies a universal solution for the gravothermal evolution of dark matter halos in SIDM.

We are particularly interested in how the the conductivity in SIDM halos depends on the cross section $\sigma/m$. 
While dark matter halos can be modeled using fluid equations, their properties differ from ordinary fluid in several crucial aspects, among which one of the most important differences is the sparsity in the number of scatterings. In the bulk region of realistic halos, the scattering rate is suppressed by the dark matter density, making their mean-free-path longer than the scale height $H=\sqrt{\nu^2/(4\pi G\rho)}$ of the halos.
Consequently, a post-scattering particle can orbit around the halo center multiple times before colliding again.
It implies that the energy transferred through the collision is diluted by the orbital evolution, reducing the effect of heat conduction.
In the literature, this scenario is called the long-mean-free-path (LMFP) regime, where the heat conductivity is proportional to the scattering cross section and can be expressed as~\cite{pollack:2014rja,choquette:2018lvq,feng:2020kxv,feng210811967}
\begin{eqnarray}
\kappa_{\rm LMFP} \propto \frac{\sigma}{m} \frac{\rho \nu^3}{m G}.
\end{eqnarray}
This can be contrasted to the ordinary fluid, where the short-mean-free-path (SMFP) inhibits heat conduction, resulting in a conductivity $\kappa_{\rm SMFP}\propto \nu/\sigma$, being inversely proportional to the cross section.
We see that in both the LMFP and SMFP regimes, the conductivities have simple dependences on the scattering cross section and universality can be achieved. 

In practice, SIDM halos that conform to astrophysical constraints predominantly exist within the LMFP regime across most of their volume and throughout much of their evolutionary history. Gravothermal core collapse can intensify their inner densities, potentially initiating the SMFP regime; however, this occurs in the inner regions, which largely evolve independently of the outer regions. 
These decoupled inner regions exhibit a self-similar density profile described by $r^{-u}$ with $u$ ranges between $2$ and $2.5$. Numerical simulations suggest a preferred value of $u=2.21$~\cite{Lynden-Bell:1980xip}.

\subsection{A universal form factor for normalizing the evolution time with baryons}

We have demonstrated the universality in the evolution of DM-only halos. 
In the presence of baryons, the contraction in response to the potential necessarily distorts the halo density profiles, violating the universality in the spatial dimension. 
Despite this, the underlying principle that the effects of SIDM can be integrated into the evolution timeline remains valid. 

We assume baryonic components, particularly stellar ones, interact predominantly through gravity, blending with dark matter in phase space to achieve a quasi-hydrostatic equilibrium. 
In this setup, the fluid equations are still applicable, with the density and velocity dispersion including contributions from both dark matter and baryons. 
However, because the baryons do not participate in this self-scatterings, the heat conductivity is adjusted to account only for the dark matter density. 
This adjustment does not affect the form of Eq.~(\ref{eq:trans}) and the SIDM dependence of the conductivity can still be absorbed into the evolution time through the same redefinition.
In the LMFP regime, we have
\begin{eqnarray}
\label{eq:transb}
\frac{\partial}{\partial r} \left( r^2 \frac{\rho_{\rm DM} \nu_{\rm tot}^3}{ G}   \frac{\partial \nu_{\rm tot}^2}{\partial r} \right) \propto r^2 \rho_{\rm tot} \nu_{\rm tot}^2 \frac{D}{D (t \frac{\sigma}{m})} \ln \frac{\nu_{\rm tot}^3}{\rho_{\rm tot}}, 
\end{eqnarray}
where $\rho_{\rm DM}$ refers to the dark matter density and $\rho_{\rm tot}$, $\nu_{\rm tot}$ represent the total density and velocity dispersion of the baryon and dark matter components.

\begin{figure}[htbp]
  \centering
  \includegraphics[width=8.2cm]{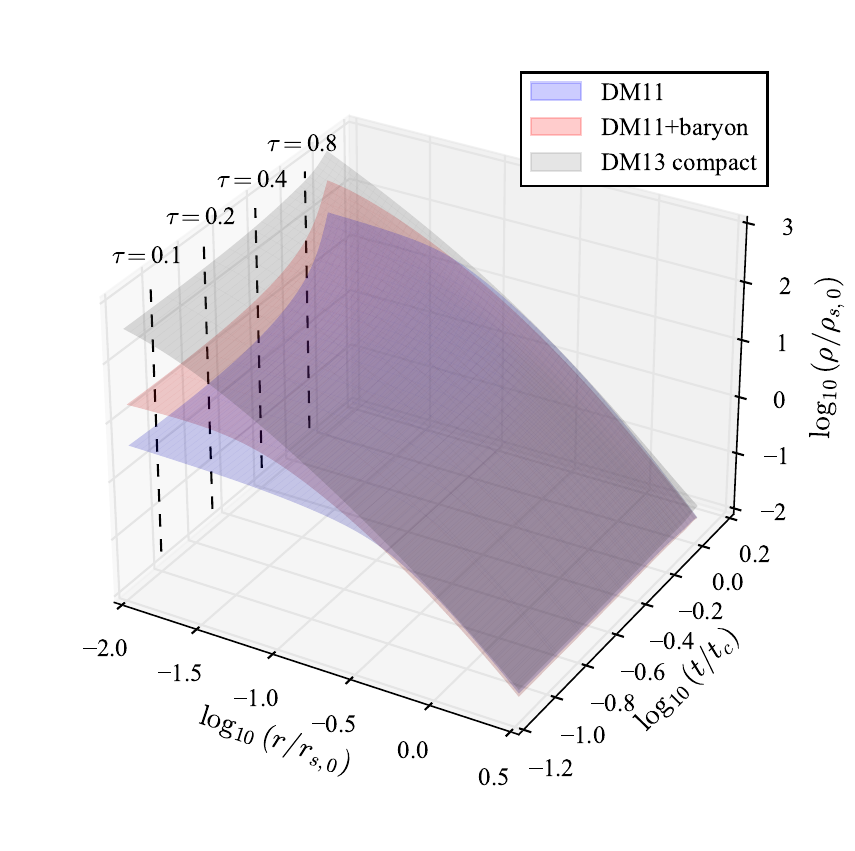}
  \caption{\label{fig:univ} The unbroken universality in the evolution timeline with baryons, depicted through the evolution of normalized density profiles.
  For clarity, three benchmark cases---DM11 (blue), DM11+baryon (red), and DM13 compact (gray)---from Table~\ref{tab:simbm} are illustrated to demonstrate the predictions of the parametric model.
  The $\rho_{s,0}$ and $r_{s,0}$ correspond to the initial NFW scale density and radius. The core collapse time $t_c$ is introduced in Eqs.~(\ref{eq:tcdmo}) and (\ref{eq:tcb}) for the dark matter-only case and dark matter plus baryon case, respectively. 
  All the cases evolve into a fixed configuration as $t/t_c$ approaching $1.1$. For more details, refer to Section~\ref{sec:valid} and Fig.~\ref{fig:valid2}.  }
\end{figure}

Before delving into the specifics, we illustrate in Fig.~\ref{fig:univ} the aspects just discussed through the normalized evolution of density profiles with and without the presence of baryons. All three axes are normalized to depict universality and its breaking clearly. 
From this figure, it is evident that all scenarios follow a unified evolution timeline, represented as $t/t_c$ and ranging from $0$ to $1.1$. We define each point along this timeline as a ``gravothermal phase.''
The density profiles, conversely, vary according to the amount of baryonic matter and show different deviations from the DM-only scenario. Subsequently, we will discuss methods to normalize the evolution time and adjust the DM density profile for a parametric model.

Equation~(\ref{eq:transb}) suggests that the SIDM dependence can be absorbed into the evolution time. Identifying an equation for the core collapse time would allow us to normalize this timeline effectively. We introduce the core collapse time equation here and will discuss the modeling of density profiles in the following section.

In the literature, the core collapse time for dark matter-only halos with NFW profiles has been estimated as follows~\cite{balberg:2002ue,pollack:2014rja,essig:2018pzq}
\begin{eqnarray}
\label{eq:tcdmo}
t_{c,0} = \frac{150}{C} \frac{1}{\frac{\sigma}{m}\rho_s} \left( \frac{1}{4\pi G \rho_s r_s^2} \right)^{\frac{1}{2}}, 
\end{eqnarray}
where $C\approx 0.9$ is a normalization constant chosen to match the core collapse time of the simulated DM-only benchmark in Table~\ref{tab:simbm} and it differs from the $0.75$ used in Ref.~\cite{yang:2023jwn}.
This approximately $20$\% difference is at the same level of the uncertainty associated with the core collapse time estimated from N-body simulations. 

Equation~(\ref{eq:tcdmo}) can be derived assuming the core collapse time to be proportional to the collisional relaxation time $t_r\approx 1/(n\sigma v)$, where $n=\rho/m$ and $v$ approximately scales as the square root of the inner halo potential $\lim_{r\to 0}\Phi_{\rm NFW}(r) = 4\pi G \rho_s r_s^2 \equiv \Phi_{0,\rm NFW}$. 
However, the scattering rate does not always linearly represent the SIDM effect on the halo density profile, and a satisfactory extension to incorporate baryons has been found challenging.
To compute the relaxation time more accurately, we consider the following equation from the perspective of energy transport
\begin{eqnarray}
\label{eq:tR}
t_R(r) \propto \frac{M(r) |\Phi(r)|/2}{4\pi r^2 \kappa |\nabla T|}, 
\end{eqnarray}
where the numerator represents the total energy enclosed in radius $r$ and the denominator computes the absolute luminosity at $r$.
In the dark matter only case, one can estimate $t_R$ as
\begin{eqnarray}
t_R &\propto& \frac{4\pi \rho_s r_s^3 |\Phi_{0,\rm NFW}|/2}{4\pi r_s^2 \rho_s \frac{\sigma}{m} \frac{|\Phi_{0,\rm NFW}|^{3/2}}{G} |\lim_{r\to 0} \nabla \Phi_{\rm NFW}(r)|}, \\ \nonumber
    &\approx& \frac{1}{4\pi \rho_s \frac{\sigma}{m}} \left( \frac{1}{4\pi G \rho_s r_s^2} \right)^{\frac{1}{2}} \propto t_{c,0}
\end{eqnarray}
which is consistent with the equation derived based on the scattering rate in Eq.~(\ref{eq:tcdmo}).

In the presence of baryons, we observe that the $\nabla \Phi_{\rm Hern}(r) = (2 \pi G \rho_H r_H^3)/(r + r_H)^2$ for the Hernquist profile has a large dynamical range between $r_H$ and $r_s$.
Therefore, it is crucial to find an effective radius where the relative contributions of the two potentials are appropriately accounted for in estimating the core collapse time.
For this purpose, we introduce
\begin{eqnarray}
r_{\rm eff} &=& \frac{r_s \Phi_{0,\rm NFW} + \alpha r_H \Phi_{\rm Hern}(0)}{\Phi_{0,\rm NFW} + \alpha \Phi_{\rm Hern}(0)} \\ \nonumber
            &=& \frac{\rho_s r_s^3 + \alpha \rho_H r_H^3/2}{\rho_s r_s^2 + \alpha \rho_H r_H^2/2} \\ \nonumber
            &=& r_s \frac{1+\alpha \hat{\rho}_H \hat{r}_H^3/2}{1+\alpha \hat{\rho}_H \hat{r}_H^2/2} \equiv r_s \hat{r}_{\rm eff},
\end{eqnarray}
where $\alpha=1.6$ is a parameter tuned against simulations with baryons, $\hat{\rho}_H\equiv \rho_H/\rho_s$ and $\hat{r}_H\equiv r_H/r_s$ are dimensionless scale density and radius introduced to parameterize the form factor later. In our discussion, all quantities shown with hats will be dimensionless. 

Base on this $r_{\rm eff}$, we estimate the $t_R$ as 
\begin{widetext}
\begin{eqnarray}
\label{eq:tcb}
t_R &\propto& \frac{r_{\rm eff} G}{\frac{\sigma}{m} \sqrt{\Phi_{0,\rm NFW} + \alpha \Phi_{\rm Hern}(0)} (\lim_{r\to 0}\nabla\Phi_{\rm NFW}+\gamma (\nabla \Phi_{\rm Hern})(0)) } \\ \nonumber
    &\propto& \frac{150}{C} \frac{1}{\frac{\sigma}{m} \left(\frac{\rho_s r_s}{r_{\rm eff}} + \gamma \frac{\rho_H r_H^3}{r_{\rm eff}(r_{\rm eff}+r_H)^2} \right)} \left( \frac{1}{4\pi G (\rho_s r_s^2+\alpha \rho_H r_H^2/2)} \right)^{\frac{1}{2}} \equiv t_{c,b}, 
\end{eqnarray}
\end{widetext}
where $\gamma=20$ is introduced to control the relative importance of baryons and dark matter to the temperature gradient. In the last line, we posited that the core collapse time in the presence of baryons, denoted as $t_{c,b}$, is proportional to $t_R$. 

A comparison between Eqs.~(\ref{eq:tcb}) and (\ref{eq:tcdmo}) reveals their similar structures. In particular, the removal of baryonic components from $t_{c,b}$ automatically simplifies the equation to the DM-only case. 
This relation is made explicit by writting $t_{c,b} = t_{c,0} {\cal F}_t(\hat{\rho}_H,\hat{r}_H)$, where 
\begin{eqnarray}
{\cal F}_t &=& \left( \frac{1}{\hat{r}_{\rm eff}} + \frac{ \gamma\hat{\rho}_H \hat{r}_H^3}{\hat{r}_{\rm eff}(\hat{r}_{\rm eff}+\hat{r}_H)^2} \right)^{-1} \left( 1+\alpha\frac{\hat{\rho}_H \hat{r}_H^2}{2} \right)^{-\frac{1}{2}}
\end{eqnarray}
is a form factor that only depends on the hatted quantities and contains all the baryon dependencies.

\begin{figure}[htbp]
  \centering
  \includegraphics[width=8cm]{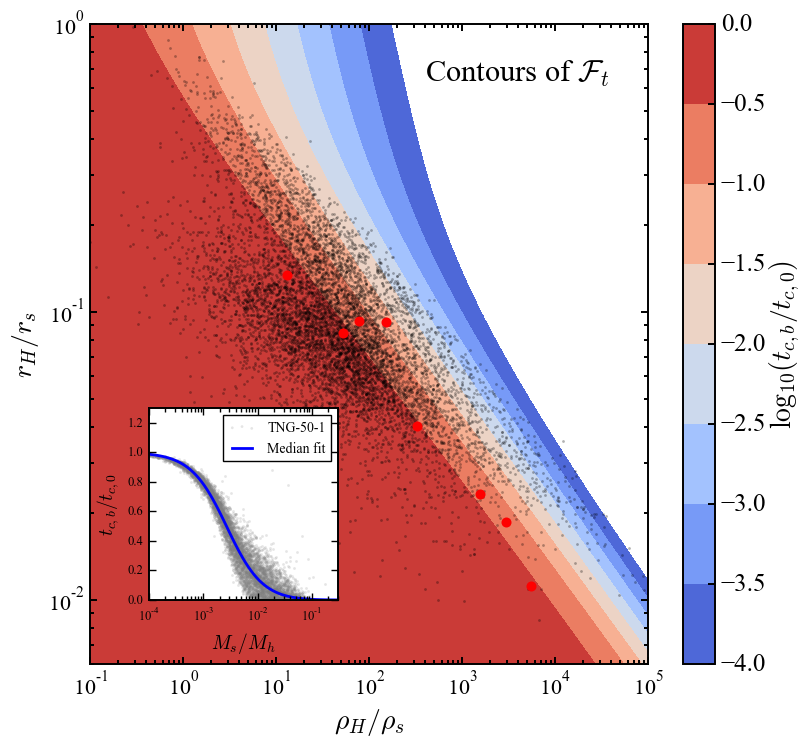}
  \caption{\label{fig:calF} 
  Contours of ${\cal F}_t\equiv t_{c,b}/t_{c,0}$ on the $\rho_H/\rho_s$ ($\hat{\rho}_H$) versus $r_H/r_s$ ($\hat{r}_H$) plane (main panel), depicted together with the isolated halos in the TNG-50-1 simulation (black dots)~\cite{nelson:2018uso,nelson:2019jkf,pillepich:2019bmb} and the ones we utilized for calibration (red dots, detailed in Table~\ref{tab:simbm}).
The inset panel presents the median of a grid of ${\cal F}_t \equiv t{c,b}/t{c,0}$ over the $\hat{\rho}_H - \hat{r}_H$ plane, projected onto $M_s/M_h$ (red band), with the best fit shown in blue. Analogous results from the TNG-50-1 data are represented by black dots.
  }
\end{figure}

In Fig.~\ref{fig:calF} (main panel), we plot contours of ${\cal F}_t$ in the $\hat{\rho}_H-\hat{r}_H$ plane and overlay isolated TNG-50 halos and those used to calibrate the $\alpha$ and $\gamma$ parameters to delineate the parameter region of interest. Interestingly, the contours align well with the median behavior of the simulated halos. Halos that scatter towards the upper-right can have core collapse times several orders of magnitude shorter than their dark matter-only counterparts. Conversely, towards the lower-left, a large population of halos remains nearly unaffected by baryons.
The broad range of ${\cal F}_t$ values covered by the TNG-50 halos demonstrates that SIDM can significantly amplify diversity in inner halo structures. Since the SIDM dependence does not manifest in ${\cal F}_t$, our discussion is generic; physical results should be interpreted by factoring in the effective SIDM cross sections.

In Fig.~\ref{fig:calF} (inset panel), we further display ${\cal F}_t$ as a function of $M_s/M_h$, calculated based on the properties of isolated halos in the TNG-50-1 simulation, as shown in Fig.~\ref{fig:TNG}. A clear correlation is observed between the two variables, with scatter increasing at higher $M_s/M_h$ values. To facilitate a rough estimate of ${\cal F}_t$ as a function of $M_s/M_h$, we derive a median relation from a fit, yielding the following expression:
\begin{eqnarray}
{\cal F}_t(M_s/M_h) \approx \frac{e^{- 7.9(M_s/M_h)}}{1+2300 (M_s/M_h)^{1.3}},
\end{eqnarray}
which is applicable for $M_s/M_h \lesssim 0.08$.

\subsection{The gravothermal phase of an SIDM halo with baryons}

The integral approach, as detailed in Ref.~\cite{yang:2023jwn}, is built upon the assumption that each halo resides in a specific {\it gravothermal state}, which can arise from the gravothermal evolution of a {\it fictitious} progenitor of the same halo and baryon parameters in CDM, regardless of its realistic accretion history. 
In this approach, the SIDM effect in a halo is accumulated through the following equations for the $V_{\rm max}$ and $R_{\rm max}$ of the dark matter component
\begin{eqnarray}
V_{\rm max}(t)    &=&  V_{\rm max, CDM}(t) + \int_{0}^{\tau(t)} d\tau' \frac{d V_{\rm max, Model} (\tau') }{d \tau'}, \nonumber \\
&& \label{eq:int} \\
R_{\rm max}(t)    &=&  R_{\rm max, CDM}(t) + \int_{0}^{\tau(t)} d\tau' \frac{d R_{\rm max, Model} (\tau')}{d \tau'}, \nonumber
\end{eqnarray}
where the $d V_{\rm max, Model} (\tau') / d \tau'$ and $d R_{\rm max, Model} (\tau') / d \tau'$ terms are analytic equations taken from the parametric model that describe how fast the dark matter $V_{\rm max}$ and $R_{\rm max}$ evolve along the gravothermal phase $\tau(t)=t/t_c$. 
Based on the obtained $V_{\rm max}(t)$ and $R_{\rm max}(t)$, one can identify the {\it fictitious} CDM progenitor that gives rise to these values $\tau(t)$. 
This {\it fictitious} CDM halo only moderately differs from the CDM halo at time $t$; hence, replacing it with the CDM halo would only decrease a small amount of accuracy. 
In Ref.~\cite{Yang:2024uqb}, a comparison of results with and without this simplification is performed for dark matter-only halos. 
In Appendix~\ref{app:checkphase}, we verify this point through quantitative examples with baryons. Applications in this work will take advantage of this observation and use the CDM halo parameters to replace the fictitious ones. 

In practice, the integral can be calculated by summing contributions over small, incremental time steps. This method effectively yields the accretion history in SIDM as a direct result of the evaluation process.
Moreover, this method offers heuristic insights: the SIDM effect with realistic accretion history is obtained by summing over the incremental evolution of many isolated halos. 
At any time $t$ in the accretion history, the halo density profile in SIDM can be determined parametrically, taking the instantaneous CDM halo and baryon parameters, along with the gravothermal phase, as input. 
An incremental change in this evolution can be expressed as follows:
\begin{gather}
    \rho_{\rm SIDM}(r, \text{``CDM'' halo \& baryon params at } t, \tau) \nonumber \\
    \Downarrow\ \ t\to t+\delta t \nonumber \\
    \rho_{\rm SIDM}(r, \text{``CDM'' halo \& baryon params at } t + \delta t, \tau + \delta \tau) \nonumber
\end{gather}
where the use of quotes around CDM underscores the use of a {\it fictitious} CDM progenitor is theoretically most plausible and can enhance the accuracy. In the presence of baryons, the fictitious progenitor should be solved under a fixed baryon configuration at the same time. Therefore, we added baryon parameters to the argument list. 

The gravothermal phase in the integral approach can be computed independently. 
Taking the evolution time to commence at the time of the halo formation, it can be calculated as 
$$
\tau(t) = \int_{0}^{t} \frac{d t}{t_{c,b}[\sigma_{\rm eff}(t)/m,\rho_s(t),r_s(t),\rho_H(t),r_H(t)]},
$$
where $\rho_s(t)$ and $r_s(t)$ specifies an NFW halo, and $\rho_H(t)$, $r_H(t)$ specifies the baryon profile, both at the time $t$. 
The SIDM cross section can be time-dependent for a differential cross section. 
At each time, we average out the angular and velocity dependencies by computing the effective constant cross section as~\cite{yang220503392,yang:2022zkd}
\begin{eqnarray}
\label{eq:eff}
\sigma_{\rm eff} &=& \frac{2 \int d v d \cos\theta \frac{d \sigma}{d \cos\theta} \sin^2\theta v^5 f_{\rm MB}(v,\nu_{\rm eff}) }{\int d v d \cos\theta \sin^2\theta v^5 f_{\rm MB}(v,\nu_{\rm eff}) }, \\ \nonumber
\end{eqnarray}
where $v$ denotes the relative velocity between the two incoming particles, $\theta$ refers to the polar angle that takes values in $[0,\pi]$,
and
\begin{eqnarray}
f_{\rm MB} (v,\nu_{\rm eff}) \propto v^2 \exp \left[-\frac{v^2}{4\nu_{\rm eff}^2}\right], 
\end{eqnarray}
is a Maxwell-Boltzmann velocity distribution that approximates the dark matter velocity distribution.
In the presence of baryons, the effective velocity dispersion can be obtained by rescaling the dark matter-only case result ($\approx 0.64 V_{\rm max,NFW}\approx 1.05 r_s \sqrt{G\rho_s}$) by $\sqrt{\Phi_{\rm 0,NFW}+\Phi_{\rm Hern}(0)}/\sqrt{\Phi_{\rm 0,NFW}}$.

For the integral approach to yield reliable results, the gravothermal evolution must proceed slowly during a period of a merger event. 
This condition implies that the CDM halo profile is predominantly described by the NFW profile during a notable change in gravothermal evolution. 
To justify this condition, one can compare the gravitational relaxation time with a small increment in the gravothermal phase $\Delta \tau$. 
Given that these parameters vary depending on the halo and the specific SIDM model, we perform a quantitative assessment for a particularly challenging case. 
We consider the BM2 halo benchmark detailed in Ref.~\cite{yang:2023jwn}, which is characterized by a low mass of $2\times 10^7~\rm M_{\odot}$, a high concentration of $c_{200}=19.7$ assuming $z=2$, and has been used to calibrate the parametric model. By adopting an effective constant cross section of $68~\rm cm^2/g$, we anticipate this halo to undergo core collapse within $3$ Gyr. The corresponding gravitational relaxation time is estimated to be $\sqrt{3\pi/(16 G \rho_s)}\approx 0.02~\rm Gyr$, which is about $150$ times shorter than the core collapse duration, allowing for $\Delta \tau$ to be smaller than one percent in one gravitational relaxation time.

In Ref.~\cite{yang:2023jwn}, the integral approach is validated using halos from a cosmological SIDM simulation. The consistent agreement across halos of various mass scales and differing effective SIDM cross sections provides robust support for employing the idea of the gravothermal state. 
Notably, most merger events do not compromise the performance of the model because the changes in the gravothermal phases are negligible during these events. 
Even if a halo is in the collapsing phase during a merger, the effectiveness of the parametric model does not seem to decrease significantly. Such violation effect is explored in Ref.~\cite{Yang:2024uqb} using an extreme SIDM cross section simulation. 

\section{Modeling baryon effects on dark matter density profiles}
\label{sec:modeling}

When baryons are present, the contracted halo profile is more accurately described by the DZ profile than by the NFW profile. However, as SIDM thermalizes the inner regions of a halo, this initial discrepancy diminishes, allowing for the calibration of a parametric model using the cored NFW profiles from the DM-only scenario. Specifically, Ref.~\cite{zhong:2023yzk} demonstrated that the differences between an instantaneously inserted and a gradually grown baryon potential become negligible in halos undergoing core formation.

Therefore, we introduce two versions of the parametric model. The first is based on the SIDM profile from Ref.~\cite{yang:2023jwn}, hereafter $\beta4$ profile, which we adapt to include baryon effects. We name it as the Contracted $\beta4$ model. The second, based on the DZ profile and adjusted for SIDM effects, will be termed as the Cored-DZ model. This latter model is more complex but effectively accounts for baryon-induced contraction in both early gravothermal phases and halos containing massive galaxies.

\subsection{The Contracted $\beta4$ model}

When the effect of adiabatic contraction is not significant, the thermalization caused by SIDM would be most important for determining the shape of the density profile. In this case, we found the $\beta4$ profile still applies, which reads
\begin{eqnarray}
\label{eq:C4}
\rho_{\rm \beta4}(r) = \frac{\rho_s}{\frac{\left(r^{4}+r_c^{4} \right)^{1/{4}}}{r_s} \left(1 + \frac{r}{r_s} \right)^2}, 
\end{eqnarray}
and the SIDM halo profile at $\tau$ is specified by evaluating the $\rho_s(\tau)$, $r_s(\tau)$, and $r_c(\tau)$. 
For completeness, we rewrite these three trajectories in the DM-only case as $\rho_s(\tau)=\rho_{s,0}g_{\rho}(\tau)$, $r_s(\tau)=r_{s,0}g_{r}(\tau)$, and $r_c(\tau)=r_{s,0}g_{c}(\tau)$, where $\rho_{s,0}$ and $r_{s,0}$ are initial NFW parameters and
\begin{widetext}
\begin{eqnarray}
\label{eq:m0}
g_{\rho}(\tau) &=& 2.033 + 0.7381 \tau + 7.264 \tau^5 -12.73 \tau^7  + 9.915 \tau^9 + (1-2.033) (\ln 0.001)^{-1} \ln \left( \tau + 0.001 \right), \nonumber \\
g_{r}(\tau) &=& 0.7178 - 0.1026 \tau +  0.2474 \tau^2 -0.4079 \tau^3 + (1-0.7178) (\ln 0.001)^{-1} \ln \left( \tau + 0.001 \right), \nonumber \\
g_c(\tau) &=& 2.555 \sqrt{\tau} -3.632 \tau + 2.131 \tau^2 -1.415 \tau^3 + 0.4683 \tau^4. 
\end{eqnarray}
\end{widetext}
For core collapsed halos having $\tau>1.1$, we freeze the density profile taking $\tau=1.1$, assuming the post-collapse evolution to proceed in the inner halo region, which separates from the outer region evolution. 

In the presence of baryons, one only needs two simple adjustments to incorporate the baryon effect and the SIDM profile is obtained as

\begin{eqnarray}
\label{eq:m1}
\rho_s(\tau) &=& \rho_{s,0} g_{\rho}(\tau), \\ \nonumber
r_s(\tau) &=& r_{s,0} g_{r}(\tau), \\ \nonumber
r_c(\tau) &=& r_{s,0} g_{c}(\tau) ({\cal F}_t)^2, \\ \nonumber
k &=& 4 ({\cal F}_t)^{1/2},  
\end{eqnarray}
where $\tau=t/t_{c,b}$ is the gravothermal phase with baryon effect incorporated. We have adjusted the core size and its transition by the form factor ${\cal F}_t=t_{c,b}/t_{c,0}$, with the indices calibrated based on the simulations. 

In Appendix A, we demonstrate the quality of this simplified scenario in Fig.~\ref{fig:app2} and discuss the necessity of incorporating adiabatic contraction for more massive halos. 

\subsection{The Cored-DZ model}

As depicted in Fig.~\ref{fig:univ}, the dark matter density profile changes in response to the baryon potential, exhibiting nonuniversal behavior. 
To model this effect in a manner that depends on the baryon content, we utilize the Dekel-Zhao (DZ) profile, which has been demonstrated to accurately fit the density profiles of dark matter halos in cosmological hydrodynamical simulations~\cite{2020MNRAS.499.2912F}. We introduce a core modification to the DZ profile, similar to our approach in the DM-only scenario, and incorporate additional functions to ensure it aligns with the key characteristics of the DZ profile during gravothermal evolution.

The DZ profile takes the following form
\begin{eqnarray}
\rho_{\rm DZ}(r) = \frac{\rho_x}{\left(\frac{r}{r_x}\right)^{a} \left(1+\left(\frac{r}{r_x}\right)^{1/2} \right)^{2(3.5-a)} }
\end{eqnarray}
where $\rho_x$ and $r_x$ are scale density and radius that can be obtained through a concentration parameter $c= R_{\rm cut}/r_x$, with $R_{\rm cut}$ being the virial radius of the halo, and a density slope parameter $a=-\lim_{{r\to 0}}d\ln\rho/d\ln r$.
In terms of $a$ and $c$,
\begin{eqnarray}
\rho_{x} &=& \left(1 - a/3\right) c^3 \mu \bar{\rho}_{\rm h}, \\ \nonumber
r_x &=& R_{\rm cut}/c, 
\end{eqnarray}
and $\mu=c^{a-3} \left(1 + c^{1/2}\right)^{2(3-a)}$, $\bar{\rho}_{\rm h}=M_h/\left(\frac{4 \pi}{3} R_{\rm cut}^3\right)$. 

In the presence of baryons, the $a$ and $c$ parameters can be determined by fitting the contracted profile, which can be obtained following the method of Ref.~\cite{gnedin:2004cx}. 
For simplified studies, an analytic method in Ref.~\cite{2020MNRAS.499.2912F} could be used, where these quantities are computed as a function of stellar to halo mass ratio $M_s/M_h$. 
However, as we illustrated in Fig.~\ref{fig:TNG}, there is a significant spread in the $\rho_H/\rho_s$ and $r_H/r_s$ plane with increasing $M_s/M_h$ ratios, indicating the need for enhanced theoretical modeling to achieve greater accuracy. In this study, we employ the method detailed in Ref.~\cite{gnedin:2004cx} to model the contracted profiles accurately. The $a$ and $c$ parameters are extracted by fitting the contracted profiles. 

The CoredDZ profile is parameterized as
\begin{eqnarray}
\rho_{\rm CoredDZ}(r) &=& \frac{f_{\rm in}(r) \rho_x f_{\rm out}(r)}{\frac{\left(r^{k}+r_c^{k} \right)^{1/{k}}}{r_x} \left( 1+\left(\frac{r}{r_x} \right)^{1/2} \right)^{2(3.5-a)} },
\end{eqnarray}
where we have introduced two functions to reshape the inner and outer profiles 
\begin{eqnarray}
&&f_{\rm in}(r) = \left( \frac{r}{r_x} + \frac{r_c}{0.4 r_s}\left(\frac{\rho_{x,0} r_{x,0}}{\rho_s r_s+0.4\rho_H r_H }\right)^{1/(a-1)} \right)^{1-a}   \\ \nonumber
&&f_{\rm out}(r) = \left( 1+\frac{r}{R_{\rm cut}}\left(\frac{r_x}{r_{x,0}}-1 \right) \right)^{-1/2}.
\end{eqnarray}
While one can formally describe the form factor that cores the DZ profile as ``$\rho_{\rm CoredDZ}(r) = {\cal F}_r(r) \rho_{\rm DZ}(r)$,'' in practice, we have found it more convenient to use a hybrid density profile. This profile introduces a core analogous to the contracted $\beta4$ profile while preserving the second factor in the denominator as in the DZ profile. The function $f_{\rm in}(r)$ ensures that in the limit of a vanishing core, the profile reverts to the DZ profile, with an additional constant term proportional to the core size to maintain a cored configuration in regions much smaller than $r_c$.
Meanwhile, $f_{\rm out}(r)$ addresses the distortion caused by differing power slopes at large radii between the NFW and DZ profiles, ensuring minimal change in the total halo mass during gravothermal evolution. 
A key aspect of ``${\cal F}_r$'' is that it depends not only on the radius but also on the interplay between baryon and dark matter properties. This feature reflects the loss of universality in the radial direction: rather than a single profile that fits all, we need a family of profiles parametrized by factors $a$ and $c$, which are coupled to the baryon parameters.

To model gravothermal evolution with baryons based on the Cored-DZ profile, we found it sufficient to start with the trajectories of the $\beta4$ profile in Ref.~\cite{yang:2023jwn}. The effect induced by the baryons can be incorporated by adjusting the core size $r_c$ and the power index $k$ as a function of the gravothermal phase $\tau$. 
The parameter trajectories for the Cored-DZ profile are 
\begin{eqnarray}
\label{eq:mx}
\rho_x(\tau) &=& \rho_{x,0} g_{\rho}(\tau), \nonumber \\
r_x(\tau) &=& r_{x,0} g_{r}(\tau), \nonumber \\
r_c(\tau) &=& r_{s,0} g_{c}(\tau) ({\cal F}_t)^2, \\ \nonumber
k &=& 4 ({\cal F}_t)^{1/2},
\end{eqnarray}
where the $\rho_{x,0}$ and $r_{x,0}$ correspond to the initial DZ profile parameters and the $r_{s,0}$ is the initial NFW scale radius. 

This modified profile facilitates the initiation of gravothermal evolution from a profile that more closely mirrors a realistic CDM context, incorporating the contraction induced by baryons. The evolution of this response in the context of SIDM has been calibrated using a series of N-body simulations listed in Table~\ref{tab:simbm} with a fixed baryon potential. 

\section{Model validation}
\label{sec:valid}

\begin{figure}[htbp]
  \centering
  \includegraphics[width=8cm]{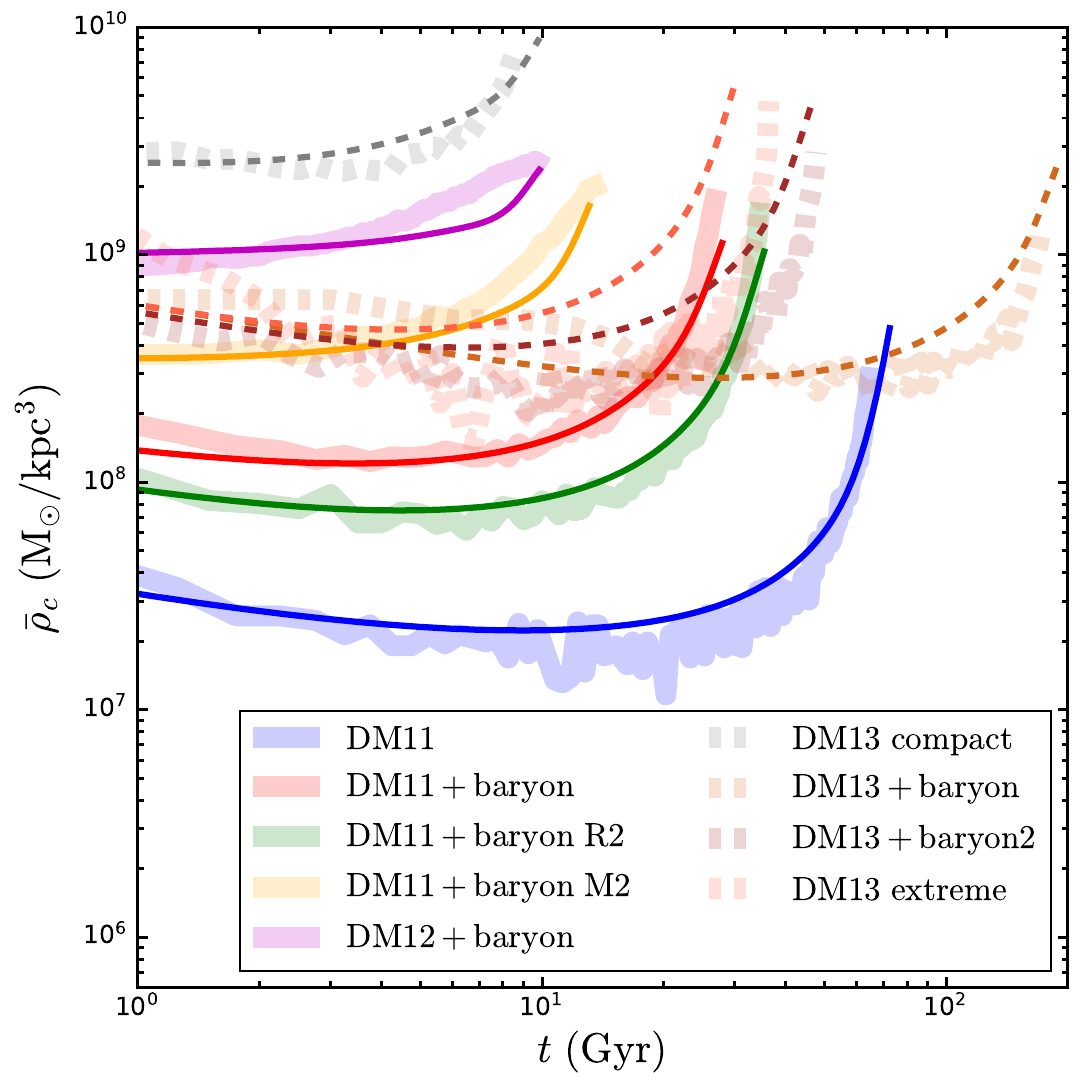}
  \caption{\label{fig:valid2}
The averaged inner halo density evolution from N-body simulations (colored bands) and model predictions (colored curves) for the benchmarks in Table~\ref{tab:simbm}. 
For clearer visualization, inner densities are calculated within $0.3$ kpc for the {\it DM11} cases and $1$ kpc for the {\it DM12} and {\it DM13} cases. Core collapse timings are determined based on an SIDM of $\sigma/m = 10~\mathrm{cm}^2/\mathrm{g}$ for {\it DM11} cases and $\sigma/m = 1~\mathrm{cm}^2/\mathrm{g}$ for the {\it DM12} and {\it DM13} cases. In the {\it DM13} cases (dashed), the presence of heavy, compact baryons leads to the compression of the innermost density profiles, whereas the outer regions are less sensitive to the baryon potential and evolve into shallower densities in N-body simulations.
}
\end{figure}

Based on the samples in Table~\ref{tab:simbm}, we adjust the $\alpha$ and $\gamma$ parameters in Eq.~\ref{eq:tcb} to obtain an overall agreement between the core collapse times predicted by the equation and derived from simulations. 
The optimal values we found are $\alpha=1.6$ and $\gamma=20$ and will be used throughout the paper.  
The additional factors, including the $({\cal F}_t)^2$ for the core size and $({\cal F}_t)^{1/2}$ for the $k$, are also introduced based on the quality of model predictions versus the simulated density profiles. 

In Fig.~\ref{fig:valid2}, we present the evolution of the averaged inner density for all eight samples, alongside the DM-only scenario ({\it DM11}, shown in blue). The results from N-body simulations are depicted with colored bands, while the predictions from our model are represented by colored curves for comparison. To enhance clarity, we have calculated inner densities within $0.3$ kpc for the {\it DM11} scenarios and within $1$ kpc for the {\it DM12} and {\it DM13} scenarios. The core collapse times have been evaluated assuming an SIDM cross section of $\sigma/m = 10~\mathrm{cm}^2/\mathrm{g}$ for the {\it DM11} cases and $\sigma/m = 1~\mathrm{cm}^2/\mathrm{g}$ for the {\it DM12} and {\it DM13} cases. 
Additionally, to incorporate the effect of adiabatic contraction, which is significant for the {\it DM11+baryon M2}, {\it DM12} and {\it DM13} cases, we fit the contracted initial conditions in the simulation using the DZ profile and used them as input for the initial condition of the Cored-DZ parametric model for the {\it DM12} and {\it DM13} cases. 
For the {\it DM11+baryon M2} case where the simulated initial condition is NFW, we use the equations for $s_1$ and $c_2$ in Ref.~\cite{2020MNRAS.499.2912F} (Eq.~45 and 46, with coefficients in Table 1) to incorporate this effect. 

As previously discussed in terms of Fig.~\ref{fig:univ}, baryon potentials alter dark matter density profiles in ways that are not universally predictable. 
Our model, which has been calibrated using controlled N-body simulations, is effective in accounting for the impact of adiabatic contraction by assuming that the contracted profiles can consistently be described by the Cored-DZ profile at all times.
Nonetheless, when the baryon content is extremely compact---indicated by high $\rho_H/\rho_s$ ratios and low $r_H/r_s$ ratios---the accuracy of the Cored-DZ profile diminishes, leading to discrepancies between the model predictions and the N-body simulations.
In Fig.~\ref{fig:valid2}, the {\it DM13} scenarios have dense and compact baryons that exert a stronger contraction on the dark matter within the inner regions of the halo, whereas the outer regions remain relatively less affected by the baryon potential. This discrepancy results in the formation of larger cores and steeper density minima in N-body simulations compared to what our parametric model predicts.
In the cases of {\it DM11+baryon M2} (orange) and {\it DM12+baryon} (magenta), the early phases of core formation are accurately captured by the Cored-DZ profile. However, as they evolve into the core collapse phase, the relatively high $\rho_H/\rho_s$ and $r_H/r_s$ values lead to late-stage distortions in the density profile during the gravothermal evolution. While the large $\rho_H/\rho_s$ compacts the inner halo earlier on, a larger $r_H$ reduces the degree of enhancement within this region.
Nevertheless, the core collapse times predicted by models align closely with those from simulations across all cases.

\begin{figure*}[htbp]
  \centering
  \includegraphics[width=5.33cm]{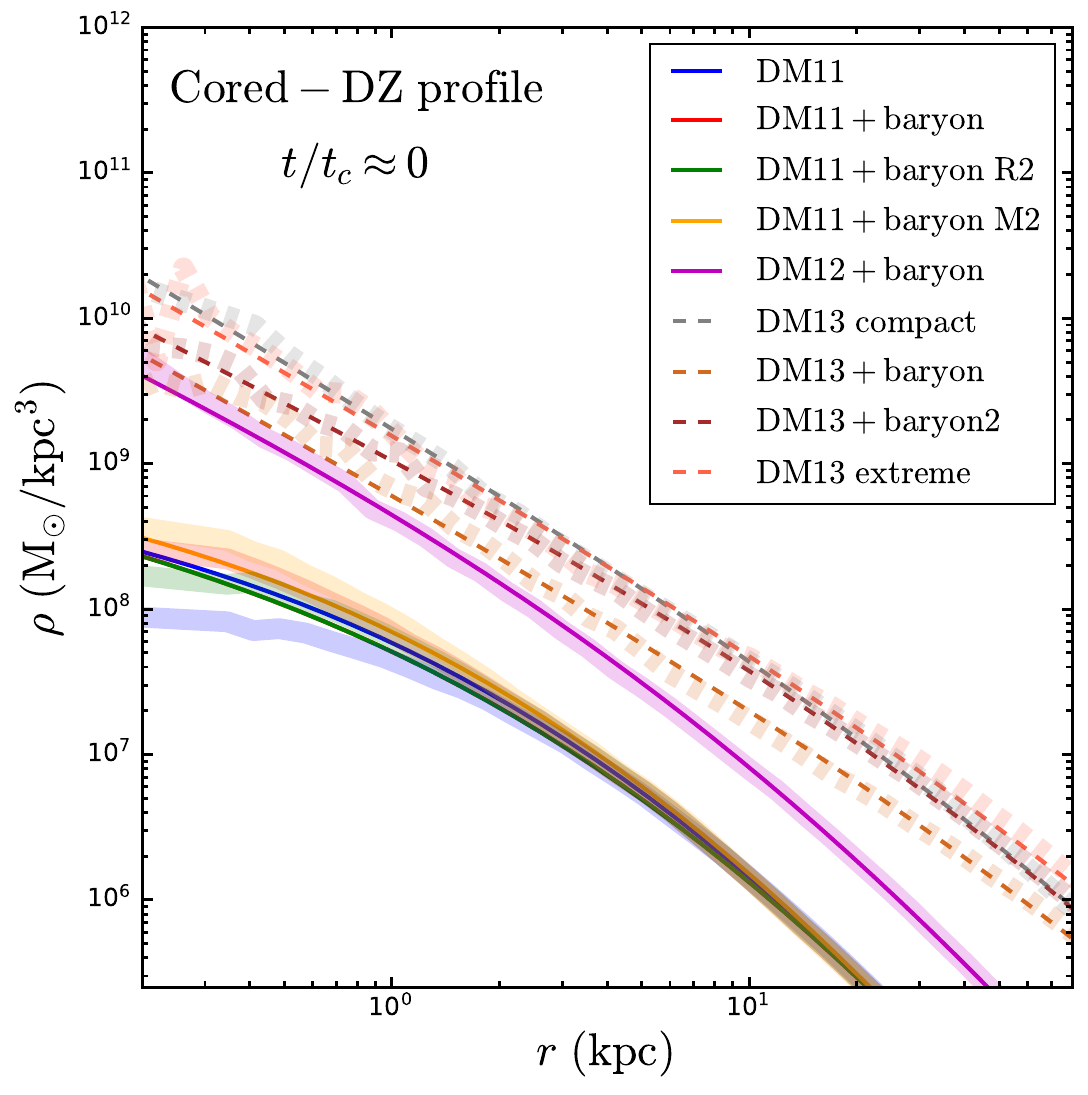}
  \includegraphics[width=5.33cm]{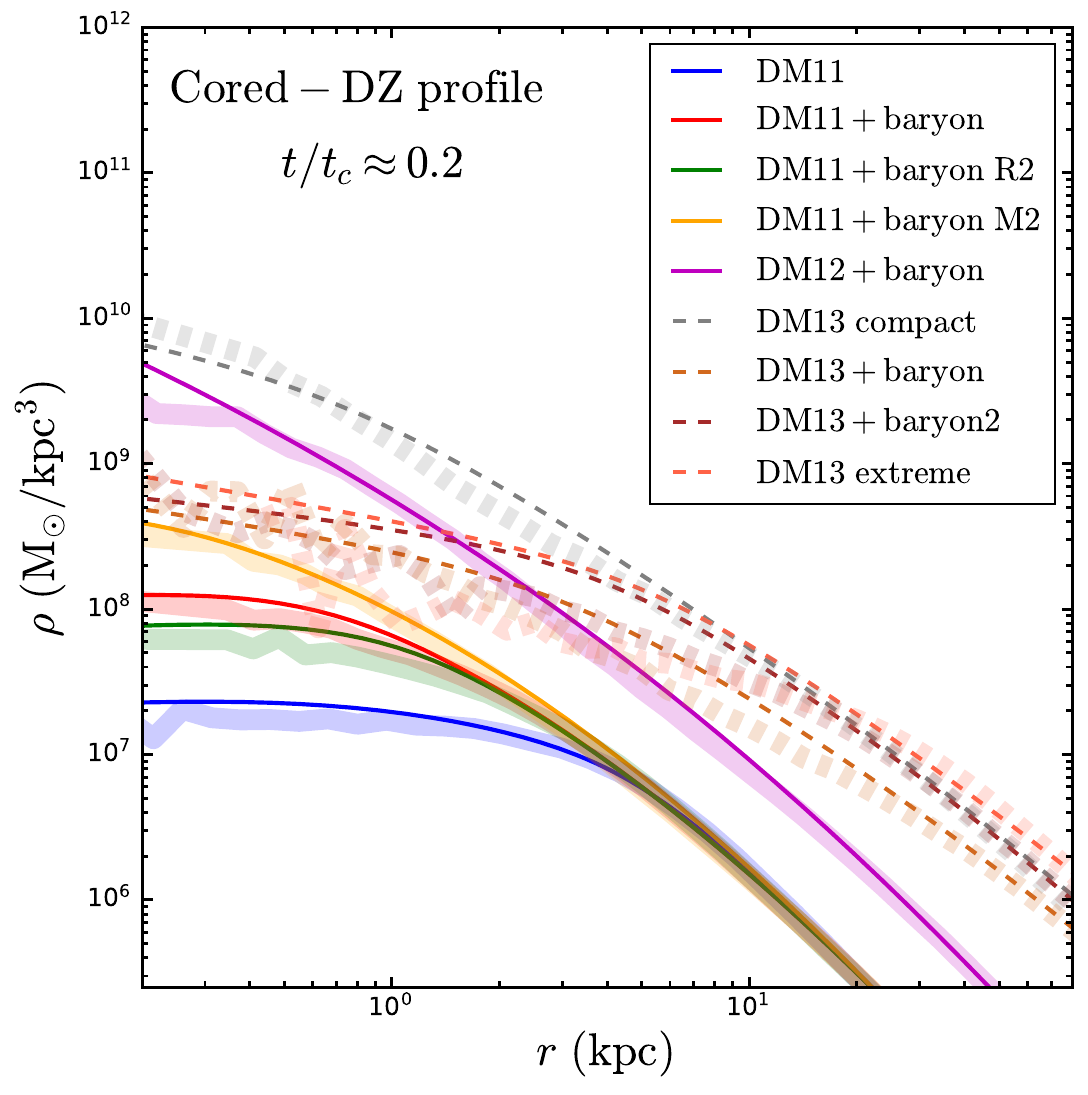}
  \includegraphics[width=5.33cm]{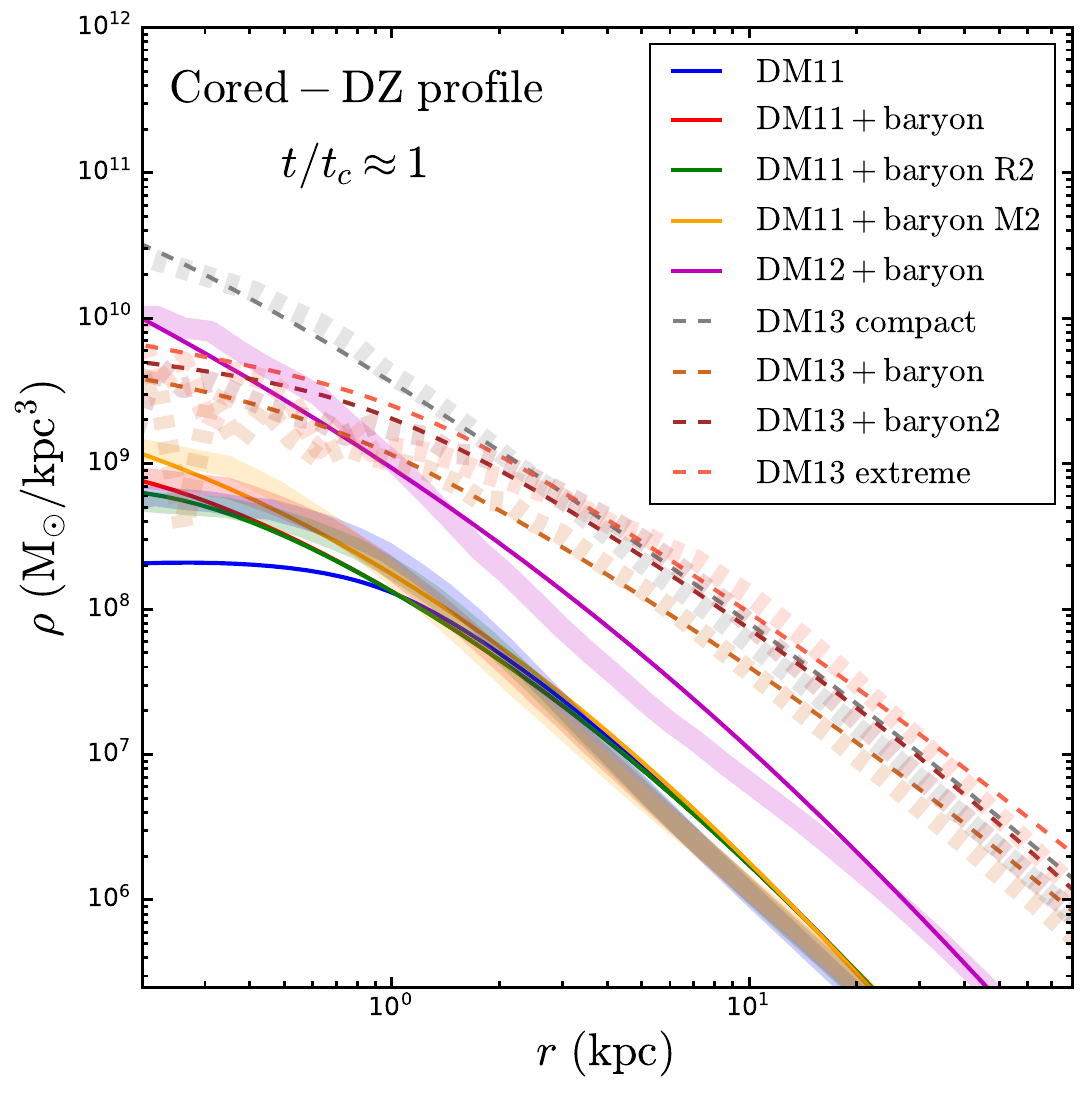}
  \caption{\label{fig:validx}
The simulated (colored bands) and Core-DZ model predicted (colored curves) halo density profiles at three representative gravothermal phases: $t/t_c\approx 0, 0.2$, and $1$. 
The {\it DM12} and {\it DM13} scenarios use a contracted CDM profile as the initial condition, whereas the {\it DM11} scenarios commence with an instant insertion method.
In the left panel ($t/t_c \approx 0$), the {\it DM11} cases are depicted at $t = 0.25$ Gyr to allow some initial evolution away from the original NFW profile.
At $t/t_c \approx 1$, the core collapse time, as calculated using Eq.~(\ref{eq:tcb}), is found to be 10\% (30\%) shorter than the simulated {\it DM13+baryon2} ({\it DM13 extreme}). To align the profiles for equivalent gravothermal phases, we adjust the timing of the simulated curves accordingly in these specific cases.
}
\end{figure*}

In Fig.~\ref{fig:validx}, we present the simulated (colored bands) and Cored-DZ model (colored curves) predicted halo profiles at three representative gravothermal phases: $t/t_c=0,0.2$, and $1$. 
In the left panel ($t/t_c \approx 0$), the {\it DM11} cases in the simulations are close to NFW. 
This is because they are simulated with an instant insertion of baryon potentials. We show the results of these cases at $t = 0.25$ Gyr to allow some initial evolution away from the original NFW profile. Nevertheless, the impact of adiabatic contraction in these instances remains minimal.
In comparison, the initial conditions of \textit{DM12} and \textit{DM13} cases are very well modeled by the DZ profile. 
In the middle panel, the agreement is good for most cases, except that the simulated {\it DM13+baryon2} and {\it DM13 extreme} cases demonstrate shallower profiles around the maximum cores, being consistent with what we discussed for Fig.~\ref{fig:valid2}. 
At $t/t_c \approx 1$, the core collapse time, as calculated using Eq.~(\ref{eq:tcb}), is found to be 10\% (30\%) shorter than the simulated {\it DM13+baryon2} ({\it DM13 extreme}). To align the profiles for equivalent gravothermal phases, we adjust the timing of the simulated curves accordingly in these specific cases. 
While distortions in the shape exist, the parametric model correctly captures most features even in this collapsed phase. 

\section{Effect of SIDM on the baryon profiles}
\label{sec:sidmfeedback}

In this work, we focus on the impact of baryons on SIDM halo density profiles. However, the gravitational influence between baryons and dark matter is mutual, and understanding the feedback of SIDM gravothermal evolution on baryonic evolution is crucial. In this section, we explore the role of such feedback in the $\rho_H/\rho_s-r_H/r_s$ plane.

\begin{figure}[htbp]
  \centering
  \includegraphics[width=8cm]{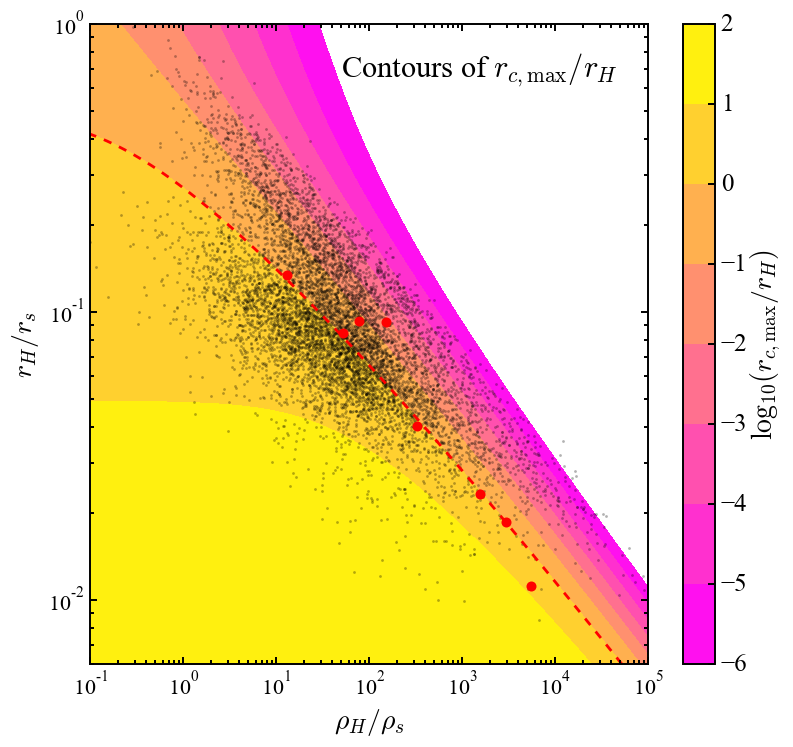}
  \caption{\label{fig:rc} Contours of the maximum core size $r_{c,\rm max}$ relative to $r_H$ on the $\rho_H/\rho_s$ ($\hat{\rho}_H$) versus $r_H/r_s$ ($\hat{r}_H$) plane, obtained using the parametric model with baryons. 
  The isolated halos in the TNG-50-1 simulation (black dots)~\cite{nelson:2018uso,nelson:2019jkf,pillepich:2019bmb} and the ones utilized for calibration (red dots, detailed in Table~\ref{tab:simbm}), are shown to indicate the region of interest. 
  The red-dashed line highlights the contour at $r_{c,\rm max}=r_H$, which is close to the median of samples. 
  }
\end{figure}

\begin{figure*}[htbp]
  \centering
  \includegraphics[width=5.4cm]{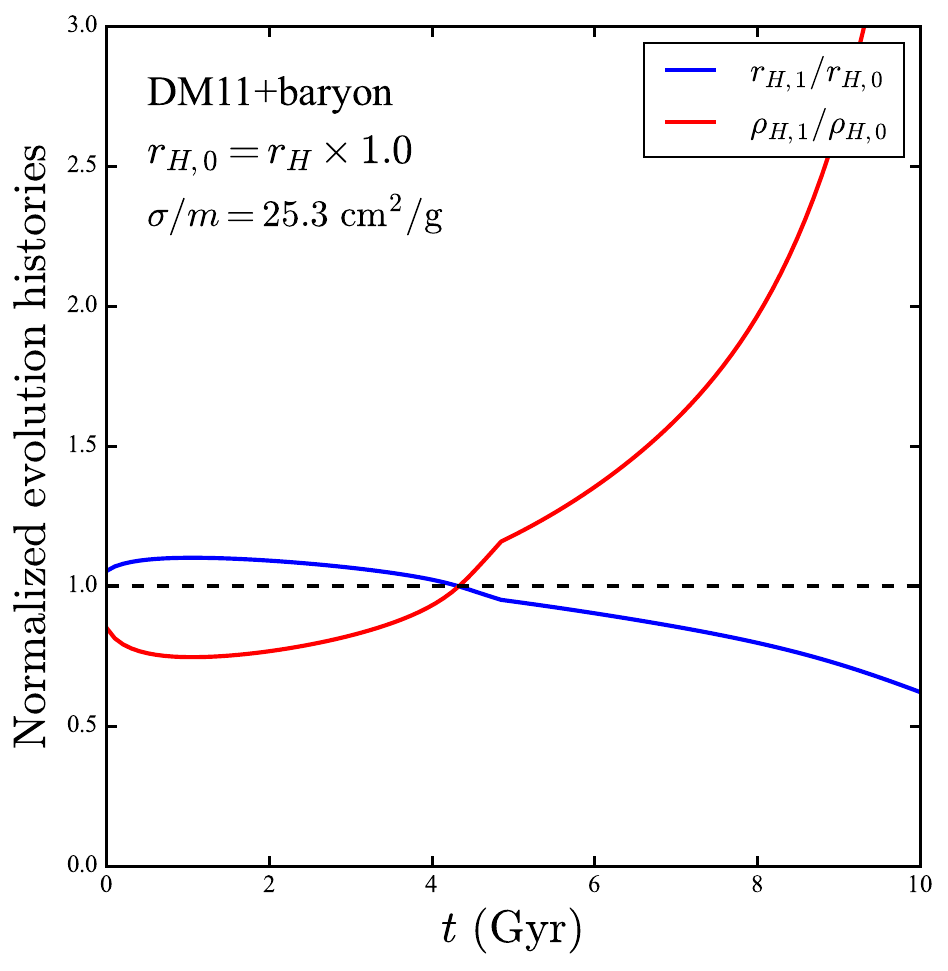}
  \includegraphics[width=5.4cm]{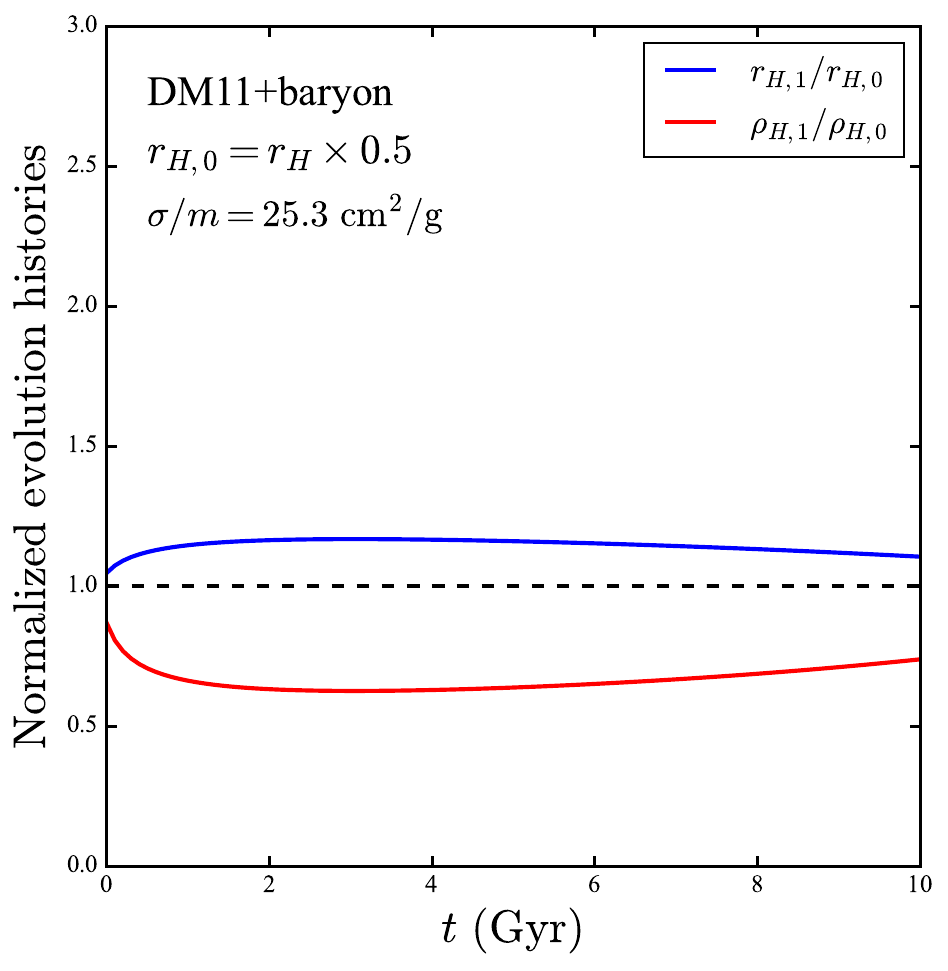}
  \includegraphics[width=5.4cm]{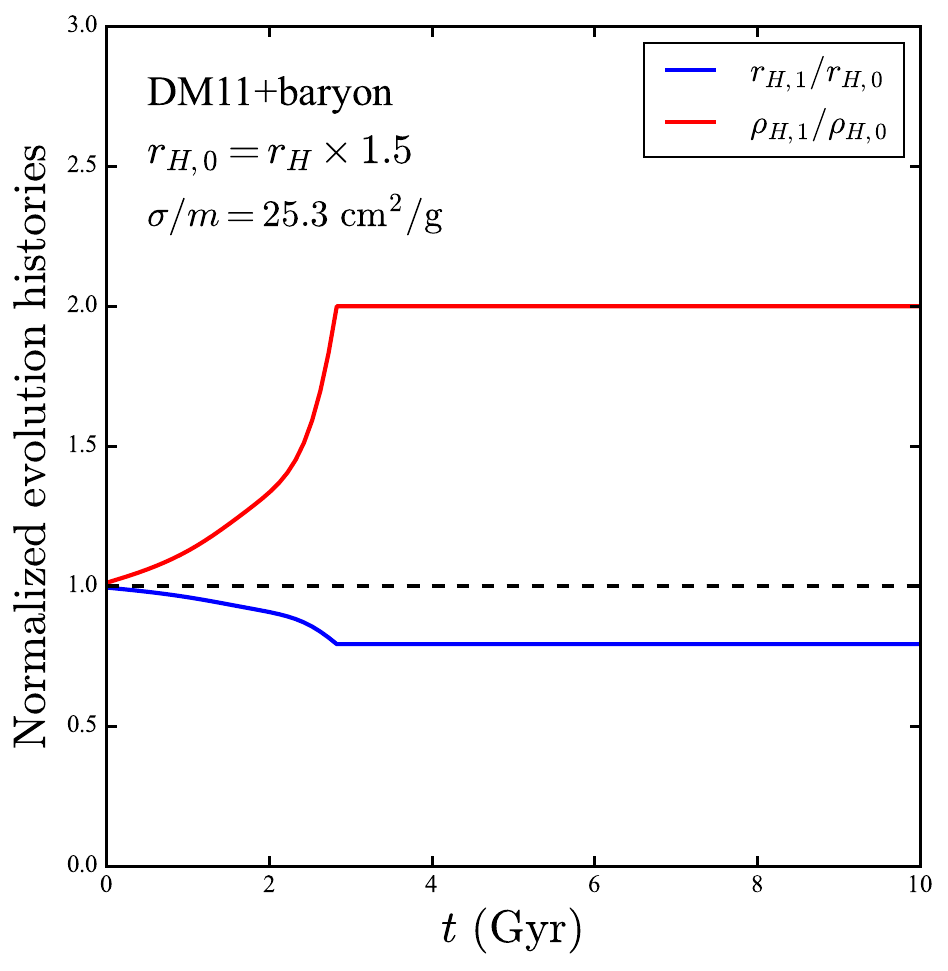}
  \caption{\label{fig:sidmfeedback} Evolution of the baryon scale radius (blue) and density (red) for the DM11 halo, depicted across three panels. The left panel corresponds to the median baryon distribution from the {\it DM11+baryon} benchmark (Table~\ref{tab:simbm}), where $r_{c,\rm max} \approx r_H$, aligning with the dashed red curve in Fig.~\ref{fig:rc}. The middle and right panels show scenarios with the scale radius decreased and increased by 50\% concerning the case on the left while maintaining the $\rho_H$. These adjustments position the middle panel below and the right panel above the dashed red curve of $r_{c,\rm max} \approx r_H$.
}
\end{figure*}

We are particularly interested in comparing the maximum core size, $r_{c,\rm max}$, with the radial scale $r_H$ of the baryon profile, as we anticipate that core expansion will make the baryon distribution more diffuse, while core collapse will compact it. In Fig.~\ref{fig:rc}, we present the contours of $r_{c,\rm max} = 0.49 r_{s,0} ({\cal F}_t)^2$ relative to $r_H$ on the $\rho_H/\rho_s-r_H/r_s$ plane. Similar to the findings in Fig.~\ref{fig:calF} for ${\cal F}_t$, the ratio $r_{c,\rm max}/r_H$ varies by several orders of magnitude, with values as low as $10^{-5}$ towards the upper-right points and exceeding $100$ towards the lower-left. These results also indicate a diversity in the effect of SIDM feedback on baryon distributions.

We also observe that the expression $r_{c,\rm max} = 0.49 r_{s,0} ({\cal F}_t)^2$ introduces dependencies on $\hat{\rho}_H$ and $\hat{r}_H$ that align well with the median data points. This alignment, which contrasts with the dark matter-only scenario's lack of scale density dependence, further justifies our choice of the $({\cal F}_t)^2$ factor. 

To better understand the influence of SIDM on baryons for varying ratios of maximum core radius to baryon scale radius $r_{c,\rm max}/r_H$, we employ the approximate conservation principle from the simple adiabatic contraction model described in Ref.~\cite{1987ApJ...318...15R,gnedin:2004cx}. We refine the integral approach to allow variation in the baryon profile while maintaining a Hernquist profile. 
Given a small timestep forward, the change in the baryon profile is solved using
\begin{eqnarray}
    [M_h(r_H) + M_s(r_H)]r_H = [M_h(r_H') + M_s(r_H')]r_H',
\end{eqnarray}
where $r_H'$ denotes the updated Hernquist radius. 
The scale density can be derived from $r_H'$ and the total baryon mass $M_s'$ as $\rho_H' = M_s'/(2\pi r_H'^3)$.

For this analysis, we take the {\it DM11+baryon} system as an example and assume the dark matter and baryon masses to both be conserved. 
We first take the initial condition as given in Table~\ref{tab:simbm}. We modify the cross section to be $25.3~\rm cm^2/g$ such that $t_{c,b}=10~$Gyr. 
Evolving this initial condition for 10 Gyr, we obtain the results in Fig.~\ref{fig:sidmfeedback} (left panel). 
We observe the expected features in this figure: the baryon scale radius (blue) increases during the core formation phase $0-2~$Gyr, after which it decreases in response to the collapse. The evolution of scale density (red) is solved from mass conservation. It looks much more significant than the size evolution because $\rho_H\propto r_H^{-3}$. 

Notably, we choose the baryon content of the DM11 system to be close to the median in both the stellar to halo mass ratio and the stellar size-mass relation, which makes $r_{c,\rm max}\approx r_H$ and resulting in a point residing on the dashed-red line of Fig.~\ref{fig:rc}. 
To explore the effect of SIDM on baryons in regions above and below the $r_{c,\rm max} = r_H$ contour (dashed red in Fig.~\ref{fig:rc}), we perform analogous studies for two cases: one with $r_H$ reduced by half and one with $r_H$ increased by half; the $\rho_H$ is fixed and hence the total stellar masses and core collapse times are modified accordingly. 

The obtained results are illustrated in the middle and right panels of Fig.~\ref{fig:sidmfeedback}. 
The middle panel case corresponds to the region below the dashed red curve. In this region, $r_{c,\rm max}> r_H$ and hence the baryon content can be more sensitive to the SIDM core evolution. 
In the simulated case, even if the stellar mass is reduced by a factor of $2^3$, the amount of size growth is comparable (slightly larger) to the case with the original $r_H$ value. 
The right panel case, instead, represents a case above the dashed red curve, where $r_{c,\rm max}< r_H$. This region is featured by a strong baryon effect on SIDM halos that can be cross-checked with contours in Fig.~\ref{fig:calF}. 
In the case that we simulate, the total baryon mass is about $3.4$ times the original case and the core collapse time is reduced to about 3 Gyr. 
Interestingly, we found the baryon size evolution to have bypassed the expansion phase and go directly into the collapsing phase. 
This result can be correlated to the behavior in the dark matter distribution. For example, the {\it DM12+baryon} case has $\rho_H/\rho_s=149$ and $r_H/r_s=0.09$, which is well above the dashed red curve. Its inner density evolution corresponds to the magenta curve in Fig.~\ref{fig:valid2}, where we see no signs of core formation. Similarly, the {\it DM11+baryon M2} case also resides above the dashed red curve and demonstrates no core formation. 
Our results align with a finding from Ref.~\cite{Correa:2024vgl}, which shows that some Milky Way mass galaxies, analogous to our {\it DM12+baryon} sample, tend to be more compact in SIDM.

\section{Applications}
\label{sec:applications}

\begin{figure*}[htbp]
  \centering
  \includegraphics[height=5.4cm]{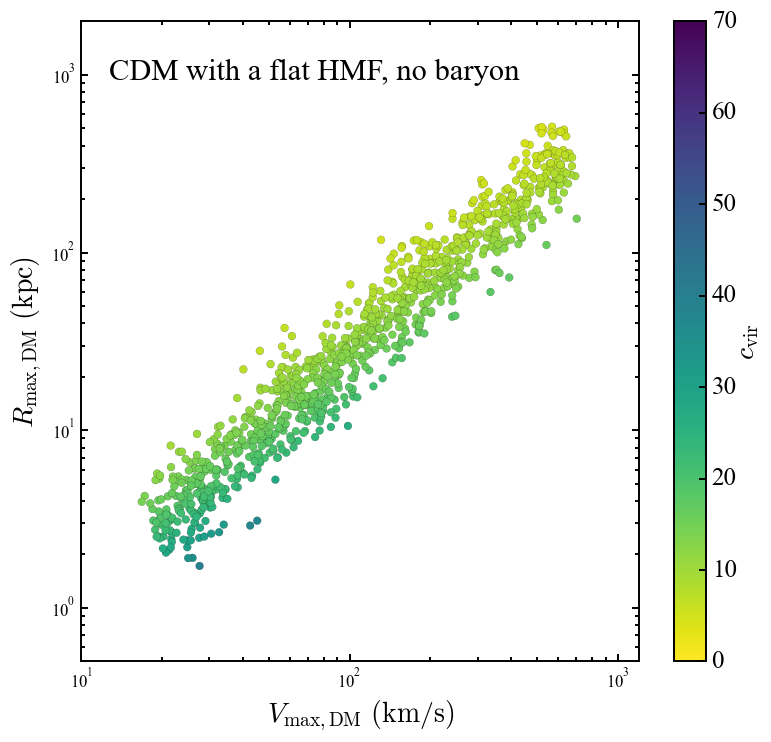}
  \includegraphics[height=5.4cm]{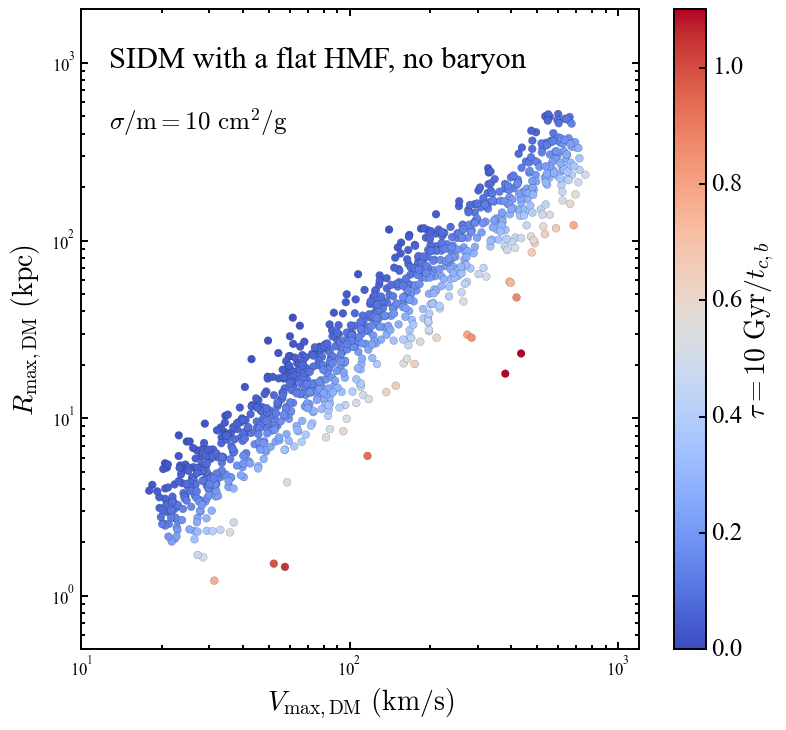}
  \includegraphics[height=5.3cm]{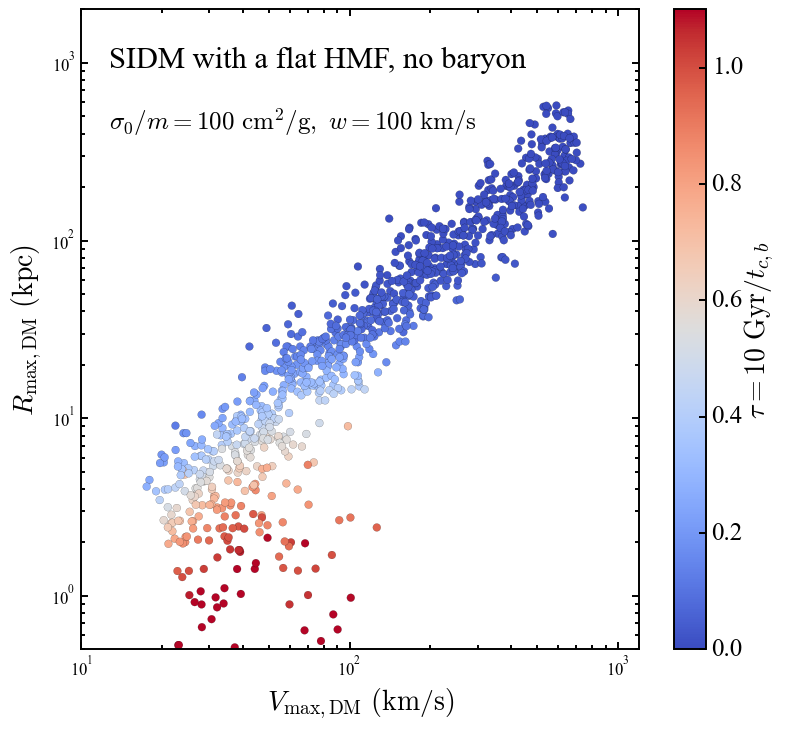} \\
  \includegraphics[height=5.4cm]{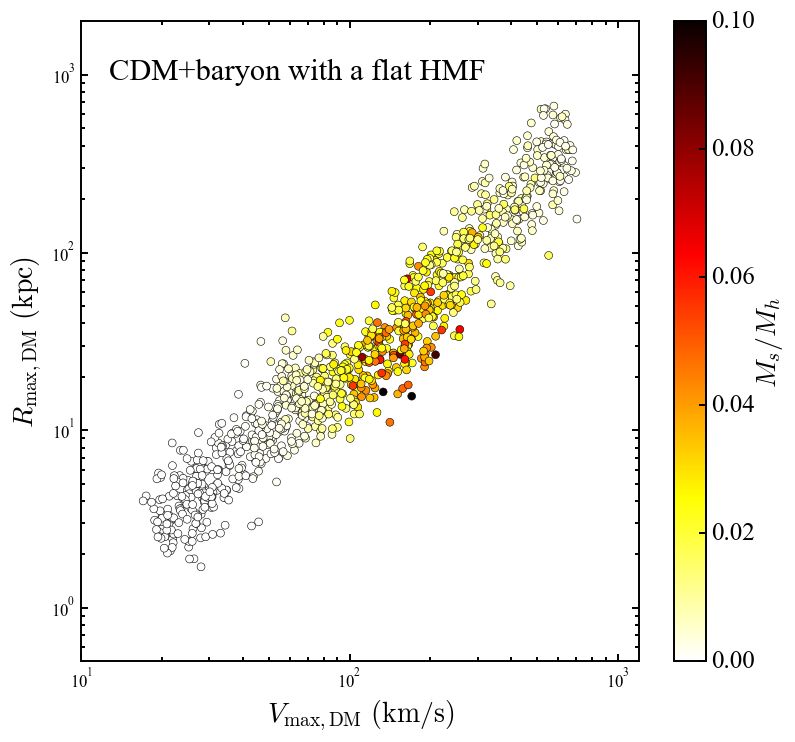}
  \includegraphics[height=5.3cm]{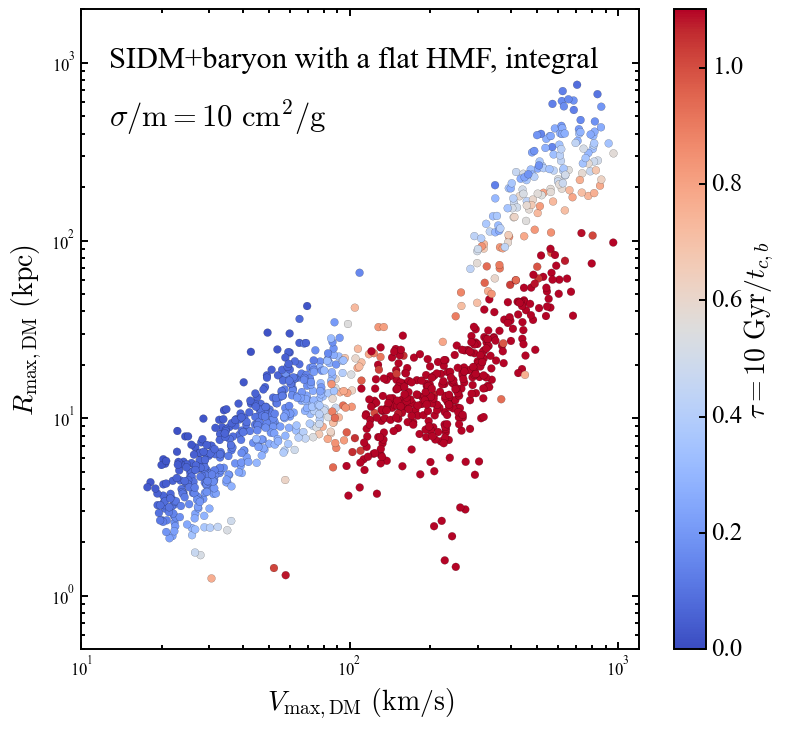}
  \includegraphics[height=5.3cm]{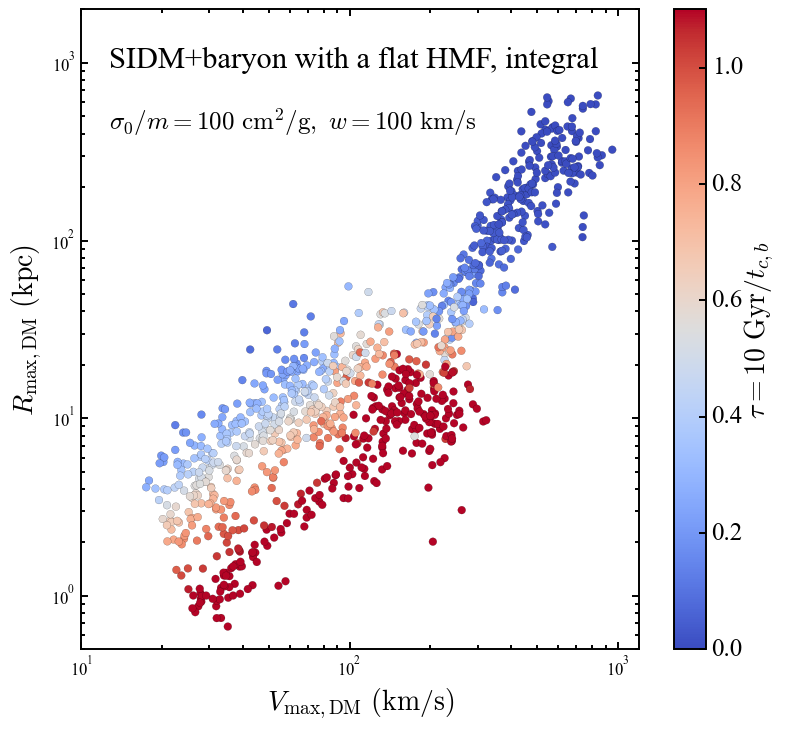} \\ 
  \includegraphics[height=5.4cm]{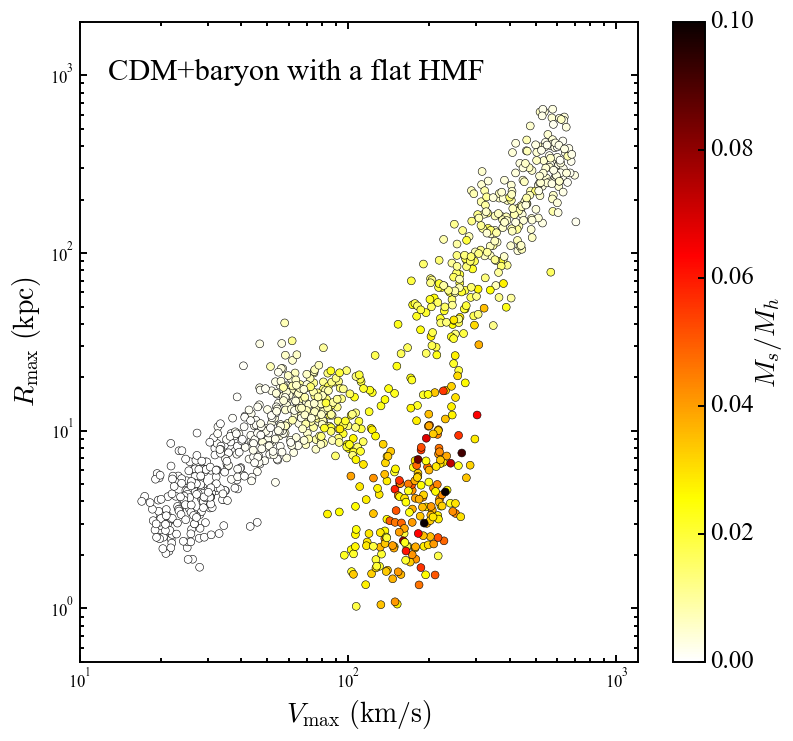}
  \includegraphics[height=5.3cm]{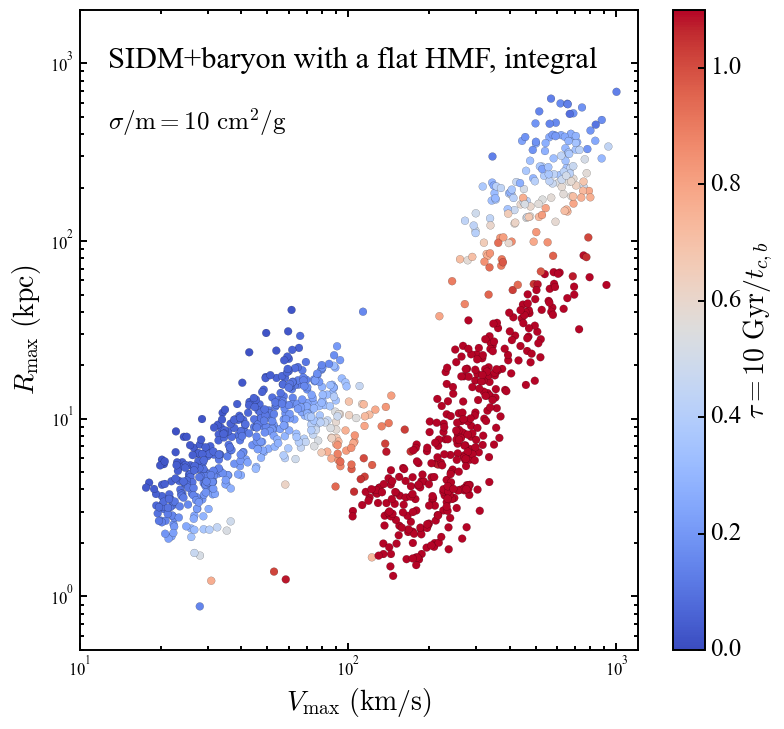}
  \includegraphics[height=5.3cm]{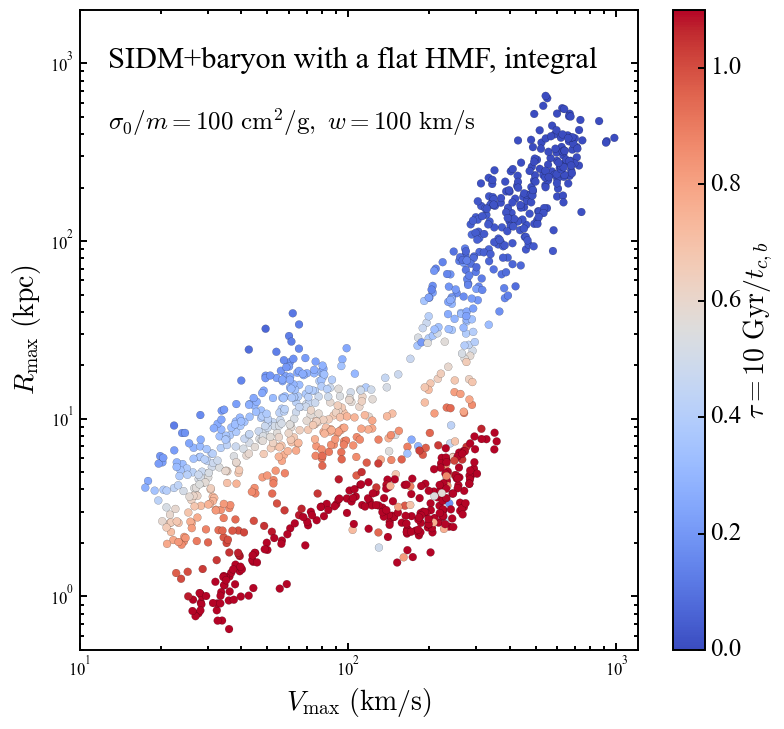} 
  \caption{\label{fig:sampled} Distributions of maximum total circular velocity ($V_{\rm max}$) and radius at which this velocity occurs ($R_{\rm max}$) for sampled halos.  The left panels depict results for the CDM model. The middle and right panels illustrate outcomes from two SIDM models: the first adopts a constant cross section per mass $\sigma/m=10~\rm cm^2/g$, and the second features a velocity-dependent cross section per mass, parameterized through a Rutherford-like cross section, with $\sigma_0/m=100~\rm cm^2/g$ and $w=100~\rm km/s$ in Eq.~(\ref{eq:rxs}). The upper row displays results in a dark matter-only scenario, while the middle and lower rows show the impacts of baryonic presence: the middle row details the $V_{\rm max,DM}$ and $R_{\rm max,DM}$ for the dark matter component alone, and the lower row includes the contributions from both dark matter and baryons. The SIDM predictions with baryons are obtained through the integral approach. All halos are uniformly sampled using a flat halo mass function over the mass range $M_h\in [10^9,10^{14}]\rm \ M_{\odot}$.
}
\end{figure*}

As with the DM-only parametric model, the baryonic extension of this model can be effectively applied to CDM simulations or semi-analytic models to forecast SIDM outcomes.
The dark matter-only version of the model has recently been implemented into the Semi-Analytical Sub-Halo Inference ModelIng ({\sc SASHIMI}) program~\cite{Ando:2024kpk}. 
Here, we demonstrate such an application using the Cored-DZ model on a sample of isolated halos. 
This application simplifies the modeling by considering generic accretion histories that do not differentiate between late and early-type galaxies. Additionally, we disregard the SIDM impacts on baryons discussed in Section~\ref{sec:sidmfeedback}, due to the absence of a calibration and validation using SIDM simulation data.
Despite this omission, the influence on the maximum circular velocity ($V_{\rm max}$) and its corresponding radius ($r_{\rm max}$) is minimal when considering combined dark matter and baryon contributions. Thus, our results focus solely on these parameters for clarity.

The parametric model also facilitates extracting the profile parameters from observed galactic rotation curves. It is advantageous because it parametrizes a halo's gravothermal state, bypassing the complexities in the accretion histories and the effect of SIDM feedback on baryon profiles.  
As an illustration, we show that SIDM parameters can be effectively retrieved given an assumed accretion history via post-processing, underscoring the model's utility in analyzing complex astrophysical phenomena. 

\subsection{Probing SIDM through sampled halos}

We demonstrate the effectiveness of the parametric model with baryons using a sample of isolated halos. These halos are constructed based on established relations in the literature for exploring the SIDM impact on the scatter within the $V_{\rm max}-R_{\rm max}$ plane.  
To efficiently explore a wide spectrum of halos, we sample the halo mass in $[10^9,10^{14}]\rm \ M_{\odot}$ using a flat mass function instead of the approximately $M^{-2}$ mass function. 
Based on the halo masses, we sample the halo concentrations based on the median $z=0$ concentration-mass relation from Ref.~\cite{dutton:2014xda}, incorporating a mass-independent lognormal scatter of 0.11 dex.
All other halo properties can be computed based on the sampled masses and concentrations. 

For baryonic components, we sample the stellar masses according to the stellar-to-halo-mass relation from Ref.~\cite{2013mnras.428.3121m}, with an intrinsic scatter of 0.15 dex. The stellar sizes are determined using the size-mass relation for red galaxies from Ref.~\cite{2019MNRAS.485..382C}, with a 0.1 dex scatter incorporated and the half-light radii ($r_e$) converted to Hernquist scale radii using $r_H = r_e (\sqrt{2}-1) 4/3$.
The inner dark matter scale densities and radii, represented through the NFW $\rho_s$ and $r_s$, are assumed to have been established 10 Gyr ago, a simplification supported by hierarchical structure formation theories (see section 3.1 of Ref.~\cite{yang:2023jwn} for associated discussion). In contrast, significant growth is assumed in stellar parameters within the last 10 Gyr, modeled using the average stellar accretion histories from Ref.~\cite{2013ApJ...770...57B}. 
We simplify the stellar size evolution considering $r_H = r_{H,z=0} (1+z)^{-0.75}$, which ignores dependencies on mass and galaxy type; for a comprehensive treatment, equations from Ref.~\cite{vanderWel:2014wba} can be employed. 

To calculate $V_{\rm max,DM}$ and $R_{\rm max,DM}$ in the presence of an accretion history, we apply the integral approach based on the discretized version of Eq.(\ref{eq:int}). In this approach, each incremental step calculates the core collapse time using the current configurations of the halo and baryons. The Cored-DZ model is applied to ensure accuracy.

When incorporating baryons using the integral approach, subtle differences emerge compared to the dark matter-only scenario. One is that the changes in $V_{\rm max,DM}$ and $R_{\rm max,DM}$ must be calculated based on the differences in the density profiles of two successive snapshots. The second is that we minimize the discrepancies between the predicted and actual values of $V_{\rm max,DM}$ and $R_{\rm max,DM}$ to derive the parameters of the fictitious CDM halo.
An efficient code for doing this is provided on the program page. 

These subtleties can significantly increase the numerical complexity of the integral approach compared with the dark matter-only case. 
Here we introduce a {\it hybrid} approach that yields results reasonably close to those obtained using the integral approach but significantly reduces the computational requirement. 
In this hybrid model, we disregard the differences between fictitious CDM halos and actual simulated CDM halos at $z=0$. Instead, we incorporate all accretion effects through the gravothermal phase at $z=0$, calculated as
$$
\tau(z=0) = \int_{0}^{10~\rm Gyr} \frac{d t}{t_{c,b}[\sigma_{\rm eff}/m,\rho_s,r_s,\rho_H(t),r_H(t)]}.
$$
Given this gravothermal phase, obtaining the density profile becomes identical to that of the basic approach. 
In Appendix~\ref{app:checkphase}, we provide detailed comparisons between the integral and hybrid approaches. Our analysis shows that the hybrid approach effectively captures the overall trends and variations associated with SIDM, with secondary effects from the integral approach being suppressed once baryonic contributions to $V_{\rm max}$ and $R_{\rm max}$ are included. 
When applying our model, we recommend starting with the hybrid approach and then assessing whether the more complex integral approach is necessary based on the problem's required precision.

In Fig.~\ref{fig:sampled}, we present the parametric model predictions for three scenarios: CDM (left column), a constant SIDM cross section of $\sigma/m = 10 , \rm cm^2/g$ (middle column), and a velocity-dependent SIDM cross section (right column) given by
\begin{eqnarray}
\label{eq:rxs}
\frac{d\sigma}{d \cos\theta} = \frac{\sigma_{0}w^4}{2\left[w^2+{v^{2}}\sin^2(\theta/2)\right]^2 },
\end{eqnarray}
where $\sigma_0/m=100~\rm cm^2/g$ and $w=100~\rm km/s$. 
Each column displays results across three cases: dark matter-only (top row), dark matter with baryon feedback (middle row), and full matter including baryons (bottom row).

In the CDM case, results for the dark matter-only scenario are color-coded by halo concentration, revealing that higher concentrations are associated with lower values of $V_{\rm max}$ and $R_{\rm max}$. When baryons are included, results are color-coded by the stellar-to-halo mass ratio to highlight the impact of stellar mass, which tends to compactify and increase the mass of systems, especially in halos around the Milky Way scale ($\sim10^{12}\rm M_{\odot}$). The middle row indicates that baryons have a moderate effect on the dark matter $V_{\rm max,DM}$ and $R_{\rm max,DM}$. This is because the NFW halos are already cuspy. 

In the SIDM models, color-coding reflects halos’ gravothermal phases at $z=0$. 
Here, baryonic effects are pronounced, widening the data spread and, in massive galaxies, inducing core-collapsing phases with decreased $R_{\rm max,DM}$ and slightly increased $V_{\rm max,DM}$. The constant cross section model ($\sigma/m = 10~\rm cm^2/g$) shows core-collapsing halos predominantly appear at high $V_{\rm max} \gtrsim 300~\rm km/s$, whereas the velocity-dependent model displays more core collapse at lower velocities ($V_{\rm max} \lesssim 60~\rm km/s$).

The middle row of Fig.~\ref{fig:sampled} presents results for the dark matter component alone, illustrating the effect of baryons on dark matter content, with and without SIDM, to address theoretical interests. 
The bottom row presents more realistic outcomes, visualizing $V_{\rm max}$ and $R_{\rm max}$ with baryonic contributions included. 
Comparing these with the middle panels reveals how baryons diminish SIDM signatures, especially around $V_{\rm max}\approx 200~\rm km/s$, where the baryon-to-dark-matter mass ratio peaks. 
Nevertheless, distinct differences remain across the three cases after accounting for baryons. 
Reducing observational uncertainty will be crucial to test analogous SIDM scenarios using variables related to rotation curves.

We have provided scripts for this analysis on GitHub, accessible at \url{https://github.com/DanengYang/parametricSIDM}. Readers can employ the codes to investigate various SIDM models.

\subsection{Deciphering SIDM models from gravothermal phases and accretion histories}
\label{sec:extraction}

\begin{figure}[htbp]
  \centering
  \includegraphics[width=8cm]{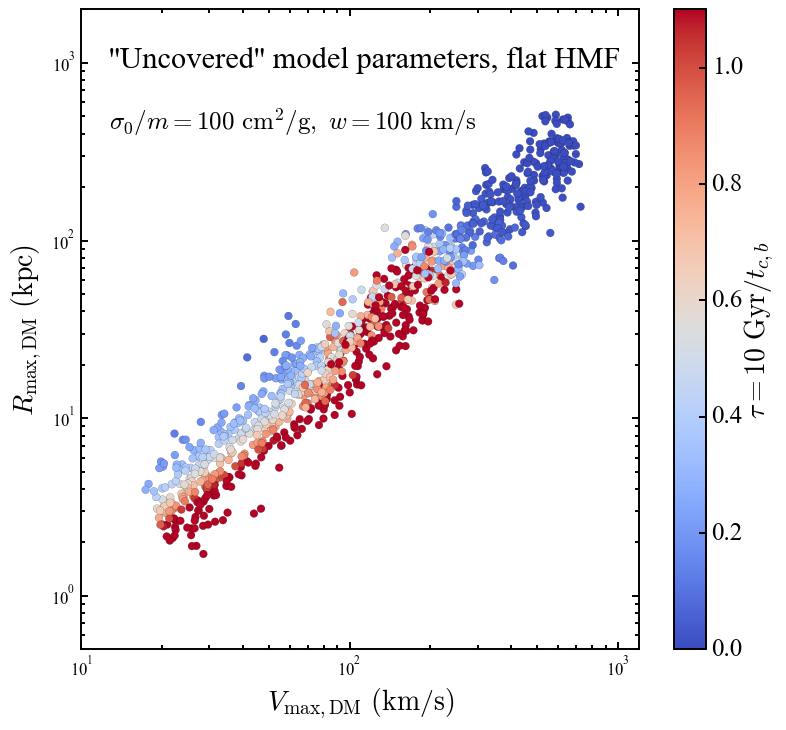}
  \caption{\label{fig:uncoverdCDM} Distributions of maximum circular velocity ($V_{\rm max,DM}$) and radius at which this velocity occurs ($R_{\rm max,DM}$) for CDM halos, color-coded by the $\tau$ for the $\sigma_0/m=100~\rm cm^2/g$ and $w=100~\rm km/s$ model obtained through the integral approach. In reality, these quantities could be uncovered by analyzing rotation curves. 
  The presented results are actual values that correspond to the analysis in the middle right panel of Fig.~\ref{fig:sampled}, where the halos are uniformly sampled using a flat halo mass function over the mass range $M_h\in [10^9,10^{14}]\rm \ M_{\odot}$.
}
\end{figure}

In Fig.~\ref{fig:sampled}, we have presented predictions from two SIDM models for the parameters $V_{\rm max}$, $R_{\rm max}$, and the gravothermal phase $\tau$. 
It appears that different SIDM models influence these quantities in varied ways. 
Assuming these parameters can be uncovered from observations, such as through fitting rotation curves, identifying the most plausible SIDM model might be feasible. However, extracting reliable data from observations and generating precise theoretical predictions pose significant challenges. A notable issue is that observations of individual galaxies capture only a single moment in their complex evolutionary histories. 
Consequently, one must test theoretical predictions across a broad range of possible accretion histories and models, which requires substantial effort.

The concept of gravothermal states facilitates contending with this challenge through a model-independent insight. 
By rewinding the current phase $\tau$ of each halo's profile---modeled using a parametric model with baryons---to zero, we effectively connect it to a fictitious CDM halo. 
This framework presents several advantages when interpreting observational data. 
First, the gravothermal phases derived from observations are not influenced by the galaxies' evolutionary histories, allowing researchers to focus solely on fitting the halos’ gravothermal states without concern for the underlying SIDM model dynamics.
Instead, the search for a plausible SIDM model is left to a post-processing process, where one can focus on the subtle correlations between SIDM models and accretion histories. 
Additionally, given our extensive knowledge about CDM halos, we could leverage them to constrain the parametric model's CDM input (fictitious CDM halos), thereby aiding the extraction of gravothermal phases from data.

In Fig.\ref{fig:uncoverdCDM}, we illustrate a heuristic representation of ``uncovered'' results, which should come from data analysis in reality, but are depicted using the predictions of the velocity-dependent SIDM model with $\sigma_0/m = 100\rm cm^2/g$ and $w=100~\rm km/s$. 
The figure demonstrates $V_{\rm max,DM}$ and $R_{\rm max,DM}$ of fictitious CDM halos, color-coded by the phase $\tau$. 
These quantities can all be derived from the halos' current gravothermal states. 
The color coding of the phase already illustrates distinguished features associated with the velocity-dependent cross section exemplified here, and the reliability and representativity can be further explored by post-processing the results for individual halos.

\begin{figure}[htbp]
  \centering
  \includegraphics[width=7.3cm]{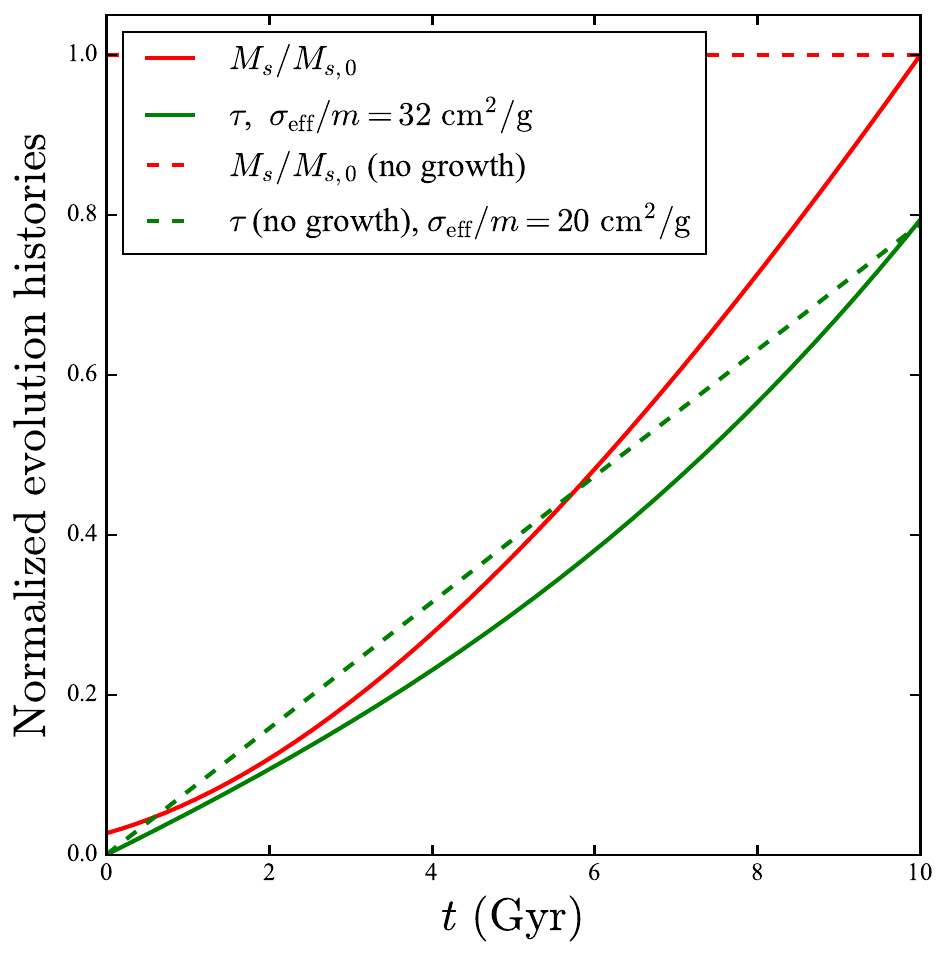}
  \caption{\label{fig:post} The normalized evolution histories of the stellar mass (red) and the gravothermal phase (green). An example post-processing is applied to extract the SIDM model for the snapshot of {\it DM11+baryon} simulation at $\tau=0.8$. The stellar mass and size evolution are modeled as $M_s(z)=M_{s,0}e^{-2z}$ and $r_H=r_{H,0}(1+z)^{-0.75}$, respectively. The evolution of $\tau$ is modeled in the integral approach to reach the value $0.8$ from the Cored-DZ fit. The case without stellar accretion is shown with dashed lines for comparison. 
}
\end{figure}

To illustrate this post-processing process, we take the snapshot of the {\it DM11+baryon} simulation at $\tau_0=0.8$ for a concrete example. 
This halo has $\rho_s=6.9\times 10^6~\rm M_{\odot}/kpc^3$ and $r_s=9.1~$kpc. 
Assuming the halo formed $10$ Gyr ago, we determine its core collapse time to be $t_c=10~{\rm Gyr}/\tau_0 \approx 12.5~$Gyr. The SIDM models capable of yielding this $t_c=12.5$ Gyr depend on the modeling of the baryon contents. In scenarios without baryons, we use Eq.~(\ref{eq:tcdmo}) to calculate the SIDM cross section, yielding $\sigma_{\rm eff}/m=53~\rm cm^2/g$. 
Alternatively, assuming the baryon potential remains fixed based on present-day observations, we utilize Eq.~(\ref{eq:tcb}) with $\rho_H=3.6 \times 10^8\rm M_{\odot}/kpc^3$ and $r_H=0.77~$kpc. This results in a reduced effective cross section of $20~\rm cm^2/g$. 

To examine how accretion history influences the interpretation, we analyze the influence of a toy accretion history. 
Here, we assume the halo to have established the $\rho_s$ and $r_s$ since 10 Gyr ago and stellar mass grow according to $M_s(z)=M_{s,0}e^{-2z}$, respectively, with subscript ``0'' indicating present day values. The stellar size evolution is modeled as $r_H=r_{H,0}(1+z)^{-0.75}$, as we choose for the sampled halos.
To determine the effective cross section, we solve for 
$$
\tau_{0} = \int_{0}^{10~\rm Gyr} \frac{d t}{t_{c,b}[\sigma_{\rm eff}/m,\rho_s,r_s,\rho_H(t),r_H(t)]},
$$
where $\rho_H$ is derived from $M_s$ and $r_H$ assuming a Hernquist profile at each time. The resulting effective cross section is $32~\rm cm^2/g$. In Fig.~\ref{fig:post}, we depict the normalized stellar mass, and gravothermal phase evolution using red, and green curves, respectively.
The case without stellar accretion is shown with dashed lines for comparison. 

Through this analysis, it becomes evident that the inclusion of a baryon potential significantly influences the derivation of an SIDM model, even when assuming a constant cross section. Neglecting the baryon effect results in an SIDM cross section approximately $66$\% larger than that from the integral model. Conversely, when incorporating the current baryon profile without considering accretion history, the derived SIDM cross section is approximately 38\% smaller.

This post-processing approach integrates specific SIDM models with various accretion histories to model gravothermal evolution. For a given SIDM model, one can evaluate the predictions using a range of accretion history scenarios to determine if the uncovered phases are consistently explained under certain scenarios. If a model outperforms others when paired with well-understood accretion histories, it becomes a viable SIDM scenario that could explain observational data and warrants further investigation, especially using cosmological simulations. 
Importantly, the post-processing of many halos under vast possibilities of accretion histories and SIDM models is most effectively conducted using a parametric approach due to its numerical efficiency.

\section{Conclusion and discussion}
\label{sec:conclusion}

In this study, we expand upon the parametric model introduced in Ref.~\cite{yang:2023jwn} to encompass the influence of baryons. Our model is based on two key elements: a formula for calculating core collapse time (Eq.~(\ref{eq:tcb})) and a density profile model (Cored-DZ) calibrated through controlled N-body simulations. These elements successfully incorporated the boost of gravothermal evolution and adiabatic contraction induced by the presence of baryons. 

Utilizing this model, we investigate the interplay between SIDM and baryonic distributions within the $\rho_H/\rho_s$ versus $r_H/r_s$ plane. As depicted in Fig.~\ref{fig:calF}, the influence of baryons on gravothermal evolution can vary significantly; halos in the upper-right region, characterized by larger $\rho_H/\rho_s$ and $r_H/r_s$ ratios, experience substantial boost in the evolution, while those in the lower-left region remain relatively unaffected. In Fig.~\ref{fig:rc}, we further examine the effect of SIDM on the baryon distribution by analyzing the ratio of maximum core size to baryon scale radius $r_{c,\rm max}/r_H$. We demonstrate that baryons corresponding to the lower-left region are more susceptible to gravothermal evolution. Above the $r_{c,\rm max}=r_H$ contour, halos may have suppressed or omitted core formation behavior, accompanied by contracted baryon profiles. 
These interactions could be systematically incorporated into the framework of the integral approach, which we propose to explore in-depth in future studies. 

\begin{figure*}[htbp]
  \centering
  \includegraphics[width=7.2cm]{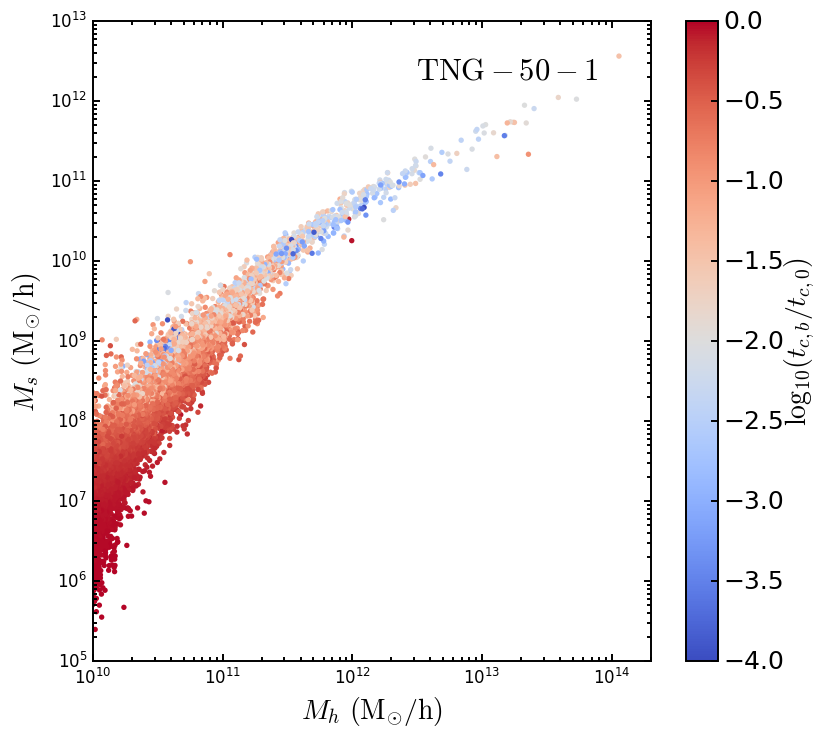}
  \includegraphics[width=7.2cm]{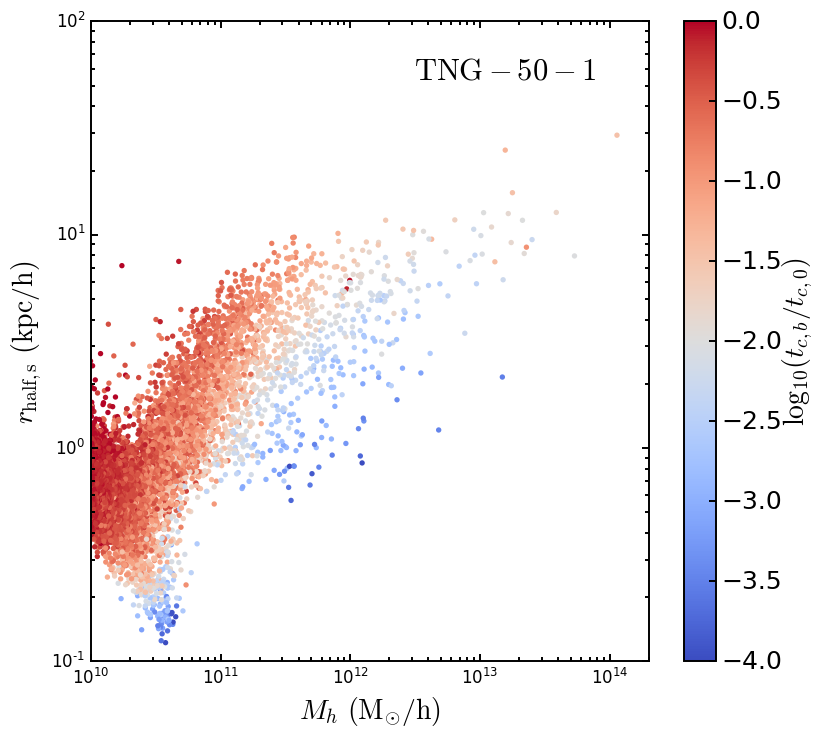} \\
  \includegraphics[width=7.2cm]{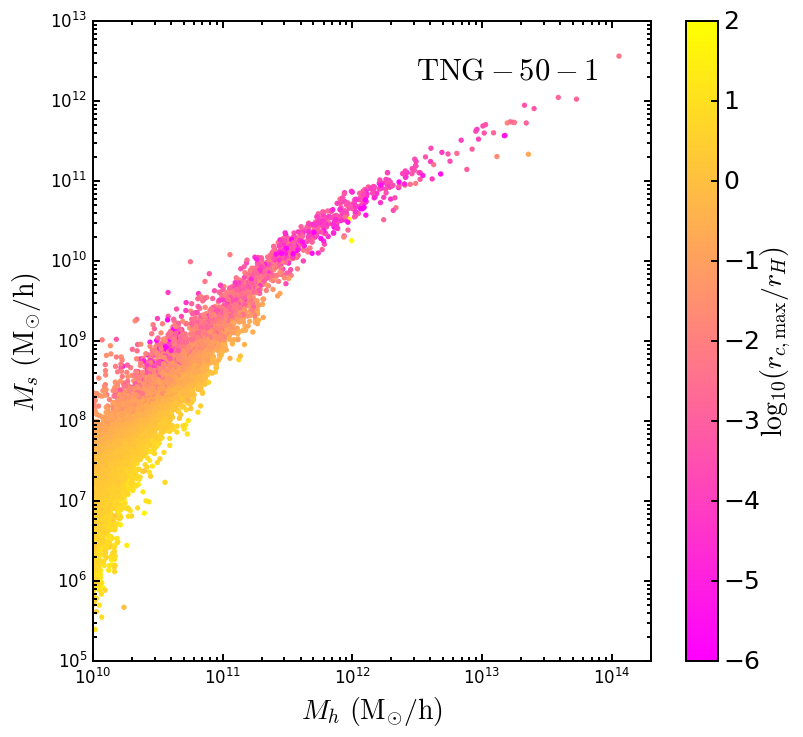}
  \includegraphics[width=7.2cm]{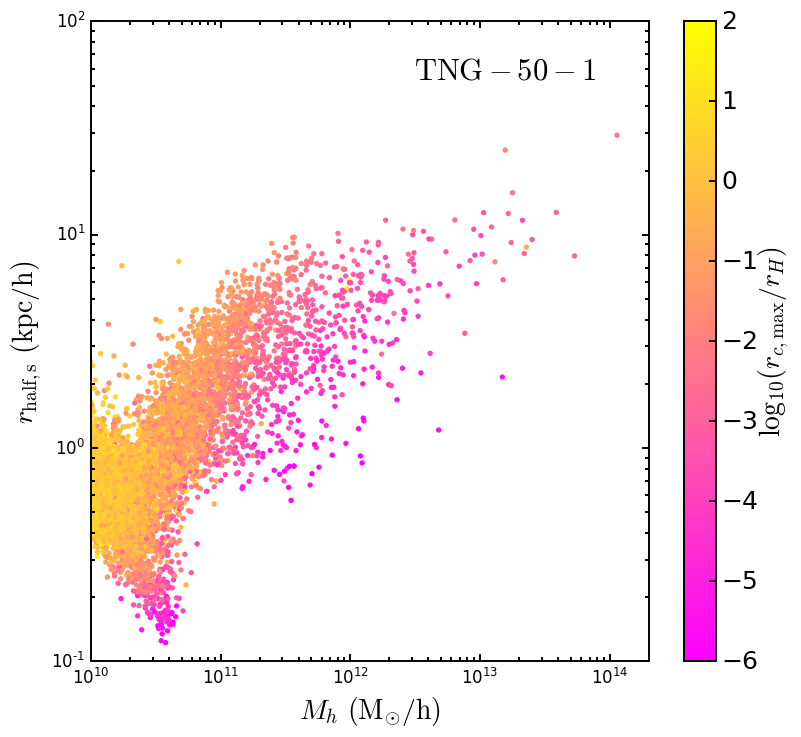}
  \caption{\label{fig:TNG2}
  The diverse influence of baryons on SIDM effects, illustrated through color-coded values of $t_{c,b}/t_{c,0}$ (top) and $r_{c,\rm max}/r_H$ (bottom) for isolated halos in TNG-50-1, displayed across the $M_h$–$M_s$ plane (left) and the stellar half-mass radius ($r_{\rm half,s}$) to halo mass ($M_h$) plane (right).
At fixed $M_h$ values, both $t_{c,b}/t_{c,0}$ and $r_{c,\rm max}/r_H$ can exhibit broad distributions, reflecting significant variations in SIDM effects.
}
\end{figure*}

Our study also illustrate the diverse influence of baryons on SIDM effect.
To visualize this diversity, we further provide in Fig.~\ref{fig:TNG2} scatter plots for isolated halos in TNG-50-1~\cite{nelson:2018uso,nelson:2019jkf,pillepich:2019bmb} on the $M_h$–$M_s$ plane (left) and the halo mass versus stellar half-mass radius ($r_{\rm half,s}$) plane (right), with points color-coded by $t_{c,b}/t_{c,0}$ (top panel) and $r_{c,\rm max}/r_H$ (bottom panel).
At fixed $M_h$ values, both $t_{c,b}/t_{c,0}$ and $r_{c,\rm max}/r_H$ exhibit broad distributions, indicating significant variations in SIDM effects.
In the $M_h$–$M_s$ panels, we observe a strong diversity at $M_h \lesssim 3 \times 10^{11}\rm M_{\odot}$. At the lower mass end, the baryonic influence on SIDM halos diminishes, yet the maximum core size, $r_{c,\rm max}$, can vary widely, being either much larger or much smaller than the stellar $r_H$.
The distribution of blue and magenta points aligns qualitatively with the stellar-to-halo mass relation, which peaks around $M_h \approx 10^{12}\rm M_{\odot}$.
In the $r_{\rm half,s}$–$M_h$ panels, this diversity extends across a wider mass range. Notably, we identify two regions where both $t_{c,b}/t_{c,0}$ and $r_{c,\rm max}/r_H$ are small: one near the peak of the stellar-to-halo mass relation ($M_h \approx 10^{12}\rm M_{\odot}$) and another around $M_h \approx 3 \times 10^{10}\rm M_{\odot}$, with $r_{\rm half,s}$ values among the lowest in the sample.

In this work, we present a simplified application by considering sampled halos in isolation. We assume that the NFW parameters $\rho_s$ and $r_s$ have remained unchanged in the sampled halos over the past $10$ Gyr. Additionally, we implement simplified accretion histories for the stellar components of all candidates. This method allows us to concentrate on the influence of baryons on gravothermal evolution. It also serves as a reasonable approximation of the behavior observed in isolated halos within cosmological simulations. 
During the hierarchical structure formation, the inner configurations of isolated halos, characterized by the $\rho_s$ and $r_s$, are established early and remain relatively stable in the context of CDM~\cite{Ludlow:2013bd,Correa:2015dva}. See Fig.~4 of Ref.~\cite{yang:2023jwn} for a visual representation. Conversely, the baryonic component can exhibit more pronounced growth, particularly in recent epochs. As depicted in Fig.~7 of Ref.~\cite{2013ApJ...770...57B}, halos within the mass range $10^{11}-10^{12}~\rm M_{\odot}$ demonstrate an increasing stellar-to-halo mass ratio ($M_s/M_h$) from redshift two until the present day. 
For more massive halos, the evolution of $M_s/M_h$ becomes more moderate, reflecting varied impacts from baryonic components.
Our simplified model features the growth of baryons atop stationary CDM halos, which is more suitable for halos in the $10^{11}-10^{12}~\rm M_{\odot}$ range. 

We also illustrate the effectiveness of the parametric model with baryons in probing SIDM models. By assuming that each halo resides in a specific gravothermal phase, retrievable through observational analysis, we illustrate how the SIDM interpretations depend on the accretion histories and SIDM models. 
Exploring the SIDM models across a population of galaxies could enable the identification of SIDM models that consistently explain all the observations. 

Crucially, the assumption that each SIDM halo is in a gravothermal state defined by $\tau$ operates independently of the parametric model. This assumption allows for the use of controlled N-body simulations to accurately model any halo's state within a cosmological simulation, as long as their $\tau$ phases match. For dark matter-only halos, the validity of this assumption is corroborated by comparisons between the model and simulations, as detailed in Refs.~\cite{yang:2023jwn} and \cite{Yang:2024uqb}, the latter of which examines the relative differences on a case-by-case basis.
In scenarios involving baryons, however, verification via N-body simulations remains to be shown. Nonetheless, as shown in Appendix~\ref{app:checkphase}, the reliability of this assumption supports theoretical explorations. Conducting dedicated N-body tests of this assumption would facilitate obtaining accurate theoretical predictions in future studies.

Applications illustrated in this work are mostly proof-of-principle. Despite their simplified nature, they provide valuable insights into the complex interactions between SIDM halos and baryonic distributions. These preliminary findings underscore the potential of further refining parametric models to deepen our understanding of the self-interacting characteristics of dark matter. Continued advancements in this direction can enhance our ability to decipher the underlying dynamics and implications of SIDM in cosmological contexts.

\section*{Acknowledgments}

I would like to thank Hai-Bo Yu, Ethan O. Nadler, Yi-Ming Zhong, Shin'ichiro Ando, and Demao Kong, for their helpful discussions. Special appreciation goes to Hai-Bo Yu for helpful comments on fitting rotation curves and SIDM's feedback on baryons. I also acknowledge the IllustrisTNG collaborations for making their simulation data publicly available. 
This work was supported in part by the U.S. Department of Energy under grant No.\ de-sc0008541, 
the National Key Research and Development Program of China (No. 2022YFF0503304), the Project for Young Scientists in Basic Research of the Chinese Academy of Sciences (No. YSBR-092), and the John Templeton Foundation under grant ID \#61884. 
The opinions expressed in this publication are those of the authors and do not necessarily reflect the views of the John Templeton Foundation.

\appendix

\section{The DM-only limit of the Cored-DZ model}

\begin{figure}[htbp]
  \centering
  \includegraphics[width=7.1cm]{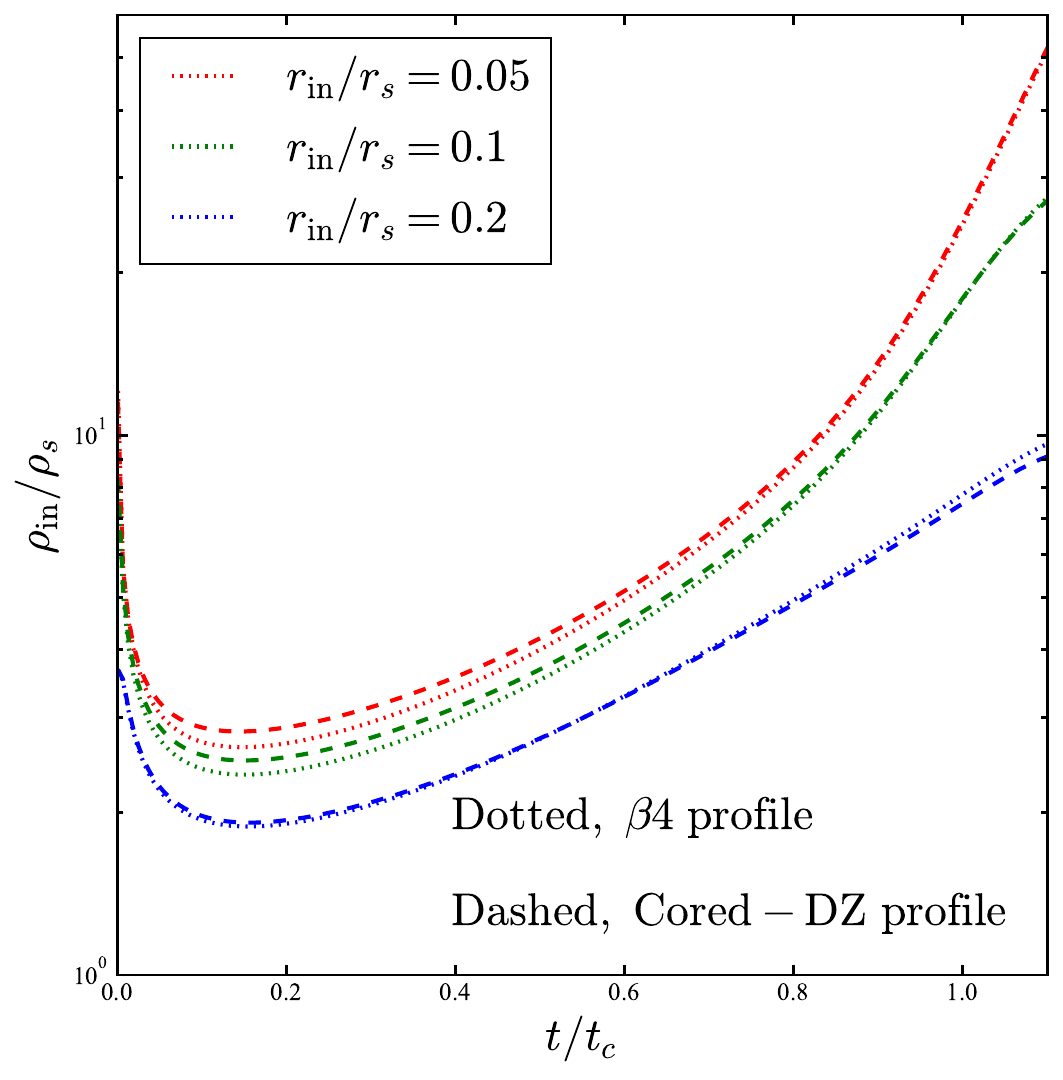} 
  \includegraphics[width=7.1cm]{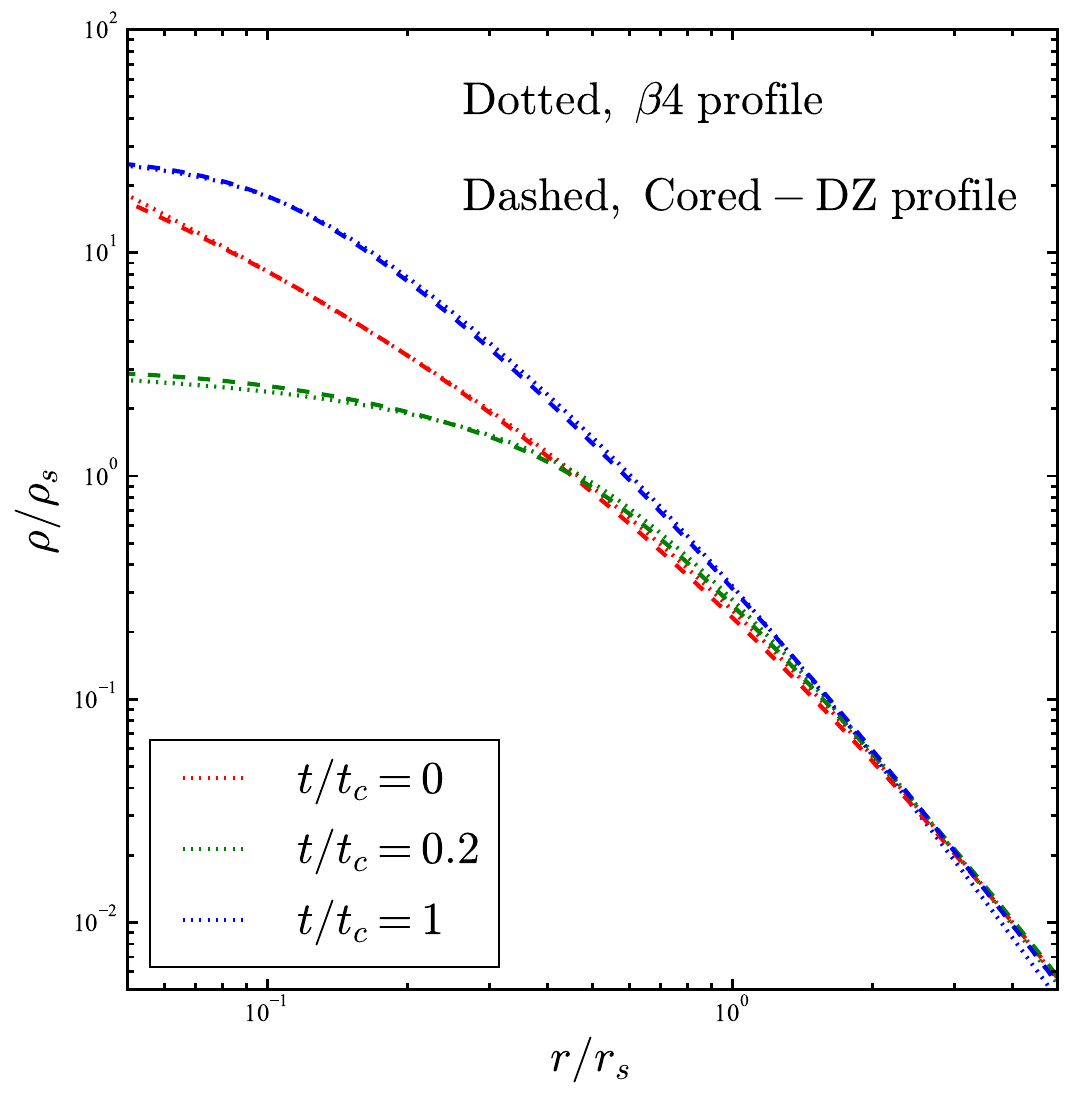}
  \caption{\label{fig:NFWDZ} The inner density evolution (top) and the DM density profiles (bottom) based on the parametric model for the $\beta4$ profile (dotted) and the Cored-DZ profile (dashed). }
\end{figure}

As a validation, we show in this appendix that the Cored-DZ model reduces to the $\beta4$ model in the DM-only limit for a large range in the density profile. 
For this purpose, we follow the equations in Ref.~\cite{2020MNRAS.499.2912F} to compute the $a$ and $c$ parameters of the DZ profile through two other parameters $s_1$ and $c_2$ as
\begin{eqnarray}
a&=&\frac{1.5 s_1-2\left(3.5-s_1\right)\left(r_1 / R_{\mathrm{cut}}\right)^{1 / 2} c_2^{1 / 2}}{1.5-\left(3.5-s_1\right)\left(r_1 / R_{\mathrm{cut}}\right)^{1 / 2} c_2^{1 / 2}} \\ \nonumber
c&=&\left(\frac{s_1-2}{\left(3.5-s_1\right)\left(r_1 / R_{\mathrm{cut}}\right)^{1 / 2}-1.5 c_2^{-1 / 2}}\right)^2
\end{eqnarray}
where $s_1$ represents the logarithmic density slope at $0.01 R_{\rm cut}$ and $c_2=R_{\rm cut}/r_{-2}$ is determined by the radius $r_{-2}$, at which the logarithmic density slope equals $-2$.

For DM-only halos, evaluate
\begin{eqnarray}
s_{1,\rm DMO} &=& -\left( r_1^2  \rho_s^{-1} r_s^{-1} \left( -2 \rho_s r_1^{-1} (1 + r_1/r_s)^{-3} \right. \right. \\ \nonumber
               && \left. \left. - \rho_s r_s r_1^{-2} (1 + r_1/r_s)^{-2} \right) (1 + r_1/r_s)^2 \right), \\ \nonumber
c_{\rm DMO} &=& R_{\rm cut}/r_s, 
\end{eqnarray}
using the NFW parameters $\rho_s$, $r_s$ and halo mass $M_h$ computed within $R_{\rm cut}$ yields a DZ profile that largely mimics the NFW profile in the range $0.05 r_s$ to $5 r_s$. 
See the red curves in Fig.~\ref{fig:NFWDZ} for a comparison of the two.

\begin{figure*}[htbp]
  \centering
  \includegraphics[height=5.4cm]{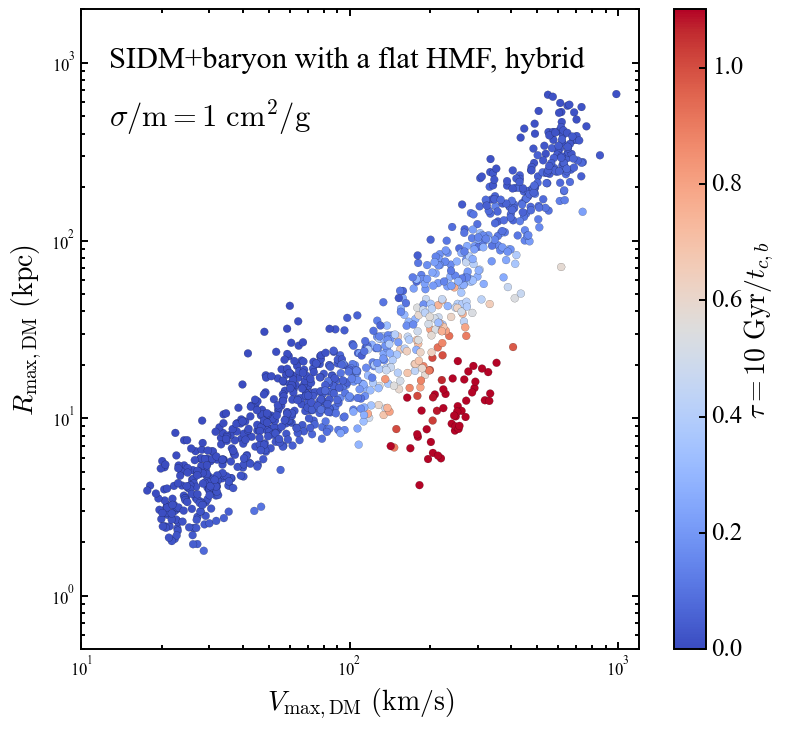}
  \includegraphics[height=5.4cm]{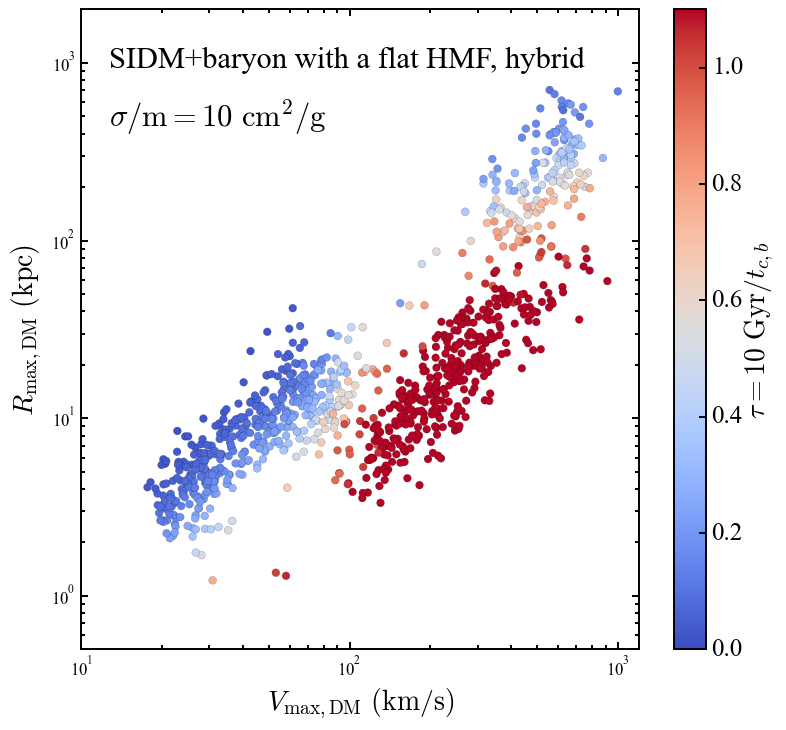}
  \includegraphics[height=5.4cm]{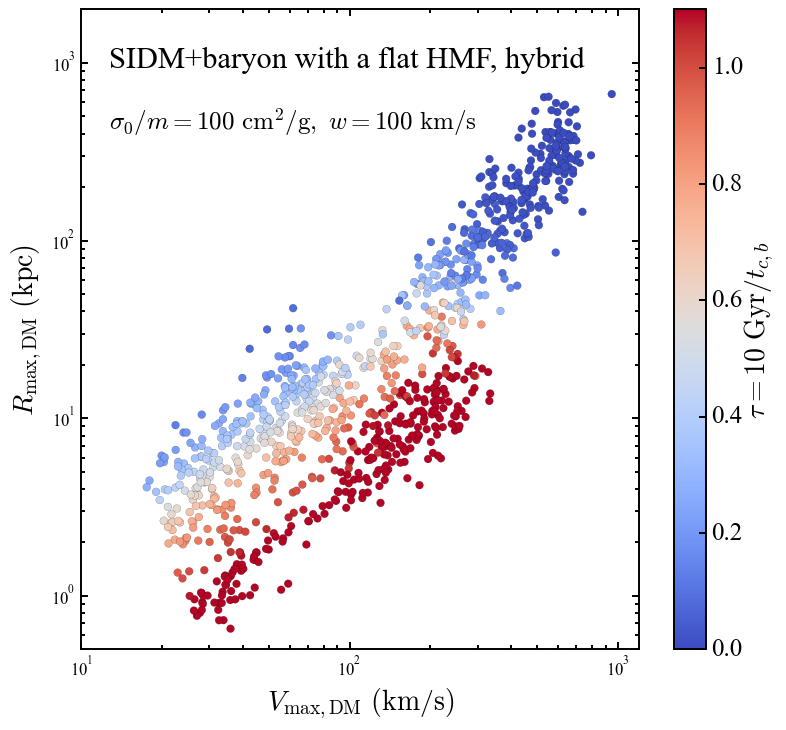} \\
  \includegraphics[height=5.4cm]{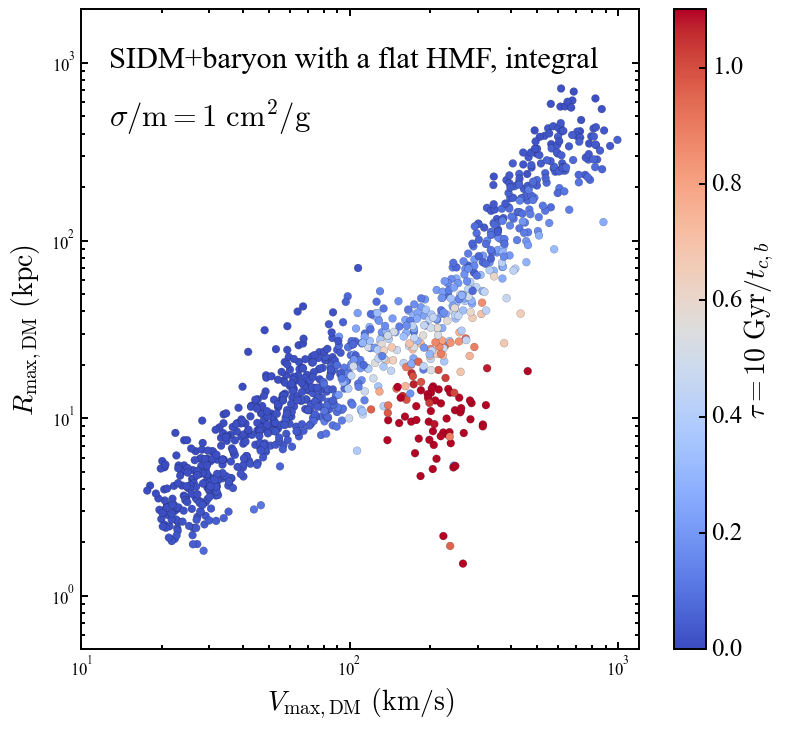}
  \includegraphics[height=5.4cm]{fig_vmax_rmax_SIDM10_baryon_flat_HMF_integral_DMO.png}
  \includegraphics[height=5.4cm]{fig_vmax_rmax_SIDM_baryon_flat_HMF_sigma0_100_w_100_integral_DMO.png}
  \caption{\label{fig:checkphase} Distributions of maximum circular velocity ($V_{\rm max,DM}$) and radius at which this velocity occurs ($R_{\rm max,DM}$) for sampled halos, considering the dark matter component under the influence of baryons. 
  The upper panels display results from the hybrid approach, while the lower panels provide results from the integral approach. 
  The predictions for three SIDM models are provided. 
  From left to right, the results correspond to cross sections per mass of $\sigma/m=1~\rm cm^2/g$, $\sigma/m=10~\rm cm^2/g$, and a velocity dependent cross section with $\sigma_0/m=100~\rm cm^2/g$ and $w=100~\rm km/s$, respectively. All halos were uniformly sampled using a flat halo mass function across the mass range $M_h \in [10^9, 10^{14}] \rm \ M_{\odot}$. 
}
\end{figure*}

\section{Testing the hybrid approach}
\label{app:checkphase}

As discussed in the main text, the hybrid approach can significantly reduce the computational complexity. Here, we present three examples to compare the performances of the hybrid and the integral approaches: two with constant cross sections per mass, $\sigma/m=1~\rm cm^2/g$, $\sigma/m=10~\rm cm^2/g$, and the other with a velocity dependent cross section specified with $\sigma_0/m=100~\rm cm^2/g$ and $w=100~\rm km/s$. 

\begin{figure*}[htbp]
  \centering
  \includegraphics[height=5.4cm]{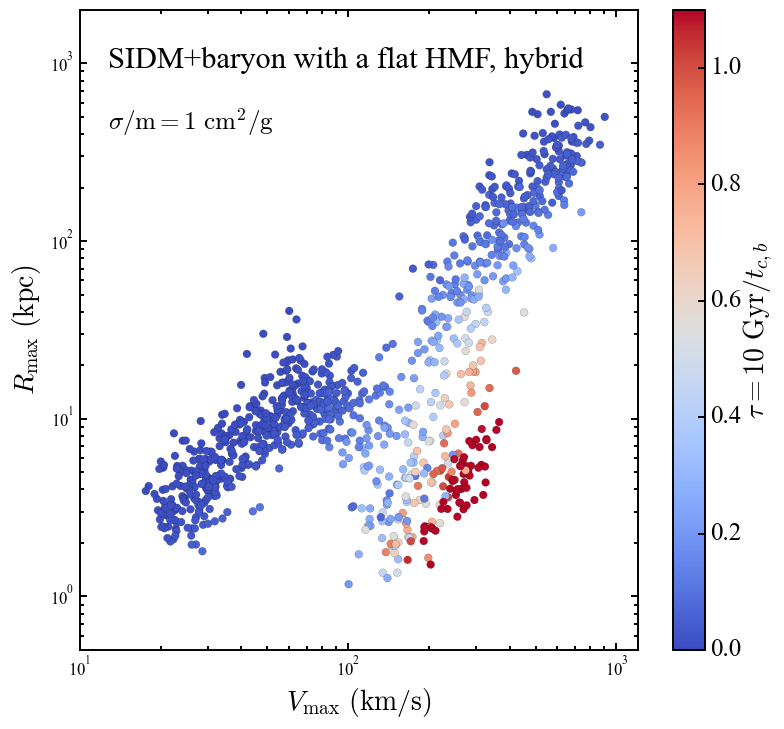}
  \includegraphics[height=5.4cm]{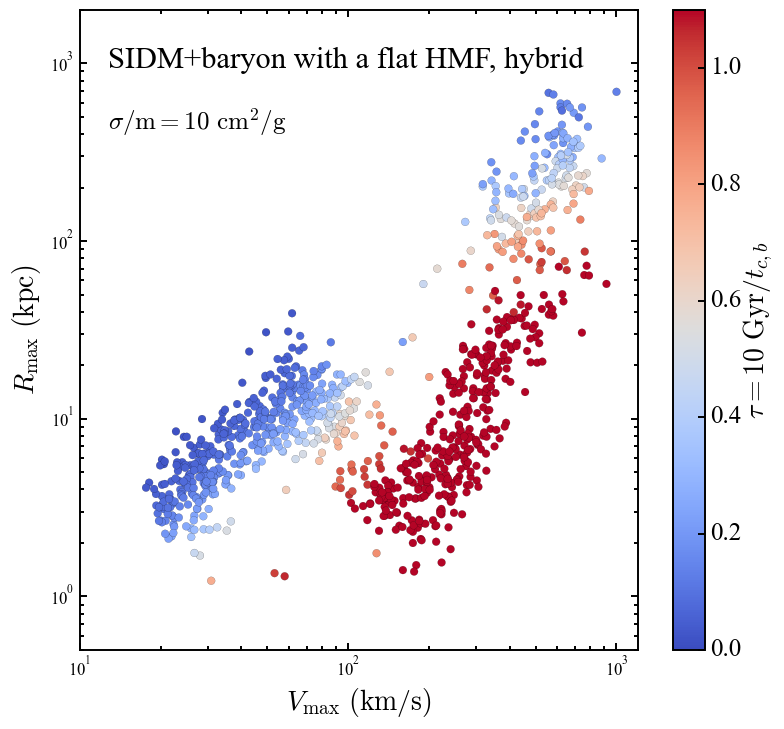}
  \includegraphics[height=5.4cm]{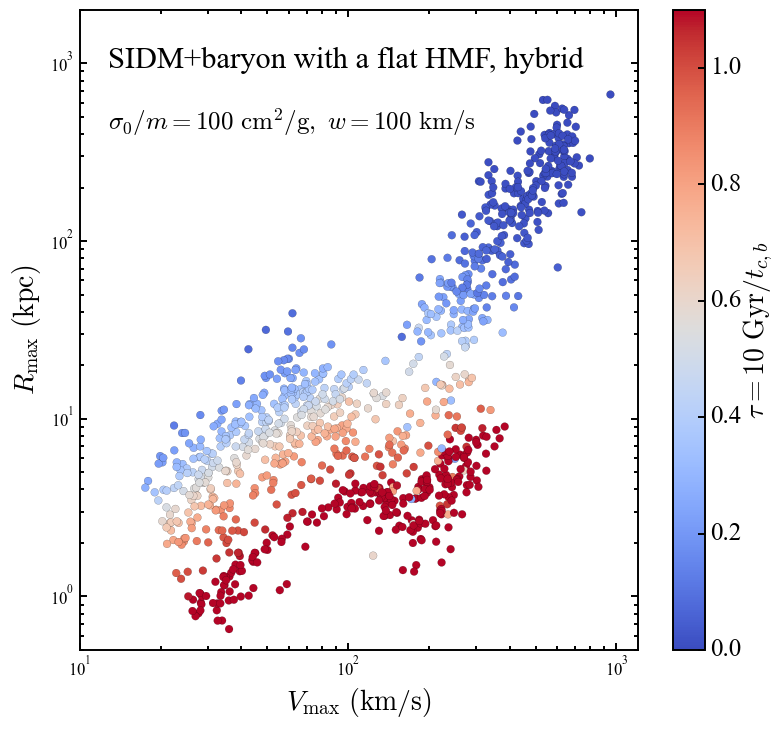} \\
  \includegraphics[height=5.4cm]{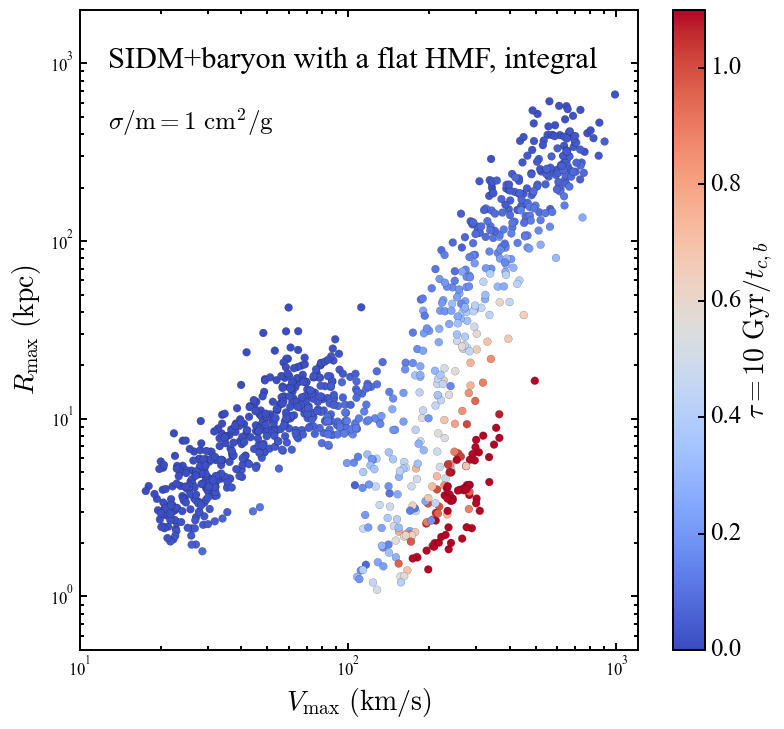}
  \includegraphics[height=5.4cm]{fig_vmax_rmax_SIDM_baryon_flat_HMF_SIDM10_integral_total.png}
  \includegraphics[height=5.4cm]{fig_vmax_rmax_SIDM_baryon_flat_HMF_sigma0_100_w_100_integral_total.png}
  \caption{\label{fig:checkphase2} Same as Fig.~\ref{fig:checkphase}, but plot the total circular velocity ($V_{\rm max}$) and radius at which this velocity occurs ($R_{\rm max}$), considering both the dark matter and baryon contributions. 
  The upper panels display results from the hybrid approach, while the lower panels provide results from the integral approach. 
  The predictions for three SIDM models are provided. 
  From left to right, the results correspond to cross sections per mass of $\sigma/m=1~\rm cm^2/g$, $\sigma/m=10~\rm cm^2/g$, and a velocity dependent cross section with $\sigma_0/m=100~\rm cm^2/g$ and $w=100~\rm km/s$, respectively. All halos were uniformly sampled using a flat halo mass function across the mass range $M_h \in [10^9, 10^{14}] \rm \ M_{\odot}$. 
}
\end{figure*}

Figure~\ref{fig:checkphase} and Fig.~\ref{fig:checkphase2} illustrate the distributions of maximum circular velocity and radius at which it occurs, for the dark matter-only contribution and the dark matter plus baryon contributions, respectively. These figures demonstrate that the hybrid approach effectively captures the overall characteristics of SIDM, particularly the spread and how it diverges from the CDM scenario. While there are notable differences in the detailed shapes, these are largely mitigated when the baryon contributions are considered in the rotation curves. Depending on the problem's complexity and the required level of precision, readers may select the appropriate method to employ.

\section{Adiabatic contraction in different models}

In this appendix, we perform a comparative analysis of three models that address baryon-induced adiabatic contraction in different ways. Alongside the Cored-DZ and Contracted $\beta4$ models, we examine an additional model that incorporates supplementary contraction effects on top of the Contracted $\beta4$ model, drawing inspiration from the approach introduced in Ref.~\cite{gnedin:2004cx}.

\begin{figure*}[htbp]
  \centering
  \includegraphics[width=5.33cm]{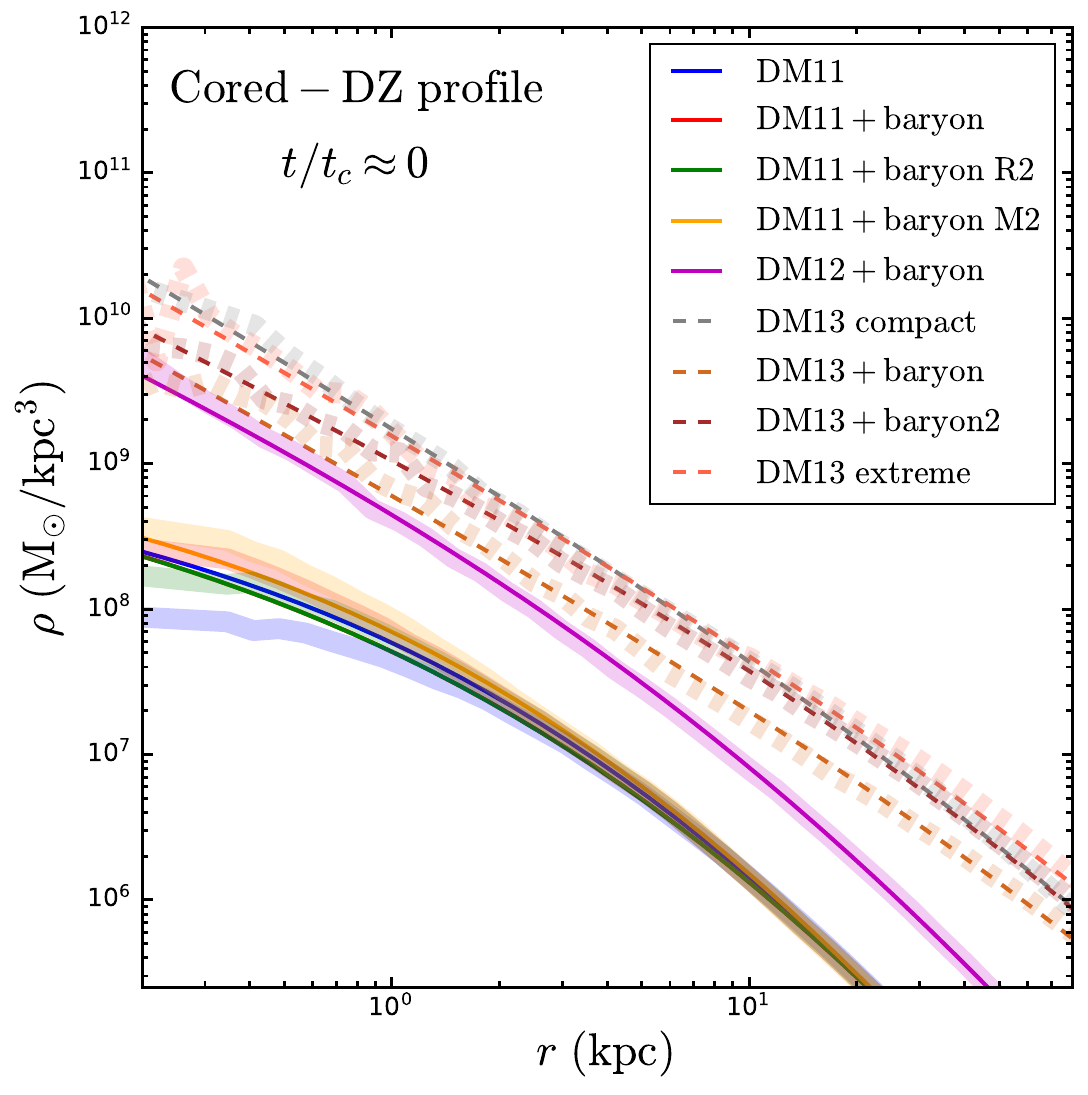}
  \includegraphics[width=5.33cm]{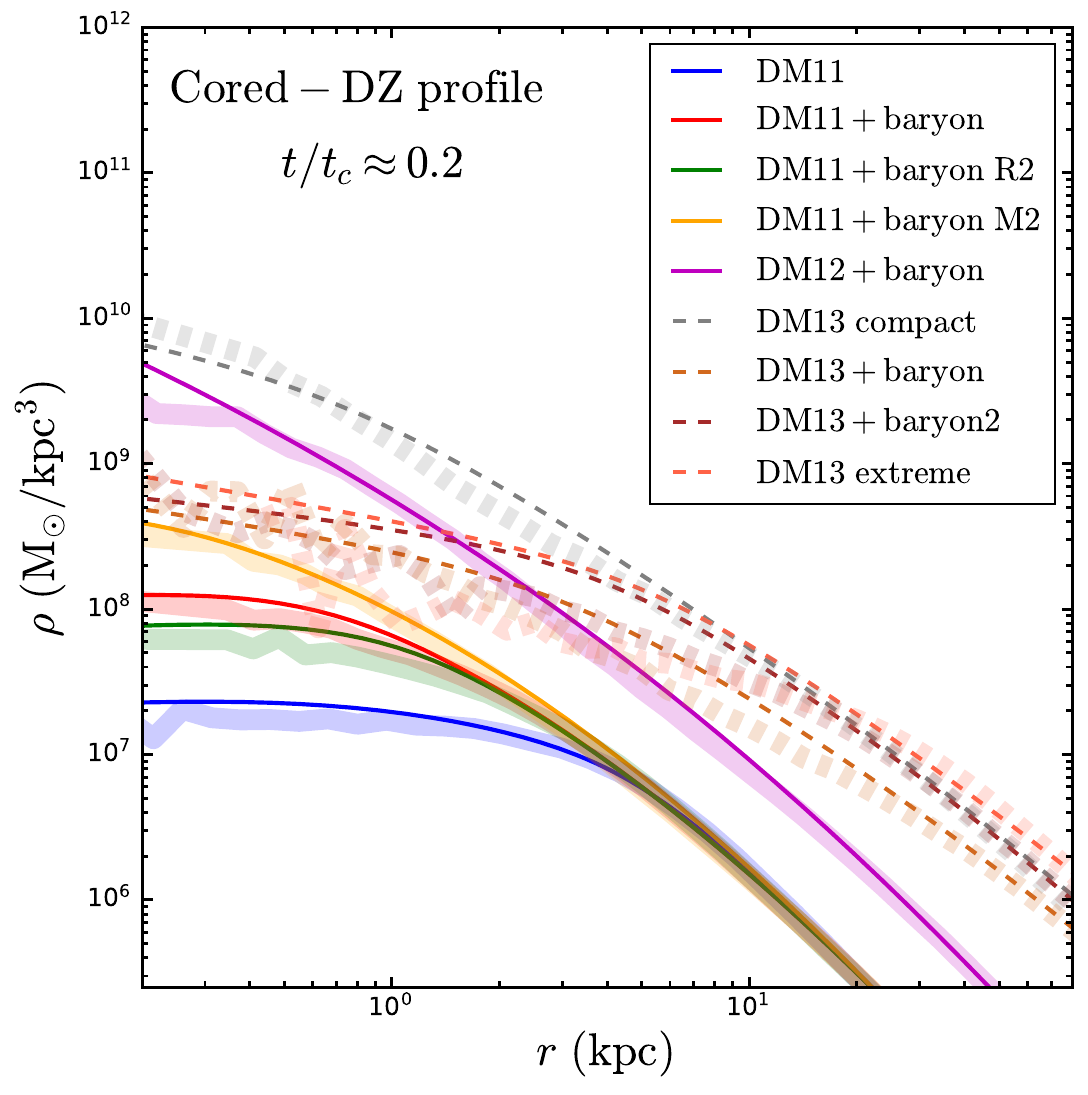}
  \includegraphics[width=5.33cm]{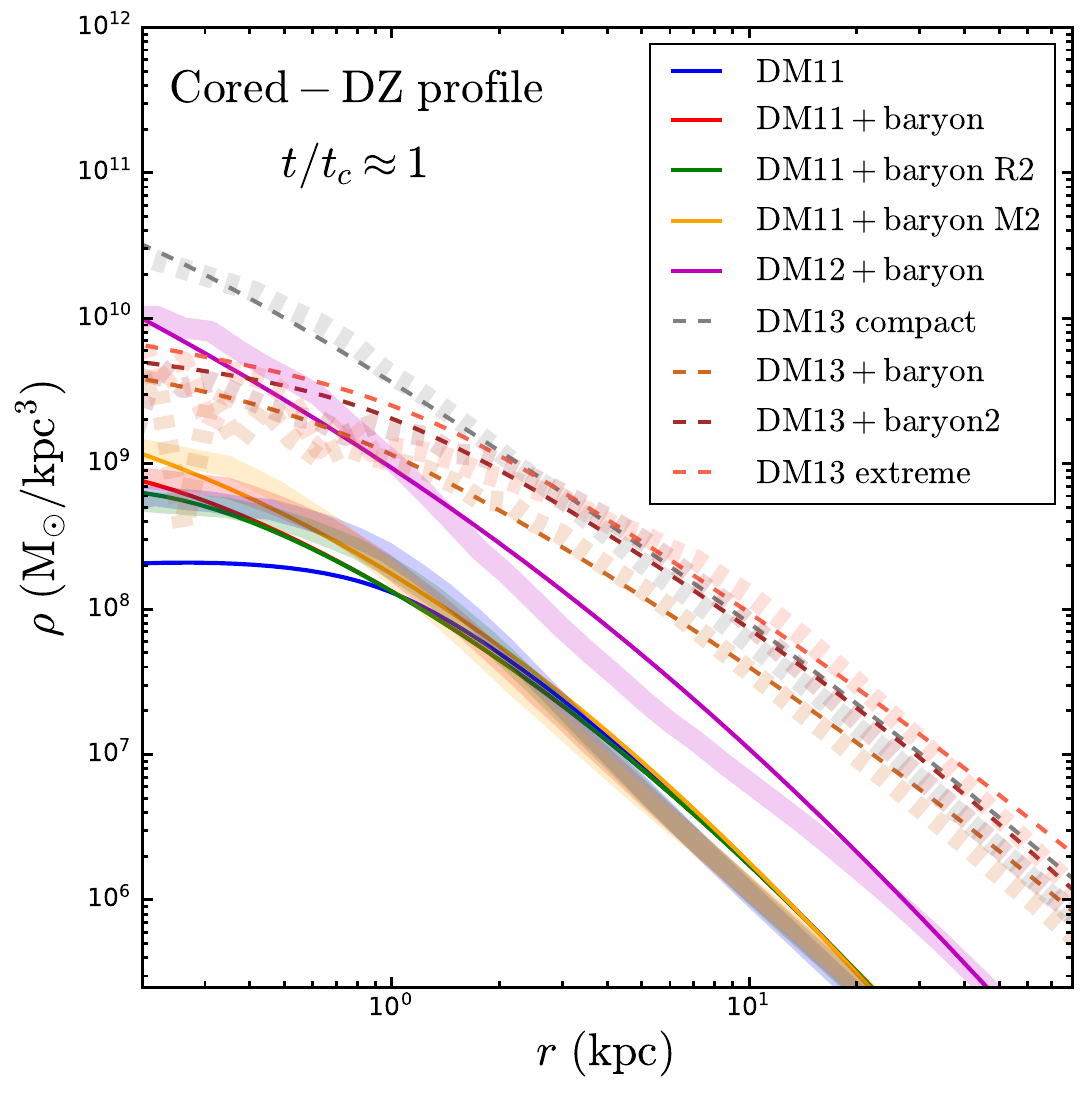}
  \caption{\label{fig:app1}
The same figure as Fig.~\ref{fig:validx}, duplicated here for an easier comparison of the other cases. The simulated (colored bands) and model predicted (colored curves) halo density profiles are presented at three representative gravothermal phases: $t/t_c\approx 0, 0.2$, and $1$. The {\it DM12} and {\it DM13} scenarios use a contracted CDM profile as the initial condition, whereas the {\it DM11} scenarios commence with an instant insertion method. In the left panel ($t/t_c \approx 0$), the {\it DM11} cases are depicted at $t = 0.25$ Gyr to allow some initial evolution away from the original NFW profile. At $t/t_c \approx 1$, the core collapse time, as calculated using Eq.~(\ref{eq:tcb}), is found to be 10\% (30\%) shorter than the simulated {\it DM13+baryon2} ({\it DM13 extreme}). To align the profiles for equivalent gravothermal phases, we adjust the timing of the simulated curves accordingly in these specific cases.
}
\end{figure*}

\begin{figure*}[htbp]
  \centering
  \includegraphics[width=5.33cm]{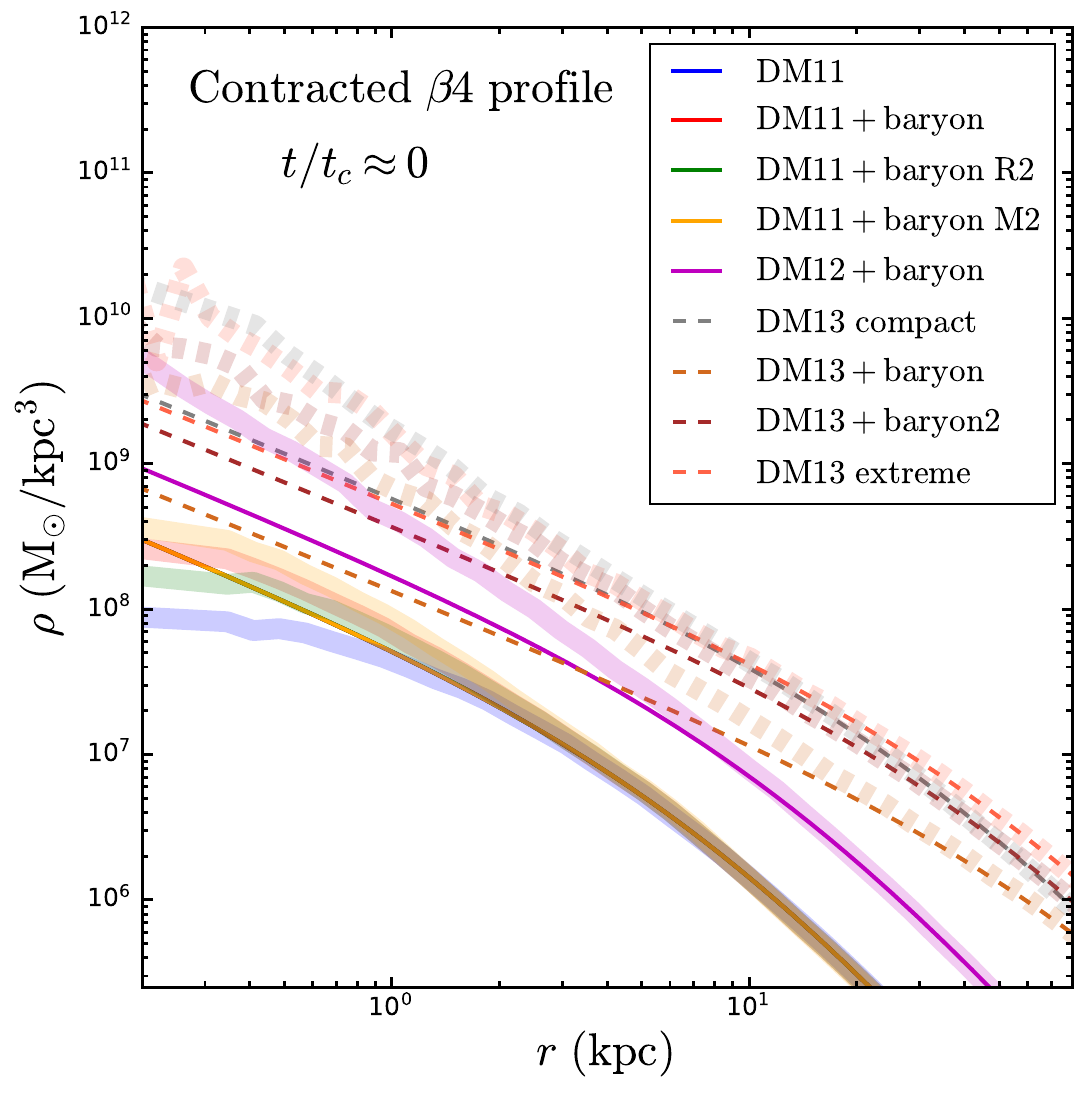}
  \includegraphics[width=5.33cm]{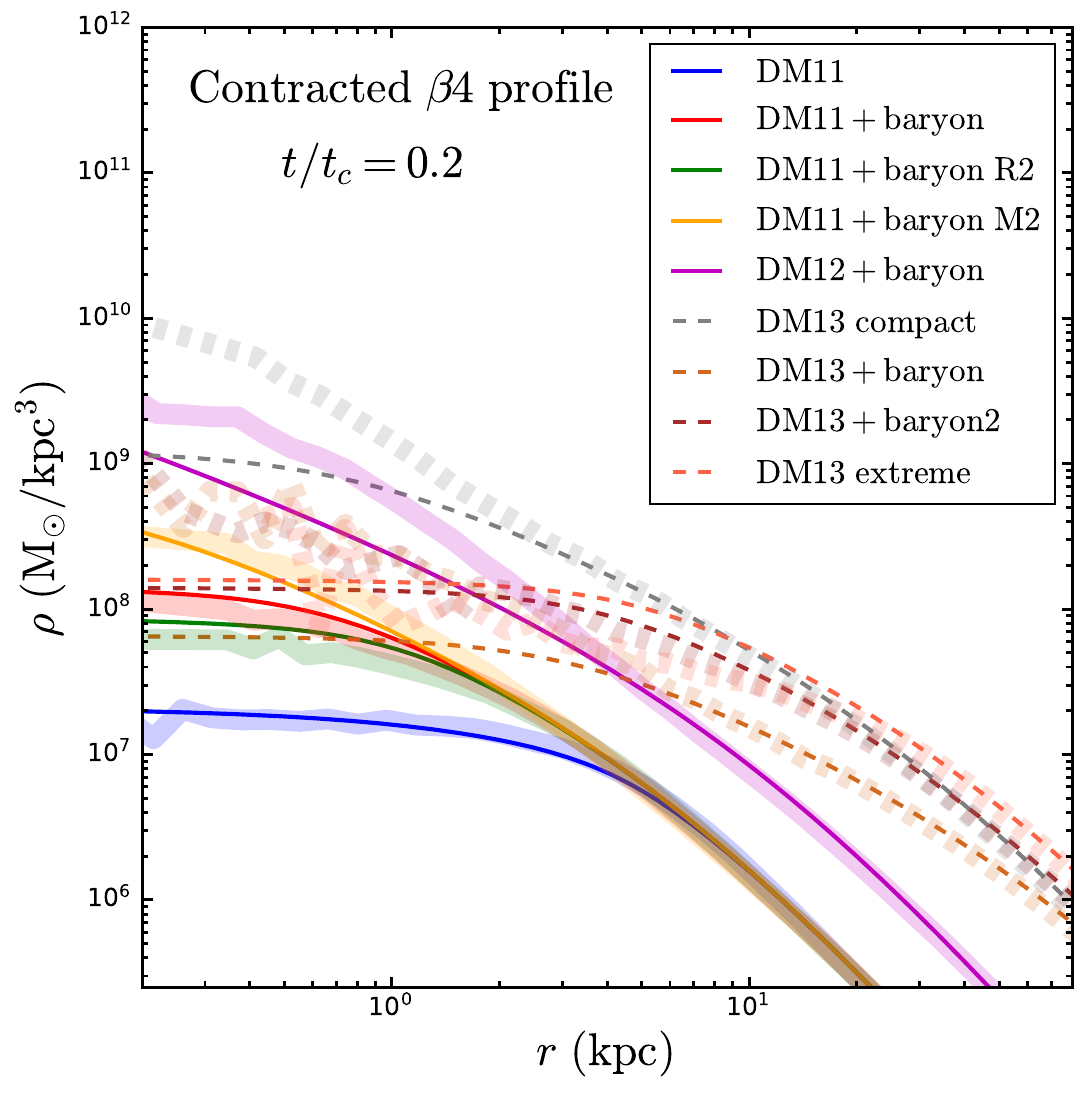}
  \includegraphics[width=5.33cm]{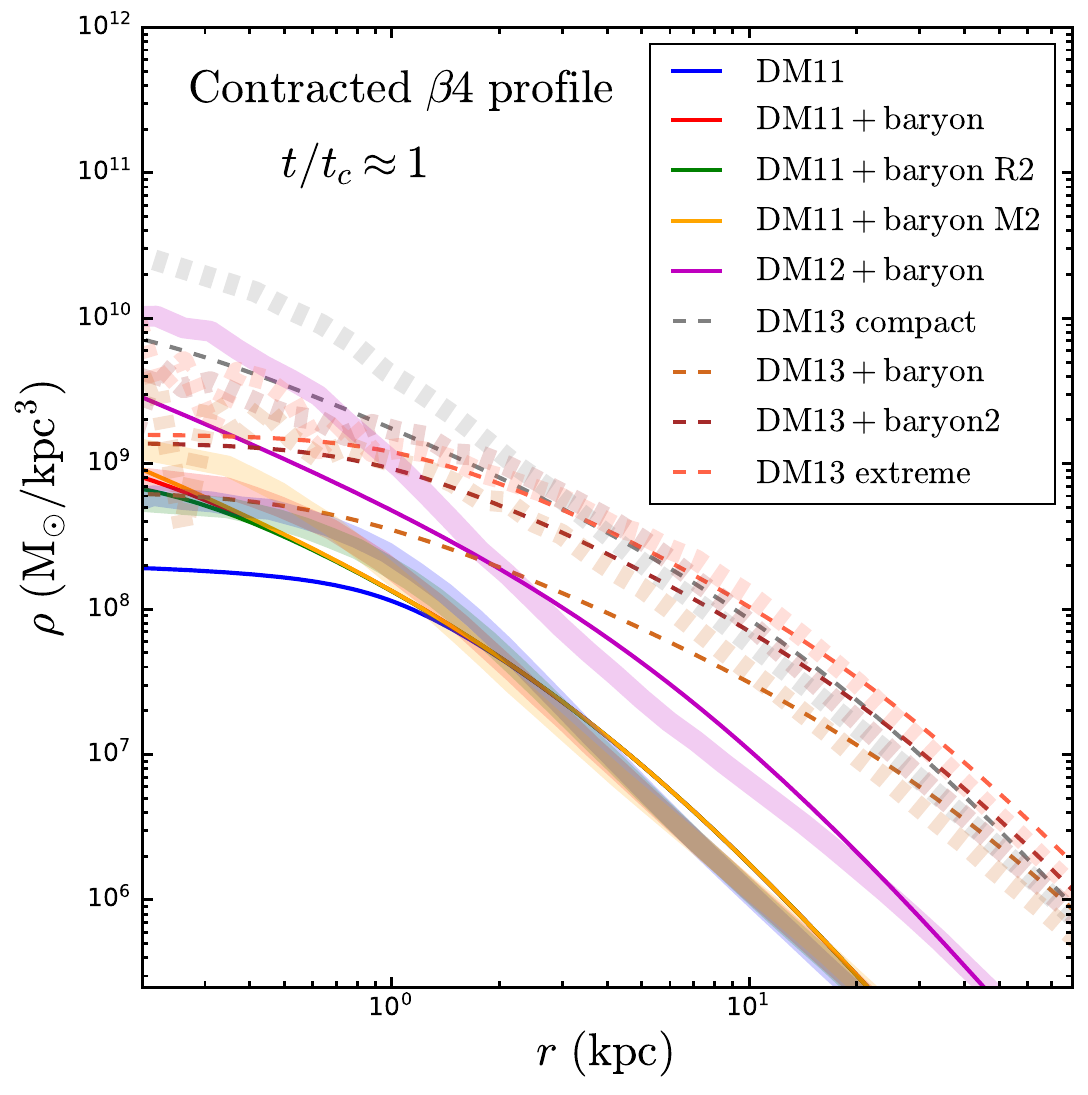}
  \caption{\label{fig:app2} Similar to Fig.~\ref{fig:app1}, but with the colored curves now representing predictions from the $\beta4$ model. 
}
\end{figure*}

By calibrating the Contracted $\beta4$ model to N-body simulations, the effect of contraction is incorporated to some levels. However, the capability of the modeling is limited by the parametric form of the density profile. In Fig.~\ref{fig:app2}, we present the Contracted $\beta4$ model results in a format consistent with that of Fig.~\ref{fig:validx}, reproduced in Fig.~\ref{fig:app1} for ease of comparison. In Fig.~\ref{fig:app2}, a comparison between the model-predicted curves and their simulated counterparts reveals a diminishing agreement towards inner halo regions of halos equal or heavier than $10^{12}~\rm M_{\odot}$, indicating inadequacies in modeling the contraction effect. 

Given this result, one could hypothesize that the variation from the Contracted $\beta4$ profile is fully induced by the part of the contraction effect that is independent of SIDM. The SIDM relevant parts are effectively accounted for in the Contracted $\beta4$ model, including the core collapse time, core size, and the transition from core to outer NFW profile.

To assess the degree of such a separation, we adopt the method outlined in Ref.~\cite{gnedin:2004cx}, incorporating an additional baryonic contraction effect into the Contracted $\beta4$ profile. 
This method employs the adiabatic invariant $M(\bar{r})r$, where $\bar{r} = 0.85 (r/R_{\rm 200})^{0.8} R_{\rm 200}$ represents an orbitally averaged radius. This radius is calibrated against hydrodynamical simulations in Ref.~\cite{gnedin:2004cx}. The choice between using $R_{\rm 200}$ and $R_{\rm vir}$ for calculating $\bar{r}$ has a negligible impact for our purposes, and we consistently use $R_{\rm 200}$ in this analysis.
Starting with a tabulated initial total mass profile $M_i(r)$, we calculate the contracted radial position $r_{f,j}$ using the formula
$$
\frac{r_j}{r_{f,j}} =1-f_b+\frac{M_{b}(\bar{r}_{f,j})}{M_i(\bar{r}_j)},
$$
where $r_j$ represents each radial position, $f_b = M_b/M_{\rm 200}$ is the ratio of baryon mass to total halo mass, and $M_b(r)$ is the baryon mass profile modeled using the Hernquist profile.
The contracted dark matter mass profile at each $r_{f,j}$ is then calculated as $M_h(r_{f,j}) = (1-f_b)M_i(r_j)$.

\begin{figure*}[htbp]
  \centering
  \includegraphics[width=5.33cm]{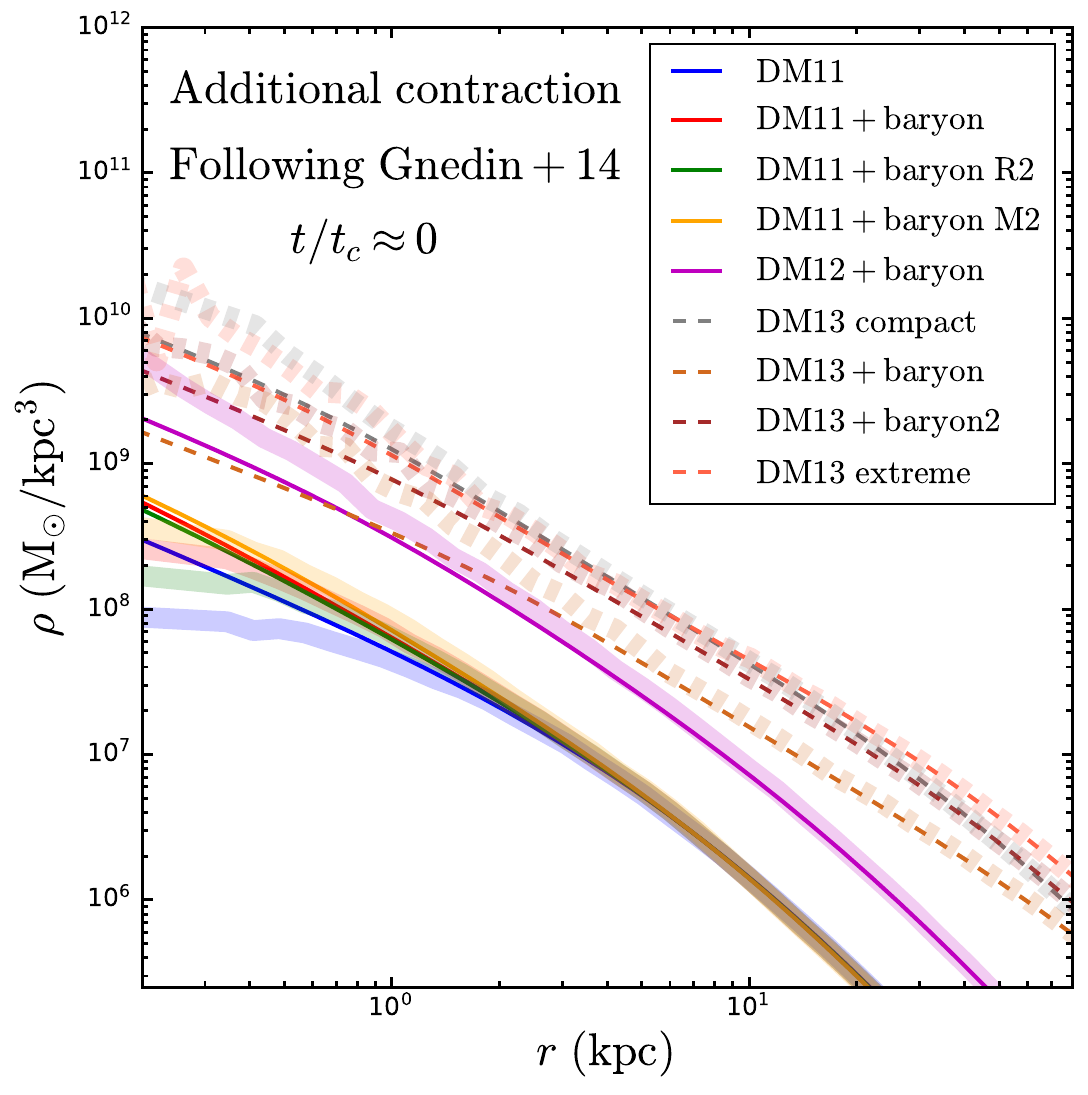}
  \includegraphics[width=5.33cm]{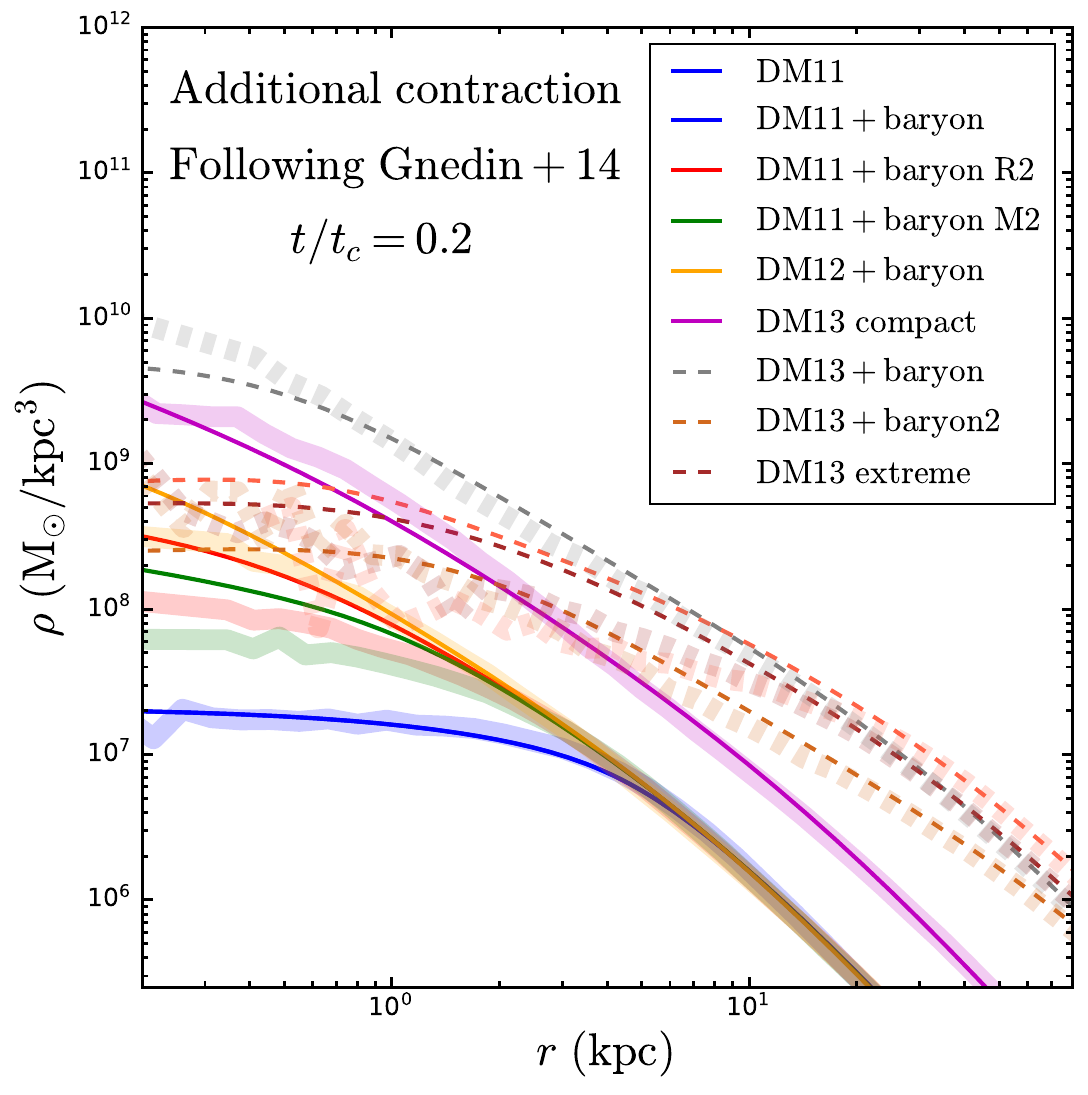}
  \includegraphics[width=5.33cm]{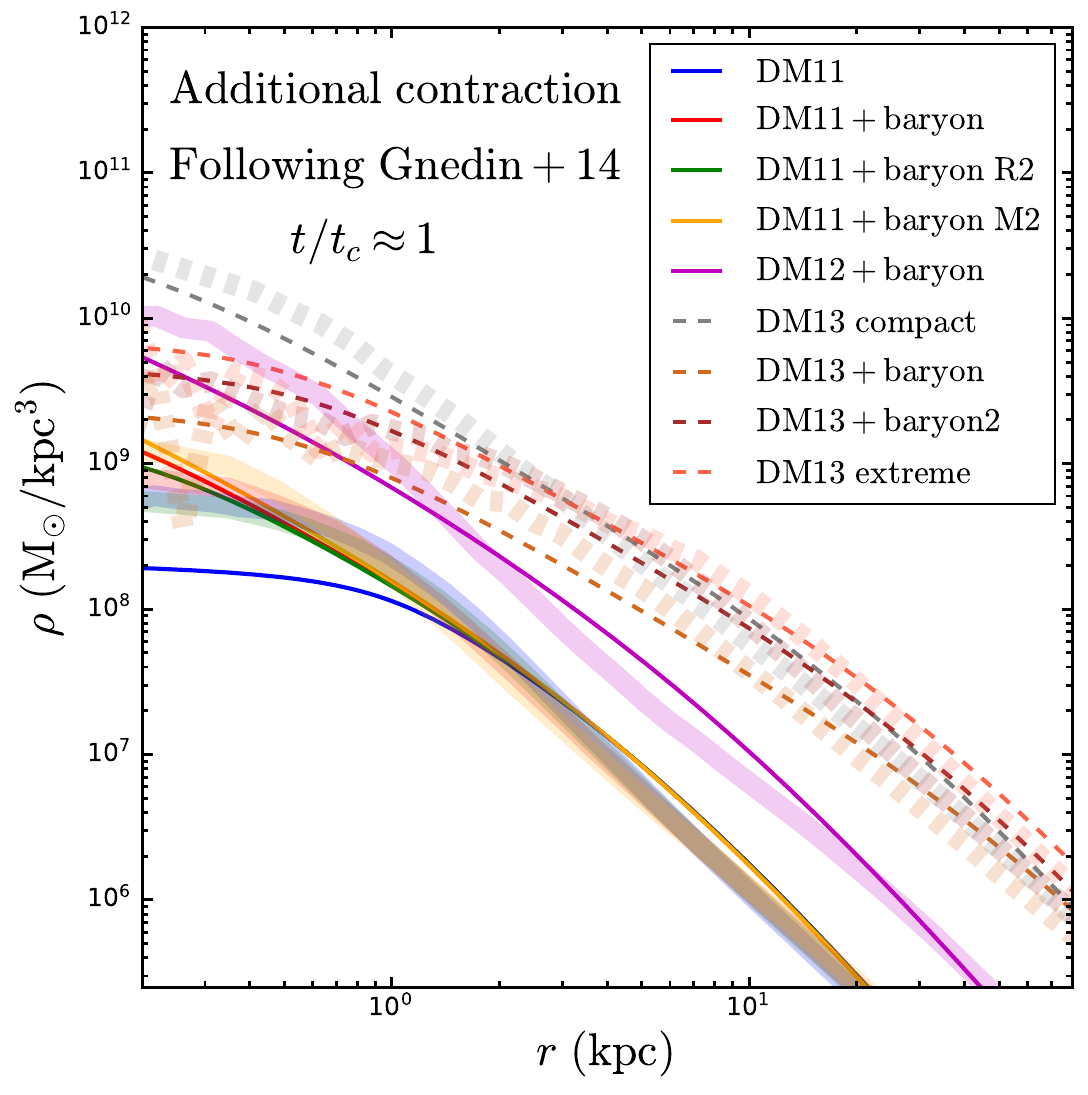}
  \caption{\label{fig:app3} Similar to Fig.\ref{fig:app1}, but here the colored curves depict predictions from the modified $\beta4$ model, which now includes an additional contraction effect following the method introduced in Ref.\cite{gnedin:2004cx}. This enhancement in the model leads to improved alignment between the model predictions and simulation results, except for the DM11 scenarios at $t/t_c \approx 0.2$, where the additional contraction effects overlap with adjustments made in Eq.~(\ref{eq:m1}). Despite these improvements, the overall quality of the model-to-simulation comparison only occasionally matches that of the Cored-DZ model.
}
\end{figure*}

The results are presented in Fig.~\ref{fig:app3}. Upon comparison with Fig.~\ref{fig:app2}, a notable enhancement in agreement is observed, particularly evident in the inner regions for halo masses exceeding $10^{12}~\rm M_{\odot}$. However, for halos with masses $10^{11}~\rm M_{\odot}$, the performance of the Contracted $\beta4$ model is already satisfactory, and the additional contraction appears to overcompensate. 
Notably, the over-estimation around the core transitions in the {\it DM13+baryon2} and {\it DM13 extreme} cases persists after accounting for the inner contraction in Fig.~\ref{fig:app3}. Therefore, the presumed separation does not work well, and the Cored-DZ model outperforms the other two models.

\clearpage

\bibliography{reference}

\end{document}